\documentclass[11pt,a4paper]{article}
\usepackage{preprint}
\newcommand{\SetFigureWidth}[2]{\expandafter\def\csname figwidth@#1\endcsname{#2}}
\let\OriginalIncludeGraphics\includegraphics
\RenewDocumentCommand\includegraphics{O{}m}{%
  \ifcsname figwidth@#2\endcsname
    \OriginalIncludeGraphics[width=\csname figwidth@#2\endcsname]{#2}%
  \else
    \OriginalIncludeGraphics[#1]{#2}%
  \fi}
\SetFigureWidth{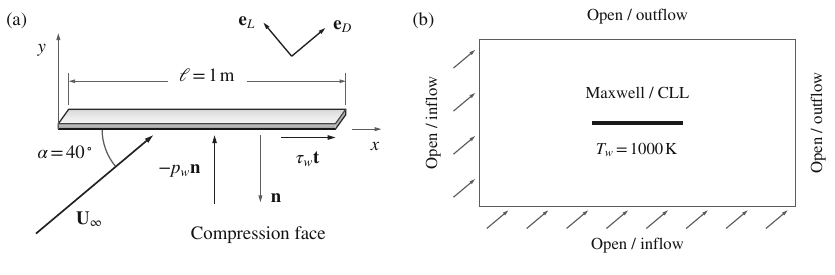}{\textwidth}
\SetFigureWidth{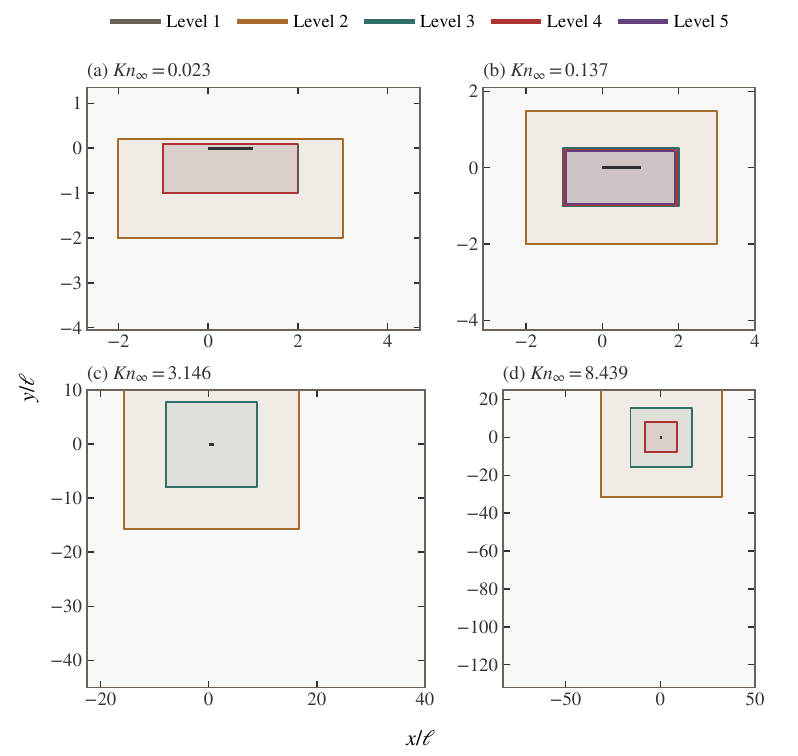}{.68\textwidth}
\SetFigureWidth{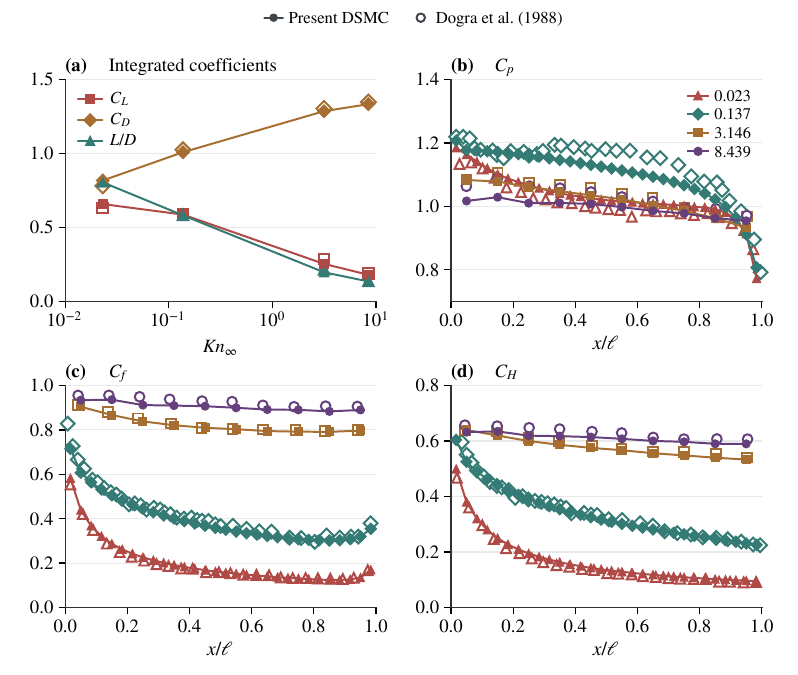}{.83\textwidth}
\SetFigureWidth{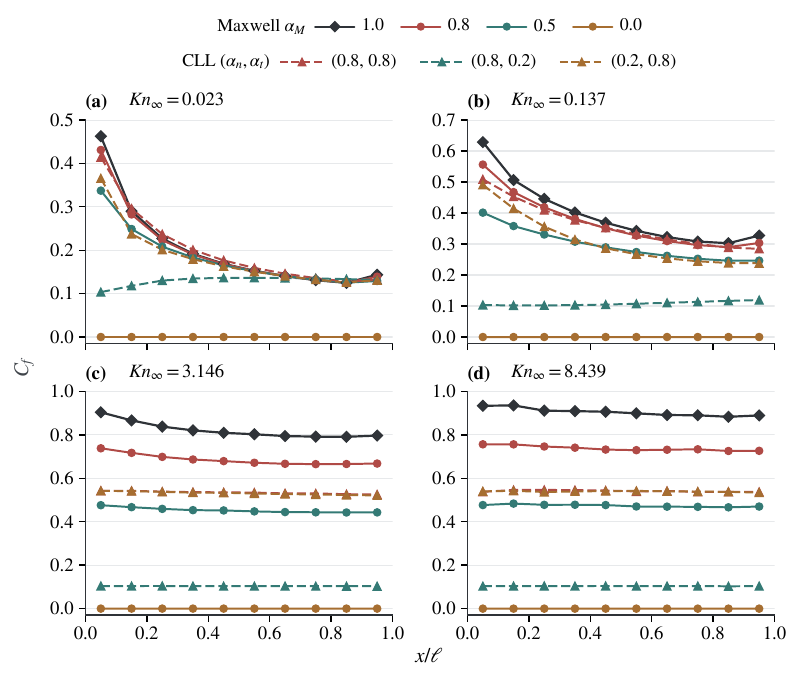}{.82\textwidth}
\SetFigureWidth{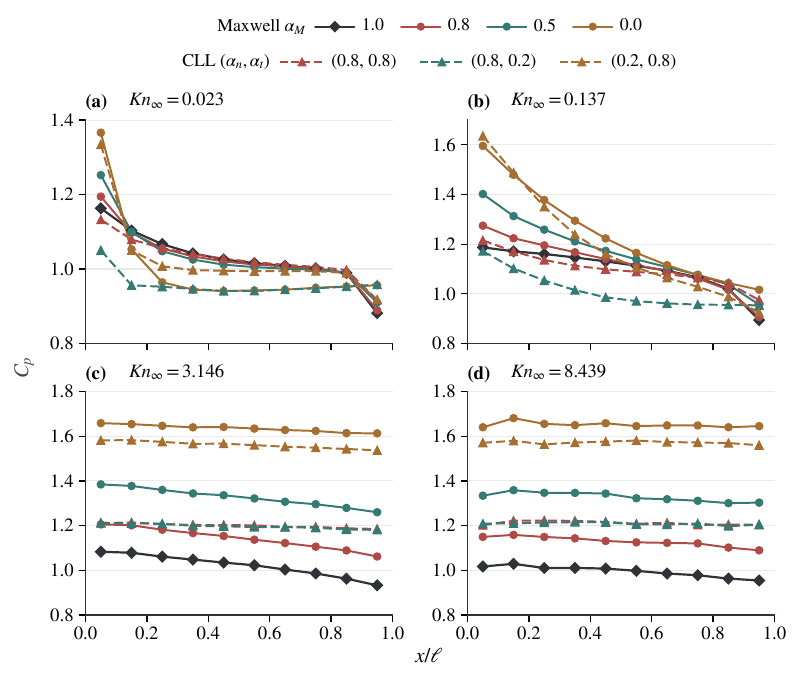}{.82\textwidth}
\SetFigureWidth{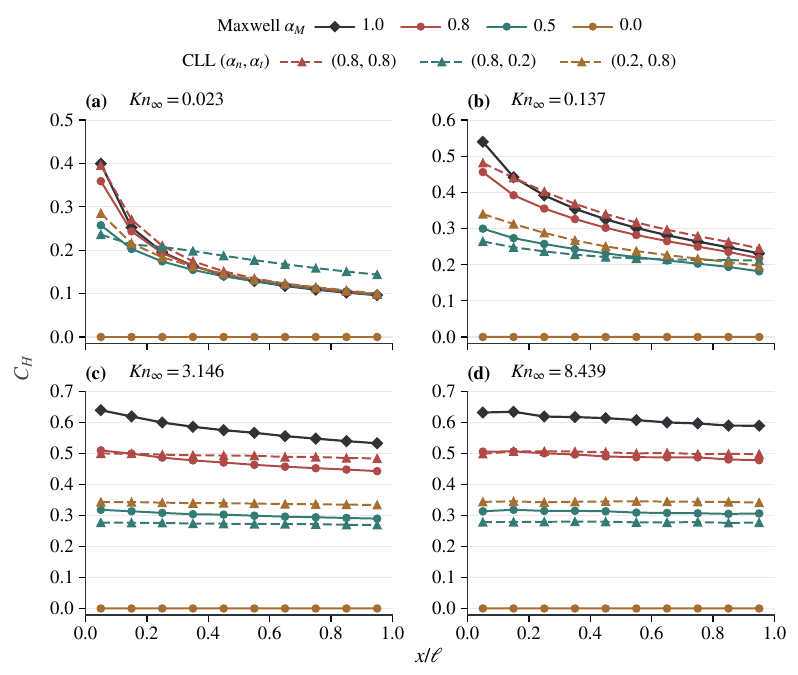}{.82\textwidth}
\SetFigureWidth{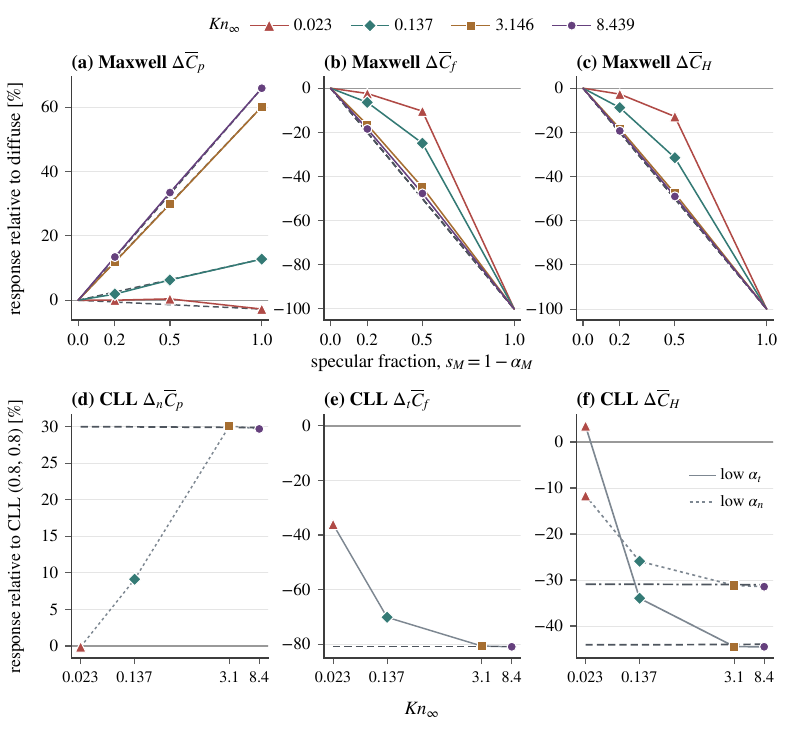}{.90\textwidth}
\SetFigureWidth{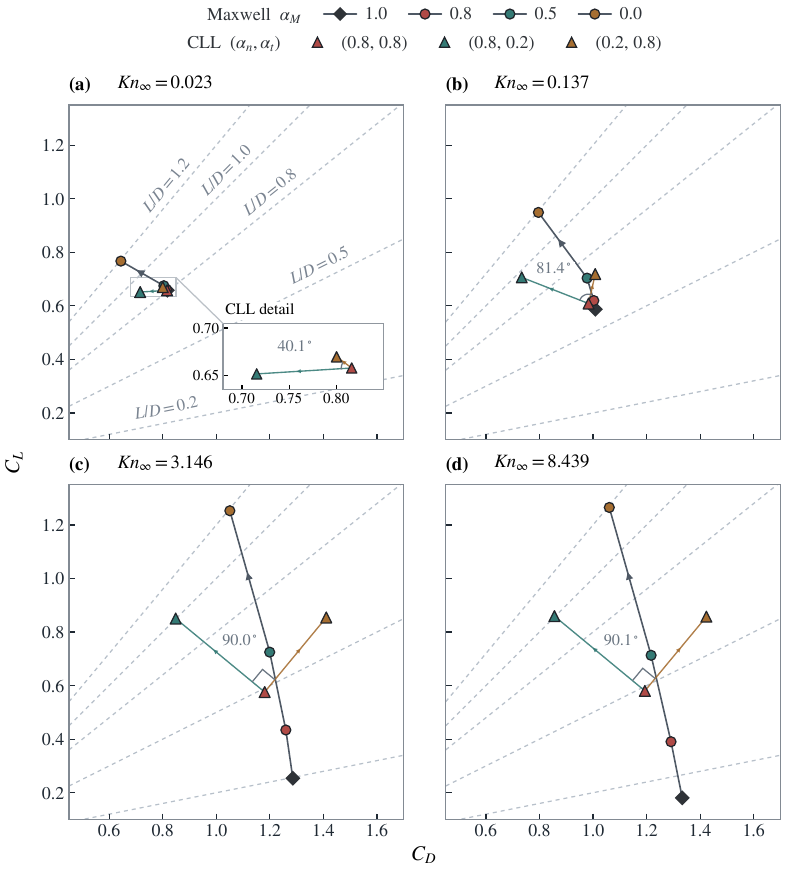}{.78\textwidth}
\SetFigureWidth{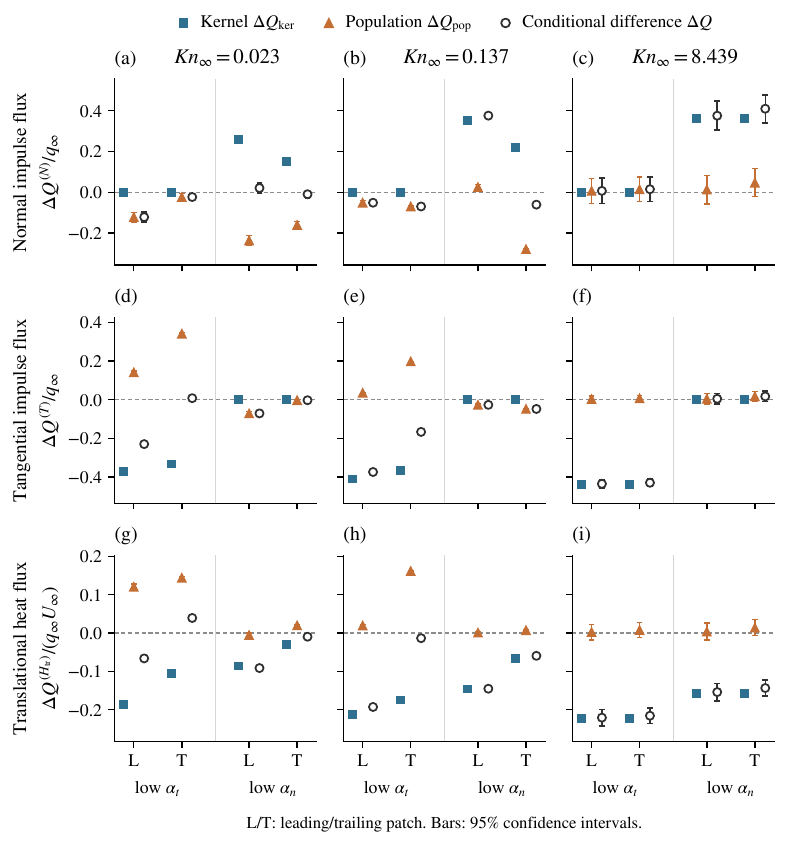}{.88\textwidth}
\SetFigureWidth{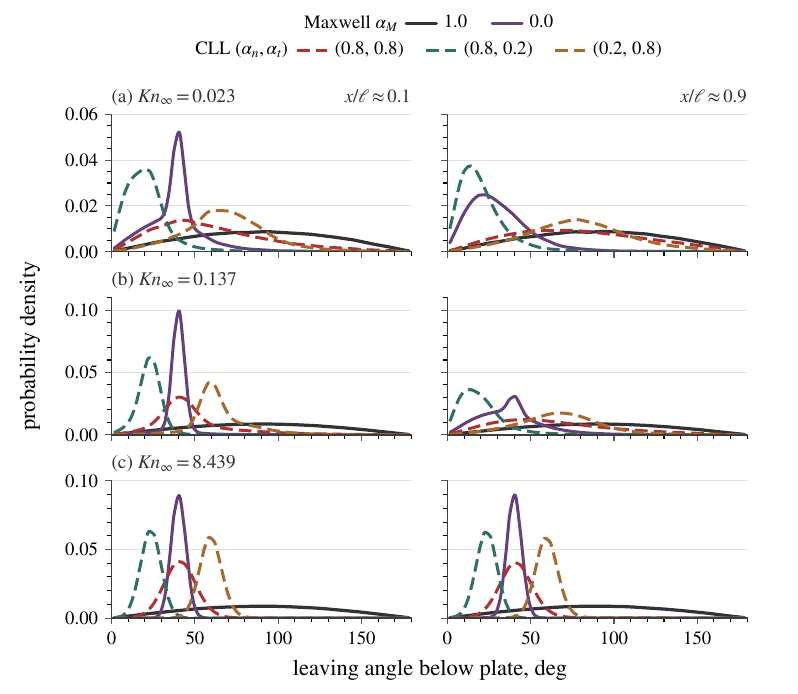}{.85\textwidth}

\newcommand{\PreprintBreak}{\par}

\newenvironment{compactmath}{\par\begingroup
  \setlength{\abovedisplayskip}{4pt plus 1pt minus 1pt}%
  \setlength{\belowdisplayskip}{4pt plus 1pt minus 1pt}%
  \setlength{\abovedisplayshortskip}{3pt plus 1pt}%
  \setlength{\belowdisplayshortskip}{4pt plus 1pt minus 1pt}%
}{\par\endgroup}

\hypersetup{
  pdftitle={Gas–surface scattering and collisional feedback in rarefied hypersonic flow past an inclined flat plate},
  pdfauthor={Einar Nævdal and Sebastian A. Altmeyer},
  pdfsubject={Author preprint: rarefied gas dynamics and gas-surface interaction},
  pdfkeywords={rarefied gas dynamics, direct simulation Monte Carlo, gas-surface interaction, hypersonic flow}
}

\begin{document}
\thispagestyle{opening}
\AddToShipoutPictureBG*{\AtPageUpperLeft{\raisebox{-\height}{\includegraphics[width=\paperwidth]{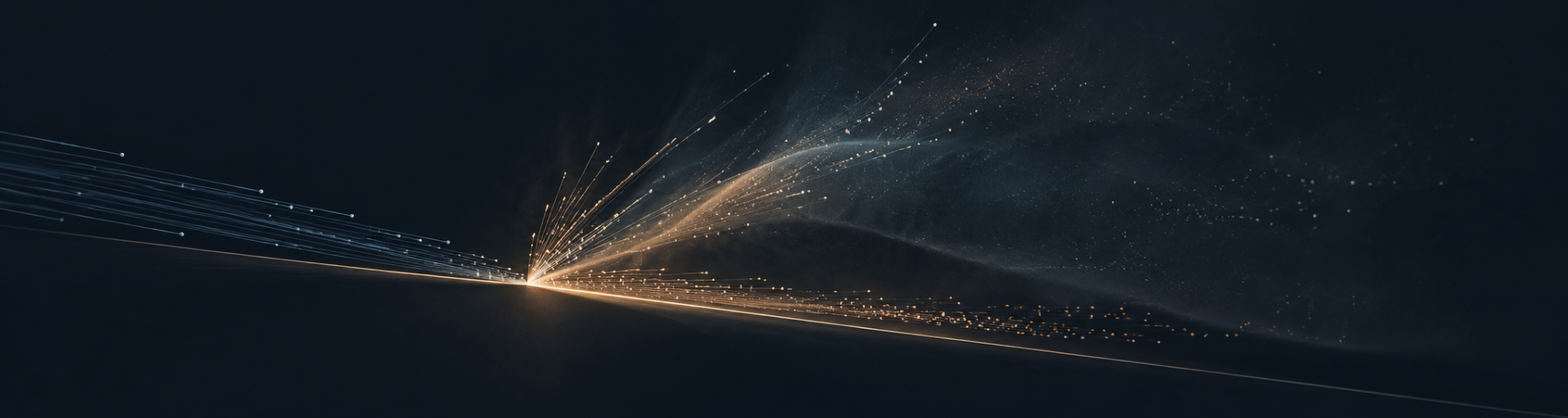}}}}
\begingroup
\vspace*{33mm}
\noindent{\fontfamily{CormorantGaramond-LF}\fontseries{sb}\fontshape{n}\fontsize{22.25}{24.6}\selectfont
\color{ink}\raggedright
Gas--surface scattering and collisional feedback\newline
in rarefied hypersonic flow past an inclined flat plate\par}
\vspace{4mm}
\noindent{\sffamily\fontseries{sb}\fontsize{12}{15}\selectfont\color{ink}%
  \mbox{\orcidlink{0000-0001-6549-4353}\hspace{0.35em}Einar N{\ae}vdal\textsuperscript{\dag}}%
  \hspace{1.7em}%
  \mbox{\orcidlink{0000-0001-5964-0203}\hspace{0.35em}Sebastian A. Altmeyer\vphantom{\textsuperscript{*}}}\par}
\vspace{2.4mm}
\noindent
\begin{minipage}[t]{0.645\textwidth}
  \sffamily\fontsize{9}{11.5}\selectfont\color{muted}\raggedright
  \noindent Departament de F\'isica, Universitat Polit\`ecnica de Catalunya--BarcelonaTech (UPC),\newline
  08034~Barcelona, Spain\par
  \vspace{1.8mm}
  \noindent\textsuperscript{\dag}\hspace{0.12em}Current address: European Space Agency (ESA), European Space Research and Technology Centre (ESTEC), Noordwijk, The Netherlands.\par
\end{minipage}\hfill
\begin{minipage}[t]{0.32\textwidth}
  \sffamily\fontsize{9}{11.5}\selectfont\color{muted}\raggedleft
  {\fontseries{sb}\selectfont\color{ink}Correspondence}\par
  Sebastian A. Altmeyer\par
  \href{mailto:sebastian.andreas.altmeyer@upc.edu}{sebastian.andreas.altmeyer@upc.edu}\par
\end{minipage}\par
\vspace{3.5mm}
\noindent{\color{hairline}\rule{\textwidth}{0.35pt}}\par
\vspace{6mm}
\noindent{\bfseries\fontsize{11.5}{14}\selectfont\color{ink}Abstract}\par
\vspace{2mm}
{\fontsize{11.3}{15.4}\selectfont\noindent Through intermolecular collisions, a gas--surface scattering law can reshape the incident molecular population on which it acts. We examine this feedback using direct simulation Monte Carlo for hypersonic flow past a $40^\circ$ inclined flat plate at four benchmark conditions spanning $0.023\leq Kn_\infty\leq8.439$. Maxwell and Cercignani--Lampis--Lord (CLL) scattering are assessed through fixed-incidence moment references and evaluation of alternative kernels on the same recorded non-reactive incident populations. At $Kn_\infty=0.023$, Maxwell shear and heat transfer depart from diffuse--specular endpoint interpolation by up to $40\%$ of their diffuse-reference values. In the trailing patch, the population change induced by reduced tangential CLL accommodation overturns the fixed-incidence reductions in both shear and translational heat transfer. At the two highest conditions, the CLL momentum responses become nearly componentwise, producing nearly orthogonal displacements in the drag--lift plane. Nevertheless, a total-energy constraint required by common incidence across wall laws retains a defect of $3.1\%$ of the diffuse heat-transfer coefficient at $Kn_\infty=8.439$. Recovery of selected response relations therefore does not establish common incidence or a free-molecular flow. A cold-beam symmetry makes lift invariant under exchange of the normal and tangential CLL accommodation coefficients, but not drag. This provides a limiting explanation for the simulated asymmetric cases: at $Kn_\infty=8.439$, they yield nearly equal lift while one produces $66\%$ more drag. Population feedback and force projection thus determine whether the conditional scattering response is reinforced, obscured or reversed in the macroscopic observables.\par}
\vspace{7mm plus 1fill}
\noindent{\color{hairline}\rule{\textwidth}{0.35pt}}\par
\vspace{2mm}
\begingroup
\fontsize{9.3}{12.8}\selectfont\color{ink}
\noindent{\bfseries Contents}\hfill {\itshape Page}\par
\vspace{1.2mm}
\newcommand{\OpeningContentsEntry}[2]{%
  \noindent\hyperref[#1]{\makebox[1.5em][l]{\ref*{#1}.}#2}%
  \nobreak\hspace{0.6em}{\color{muted}\leaders\hbox to 5pt{\hss.\hss}\hfill}%
  \nobreak\hspace{0.6em}\hyperref[#1]{\pageref*{#1}}\par}
\OpeningContentsEntry{SEC:intro}{Introduction}
\OpeningContentsEntry{SEC:meth}{Methodology}
\OpeningContentsEntry{SEC:res}{Results and discussion}
\OpeningContentsEntry{SEC:conclu}{Conclusions}
\vspace{3mm}
\noindent{\fontsize{8.8}{11}\selectfont\color{muted}\textit{Data, code and movies:}\enspace\href{https://doi.org/10.5281/zenodo.21360802}{doi:10.5281/zenodo.21360802}}\par
\vspace{3mm}
\endgroup
\endgroup
\clearpage

\section{Introduction}
\label{SEC:intro}

In a rarefied flow, changing how molecules leave a surface can also change how they reach it. The gas--surface interaction (GSI) kernel specifies a reflected distribution conditional on the incident molecular state, whereas pressure, shear and heat transfer follow from averages of momentum and energy exchange over the incident flux \citep{Cercignani1971KineticModels,Lord1991ExtensionsCLL}. When intermolecular collisions are appreciable, reflected molecules can modify the surrounding gas and hence subsequent incidence at the surface \citep{Lord1992IncompleteAccommodation}. Changing accommodation can therefore alter both the conditional exchange at the wall and the rate and distribution of surface impacts. The difficulty is not only to predict a load, but to identify which physical assumption produced it.

This distinction matters both to spacecraft design and to the interpretation of flight measurements. In very low Earth orbit (VLEO), lower altitude can improve Earth-observation resolution or reduce payload size, but stronger atmospheric drag constrains orbital lifetime \citep{Crisp2020BenefitsVLEO,Jiang2023VLEODrag,Hild2022VLEOOptimisation}. GSI modelling also enters atmospheric inference through the drag coefficient used to recover density from measured acceleration. For Swarm A and C, revising the adopted energy-accommodation coefficient changed the recovered densities by approximately $4$--$5\%$ \citep{March2021GSIInfluence}. The Satellite for Orbital Aerodynamics Research (SOAR) concept was designed to recover drag and lift coefficients for different materials and incidence angles using orbit, attitude and in-situ atmospheric measurements \citep{Crisp2021SOAR}. In atmospheric-entry research, flight validation also motivated the European Experimental Reentry Testbed (EXPERT) programme and comparisons of predicted Apollo afterbody heating with flight measurements \citep{Muylaert2003EXPERT,Walpot2012ApolloBaseFlow}.

The Maxwell and Cercignani--Lampis--Lord (CLL) models provide complementary descriptions of molecular reflection. Maxwell reflection is an ensemble mixture of specular events and diffuse re-emission at the wall temperature \citep{HedahlWilmoth1995NASA,PadillaBoyd2009GSIAssessment}. CLL instead factors the conditional scattering law into independent normal and tangential parts, with separate translational-energy accommodation coefficients in the convention used here \citep{Cercignani1971KineticModels,Lord1991ExtensionsCLL}. Independence within the kernel does not, however, guarantee independence in the self-consistent force response: a change in normal accommodation can also change the population carrying tangential momentum towards the wall. The wall law must therefore be examined together with the incident population generated by the surrounding flow.

Real surfaces introduce a further difficulty because their scattering behaviour depends on incident energy and angle, gas species, surface material, temperature and condition. Molecular-beam measurements often exhibit finite-width scattering lobes that cannot be represented by a simple specular--diffuse mixture \citep{Livadiotti2020GSIReview,Murray2017GasSurfaceScattering}. Recent work has explored conditional statistical representations of scattering data \citep{Schuette2026EnhancedScattering} and molecular-dynamics-informed hybrid kernels for atomic oxygen \citep{ApparPoovathingal2026VLEOGSIModel}. In a rounded-wedge calculation, \citet{LiuEtAl2021MDScatteringDatabase} found that predictions obtained using CLL and a molecular-dynamics-derived scattering database diverged increasingly with rarefaction. Improving the kernel and explaining its macroscopic consequences are therefore related but distinct tasks.

\PreprintBreak
Earlier studies provide distinct precedents for this molecular-to-macroscopic comparison. \citet{CercignaniLampis1972FlatPlate} calculated flat-plate drag, lift and lift-to-drag ratio using a two-parameter scattering model in free-molecular flow. \citet{Dogra1988FlatPlate} subsequently showed that transitional-flow effects could be mistaken for a specular-reflection contribution when flight measurements were interpreted using free-molecular theory. In a zero-incidence supersonic flat-plate study, \citet{CercignaniFrezzotti1989FlatPlate} found nearly coincident density, velocity and temperature fields for selected Maxwell and Cercignani--Lampis cases, despite differences in energy fluxes. \citet{Lord1992IncompleteAccommodation} demonstrated a direct connection between incident-population changes and aerodynamic response: in a rarefied cylinder calculation, reduced normal accommodation increased normal momentum transfer per collision at the front face, but a reduction in incident mass flux largely suppressed the drag increase predicted in free-molecular flow. \citet{HedahlWilmoth1995NASA} found that Maxwell and CLL could produce qualitatively similar single-plate forces and heat transfer despite different flow fields. \citet{PadillaBoyd2009GSIAssessment} likewise assessed the models through boundary-layer velocity profiles and molecular scattering distributions. These studies establish both the importance of accommodation and the difficulty of inferring its action from macroscopic agreement alone.

The question addressed here is more specific: when do changes in pressure, shear and heat transfer follow the conditional response prescribed by the scattering law, and when are they altered by changes in the incident population? At fixed incidence, Maxwell surface moments interpolate linearly between the diffuse and specular endpoints, while CLL supplies separate momentum and translational-energy references. Under the prescribed internal-mode and catalytic treatment, an additional total-energy constraint links the three CLL cases to diffuse Maxwell reflection. Evaluating alternative kernels on each acquired non-reactive incident population gives a finite-change decomposition into kernel and incident-population contributions. Comparison with finite-temperature freestream references further distinguishes distortion already present in the baseline incident population from the additional population change associated with a wall-law perturbation.

We use direct simulation Monte Carlo (DSMC) to resolve molecular transport and intermolecular collisions without imposing a Navier--Stokes--Fourier constitutive closure \citep{Bird1994,OuChen2020BluntPlates,Plimpton2019PoF}. The entry-flow benchmark provides a canonical test: a nominally zero-thickness plate of length $\ell=1\,\mathrm{m}$ at $40^\circ$ incidence in a $7.5\,\mathrm{km\,s^{-1}}$ freestream, with a wall temperature of $1000\,\mathrm{K}$. The freestream Knudsen number is $Kn_\infty=\lambda_\infty/\ell$, where $\lambda_\infty$ denotes the mean free path; the prescribed values are $0.023$, $0.137$, $3.146$ and $8.439$. Freestream temperature and the reconstructed mixture composition also vary between benchmark states, so the campaign isolates wall-law changes within each state rather than constituting a density-only rarefaction continuation. Following reproduction of the diffuse benchmark, a Maxwell sweep from diffuse to specular reflection and three selected CLL pairs are compared while the freestream state, gas--gas collision model, catalytic wall treatment, numerical controls and statistical processing remain fixed within each condition. These are controlled scattering-law probes, not calibrated spacecraft-material models.

The analysis connects compression-side pressure, skin friction and heat transfer to incident impact rates, exchange per incident molecule, outgoing-angle distributions and near-wall density fields, before projecting the force components into drag and lift. Sampling uncertainty is propagated from correlated DSMC records. At $Kn_\infty=0.023$, the two-patch non-reactive analysis shows that population changes nearly cancel the conditional pressure increase caused by reduced normal accommodation; in the trailing patch, they overturn the shear and translational-heat reductions caused by reduced tangential accommodation at fixed incidence. At the two highest conditions, the principal CLL momentum responses approach their kernel references, yet the total-energy constraint retains a resolved defect. Recovery of selected response relations therefore does not establish common incidence or a free-molecular flow. Finally, a cold-beam exchange symmetry provides a limiting explanation for the near-equal lift and unequal drag of the asymmetric CLL cases, while finite temperature and changes in incidence can break this equality.

Section~\ref{SEC:meth} describes the configuration, scattering models, kernel references and uncertainty treatment. Section~\ref{SEC:res} presents the benchmark comparison, surface responses, molecular diagnostics and force decomposition. The conclusions are given in section~\ref{SEC:conclu}.
\section{Methodology}
\label{SEC:meth}

At each prescribed Knudsen number, only the wall-scattering law was varied; the freestream state, gas--gas collision model, catalytic treatment, numerical controls, sampling procedure and statistical analysis were held fixed. All production simulations used the 24 September 2025 release of the open-source DSMC solver SPARTA (Stochastic PArallel Rarefied-gas Time-accurate Analyzer) \citep{Plimpton2019PoF,SPARTA24Sep2025}. The accompanying data release links each case to its solver inputs, geometry, collision data, raw block records, processed results, uncertainty analysis and figure-generation routines \citep{Naevdal2026DataRelease}.

\subsection{Benchmark configuration and wall models}
\label{subsec:numerical_setup}

\subsubsection{Simulation configuration}

The configuration follows the inclined-flat-plate benchmark of \citet{Dogra1988FlatPlate}. A plate of nominal length $\ell=1~\mathrm{m}$ was represented by a closed rectangle of numerical thickness $10^{-4}~\mathrm{m}$, with 30 equal line elements on each long face. The plate tangent defines the global $x$ direction, and the freestream velocity is $\boldsymbol{U}_\infty=U_\infty(\cos\alpha,\sin\alpha,0)$, with $U_\infty=7.5~\mathrm{km\,s^{-1}}$ and $\alpha=40^\circ$. The lower-face outward normal points in the negative $y$ direction, so $\boldsymbol{U}_\infty\bcdot\boldsymbol{n}<0$ there. All reported surface distributions and integrated coefficients refer to this compression face; the opposite face and the two closing end caps are excluded from the reported loads.

\begin{figure}
    \centering
    \includegraphics[width=\textwidth]{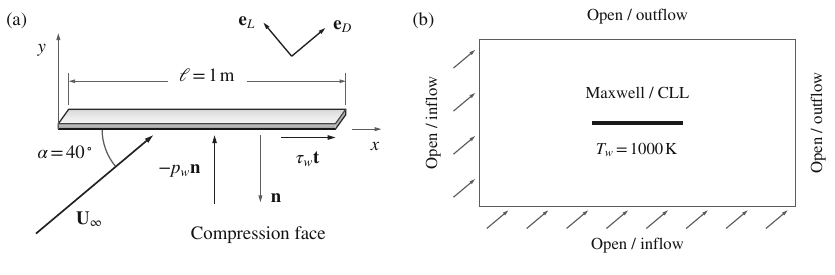}
    \caption{Schematic of the two-dimensional inclined-flat-plate configuration: (a) plate-fixed coordinates, incidence and force conventions, shown in perspective; (b) computational domain and boundary conditions. The unit vectors $\boldsymbol{n}$ and $\boldsymbol{t}$ denote the outward normal and plate tangent; $\boldsymbol{e}_D$ and $\boldsymbol{e}_L$ denote the positive drag and lift directions. All reported surface distributions and integrated loads refer to the lower compression face.}
    \label{fig:configuration}
\end{figure}

Continuous freestream inflow was imposed through the $x_{\mathrm{lo}}$ and $y_{\mathrm{lo}}$ boundaries, with the remaining in-plane boundaries open. Each production domain was initialised empty and populated by freestream injection before selection of the retained sampling interval. Figure~\ref{fig:configuration} defines the coordinate, force-sign and boundary conventions.

The geometry, incidence, freestream speed and wall temperature, $T_w=1000~\mathrm{K}$, were fixed across the four benchmark conditions, $Kn_\infty\in\{0.023,0.137,3.146,8.439\}$. These plate-length-based values, $Kn_\infty=\lambda_\infty/\ell$, are retained as prescribed benchmark labels rather than recomputed using a separate multicomponent mean-free-path convention for the reconstructed gas. The nominal altitudes and the number densities, temperatures and effective molecular masses used to reconstruct the freestream states are listed in table~\ref{tab:freestream_states}.
\begin{table}[!b]
\centering
\setlength{\tabcolsep}{7pt}
\renewcommand{\arraystretch}{1.08}
\begin{tabular}{
S[table-format=1.3]
S[table-format=3.0]
S[table-format=1.2e2]
S[table-format=3.1]
S[table-format=1.3e-2]
}
\toprule
{$Kn_\infty$}
& {$h$ [\si{\kilo\metre}]}
& {$n_\infty$ [\si{\per\metre\cubed}]}
& {$T_\infty$ [\si{\kelvin}]}
& {$m_\infty$ [\si{\kilogram}]} \\
\midrule
0.023 & 90  & 7.14e19 & 188.0 & 4.784e-26 \\
0.137 & 100 & 1.20e19 & 194.3 & 4.692e-26 \\
3.146 & 120 & 5.22e17 & 367.8 & 4.344e-26 \\
8.439 & 130 & 1.95e17 & 499.7 & 4.225e-26 \\
\bottomrule
\end{tabular}
\caption{Prescribed benchmark freestream states used to reconstruct the neutral mixture; $h$ is the nominal altitude and $m_\infty$ the effective molecular mass.}
\label{tab:freestream_states}
\vspace{16pt}
\end{table}
The benchmark does not supply a complete solver-ready species composition and collision dataset. We therefore used a neutral five-species reconstruction comprising $\mathrm{O}$, $\mathrm{N}$, $\mathrm{O}_2$, $\mathrm{N}_2$ and Ar at each prescribed temperature, rather than an exact species-level recreation of the original DSMC gas. Fractions were chosen to match the effective molecular mass while retaining O, O$_2$ and N$_2$ as the dominant constituents. Trace atomic nitrogen preserves the nitrogen surface-recombination pathway, a small Ar fraction closes the mixture mass, and He is excluded. The case-specific fractions and deterministic N$_2$ closure are archived in the data release \citep{Naevdal2026DataRelease}. Gas-phase chemistry, ionisation and radiation were omitted; the collision and internal-relaxation models are specified in section~\ref{subsec:dsmc_numerical_implementation}.

Diffuse Maxwell reflection, $\alpha_M=1.0$, is both the reproduced benchmark condition and the reference for wall-model comparisons. All cases retain the same probability-based catalytic treatment, with $\mathrm{O}\rightarrow\mathrm{O}_2$ and $\mathrm{N}\rightarrow\mathrm{N}_2$ probabilities of $4.9\times10^{-3}$ and $7.7\times10^{-3}$, respectively \citep{Dogra1988FlatPlate}. The benchmark specifies these probabilities but not a solver-ready surface-reaction mechanism. The pathways were therefore represented by SPARTA's probability-style exchange formulation, which returns O$_2$ and N$_2$ to the gas without resolving adsorbed intermediates or an explicit two-atom surface reaction.

The Maxwell sweep, $\alpha_M\in\{1.0,0.8,0.5,0.0\}$, spans diffuse to specular reflection. The CLL sweep uses $(\alpha_n,\alpha_t)\in\{(0.8,0.8),(0.8,0.2),(0.2,0.8)\}$. The values $0.2$ and $0.8$ provide well-separated interior levels of energy accommodation, so the asymmetric pairs impose equal one-parameter reductions of $0.6$ from $(0.8,0.8)$ without invoking the zero- or complete-accommodation limits.

The selected CLL pairs are controlled sensitivity probes, not calibrated descriptions of specific surface materials. Their purpose is to separate the consequences of normal and tangential accommodation under otherwise identical physical, numerical, and statistical conditions. The corresponding Maxwell and CLL scattering kernels are defined in section~\ref{subsubsec:wall_scattering_models}.

\subsubsection{Wall-scattering models}
\label{subsubsec:wall_scattering_models}

In DSMC, the wall model is a conditional mapping from an incident molecular state to a probability distribution of reflected states. Let $\boldsymbol{v}_i\bcdot\boldsymbol{n}<0$ and $\boldsymbol{v}_r\bcdot\boldsymbol{n}>0$ denote incident and reflected velocities relative to the outward wall normal $\boldsymbol{n}$. The corresponding distribution functions are related through the scattering kernel $K(\boldsymbol{v}_i\rightarrow\boldsymbol{v}_r)$,
\begin{equation}
f_r(\boldsymbol{v}_r)\left(\boldsymbol{v}_r\bcdot\boldsymbol{n}\right)=\int_{\boldsymbol{v}_i\bcdot\boldsymbol{n}<0}K(\boldsymbol{v}_i\rightarrow\boldsymbol{v}_r)f_i(\boldsymbol{v}_i)\left|\boldsymbol{v}_i\bcdot\boldsymbol{n}\right|\,\mathrm{d}\boldsymbol{v}_i,\qquad \boldsymbol{v}_r\bcdot\boldsymbol{n}>0.
\label{eq:scattering_kernel_bc}
\end{equation}
The Maxwell and CLL kernels defined below describe the conditional non-reactive translational-scattering branch. The fixed catalytic recombination pathway was applied separately through SPARTA's probability-based surface-reaction model. For non-reactive events, the selected kernel determines the complete outgoing translational distribution rather than only its mean momentum or energy exchange. The kernels considered here are non-negative, normalised over the outgoing half-space and reciprocal at thermal equilibrium \citep{Cercignani1971KineticModels,Lord1991ExtensionsCLL}.

The Maxwell model represents reflection as a binary ensemble mixture of specular reflection and diffuse re-emission,
\begin{equation}
K_M(\boldsymbol{v}_i\rightarrow\boldsymbol{v}_r)=(1-\alpha_M)\delta\!\left(\boldsymbol{v}_r-\boldsymbol{v}_{\mathrm{sp}}\right)+\alpha_M K_D(\boldsymbol{v}_r;T_w),\qquad \boldsymbol{v}_{\mathrm{sp}}=\boldsymbol{v}_i-2(\boldsymbol{v}_i\bcdot\boldsymbol{n})\boldsymbol{n},
\label{eq:maxwell_scattering_kernel}
\end{equation}
where $\delta$ is the Dirac delta and $K_D$ is the flux-weighted half-range Maxwellian at $T_w$. Each impact is either specular or fully diffuse: $\alpha_M$ sets the probability of diffuse re-emission, not partial accommodation within an individual event. Specular events retain incident rotational and vibrational states, whereas diffuse events resample those states from equilibrium distributions at $T_w$. The endpoints $\alpha_M=0$ and $1$ recover purely specular and diffuse Maxwell reflection, respectively \citep{Bird1994,HedahlWilmoth1995NASA}.

The CLL model replaces the binary mixture with a continuous scattering law whose normal and two tangential velocity transformations are independent conditional on the incident state \citep{Cercignani1971KineticModels,Lord1991ExtensionsCLL}. In the local surface frame used by SPARTA \citep{SPARTA24Sep2025},
\begin{equation}
K_{\mathrm{CLL}}(\boldsymbol{v}_i\rightarrow\boldsymbol{v}_r)=K_n(v_{i,n}\rightarrow v_{r,n})\prod_{j=1}^{2}K_t(v_{i,t_j}\rightarrow v_{r,t_j}),
\label{eq:cll_kernel_factorization}
\end{equation}
with thermal speed $c_w=\sqrt{2k_B T_w/m_{\mathrm{mol}}}$, where $k_B$ is Boltzmann's constant and $m_{\mathrm{mol}}$ the incident molecular~mass.

The component kernels are
\begin{align}
K_n&=\frac{2v_{r,n}}{\alpha_n c_w^2}\exp\!\left[-\frac{v_{r,n}^2+(1-\alpha_n)v_{i,n}^2}{\alpha_n c_w^2}\right]\mathrm{I}_0\!\left(\frac{2\sqrt{1-\alpha_n}\,|v_{i,n}|v_{r,n}}{\alpha_n c_w^2}\right),\qquad v_{r,n}>0,
\label{eq:cll_normal_kernel}\\
K_t&=\frac{1}{c_w\sqrt{\upi\alpha_t}}\exp\!\left[-\frac{\left(v_{r,t_j}-\sqrt{1-\alpha_t}\,v_{i,t_j}\right)^2}{\alpha_t c_w^2}\right].
\label{eq:cll_tangential_kernel}
\end{align}
Here $\mathrm{I}_0$ is the modified Bessel function of the first kind of order zero. Both $\alpha_n$ and $\alpha_t$ are translational-energy accommodation coefficients in the convention used here. On the branch implemented here, the tangential momentum accommodation coefficient is $\sigma_t=1-\sqrt{1-\alpha_t}$, so $\alpha_t=\sigma_t(2-\sigma_t)$ \citep{Lord1991ExtensionsCLL}. The distinction is apparent in the conditional moments:
\begin{equation}
\left\langle v_{r,t_j}\mid\boldsymbol{v}_i\right\rangle
=
\sqrt{1-\alpha_t}\,v_{i,t_j},
\qquad
\left\langle v_{r,n}^{2}\mid\boldsymbol{v}_i\right\rangle
=
(1-\alpha_n)v_{i,n}^{2}+\alpha_n c_w^{2}.
\label{eq:cll_conditional_memory_moments}
\end{equation}
At fixed incidence, the mean reflected tangential velocity scales with $\sqrt{1-\alpha_t}$, whereas the incident contribution to the mean reflected normal kinetic energy scales with $1-\alpha_n$. The limits $(\alpha_n,\alpha_t)\rightarrow(0,0)$ and $(1,1)$ recover translational specular reflection and diffuse re-emission at $T_w$, respectively; intermediate values yield finite-width lobular distributions.

For each wall event, including reactive events generated by the separately applied catalytic pathway, the molecular momentum and energy transferred to the wall are obtained from the imbalance between the incident state and all outgoing products,
\begin{equation}
\Delta\boldsymbol{p}_w
=
\boldsymbol{p}_{\mathrm{in}}
-
\sum_s\boldsymbol{p}_{\mathrm{out},s},
\qquad
\Delta E_w
=
E_{\mathrm{in}}
-
\sum_s E_{\mathrm{out},s}
+
E_{\mathrm{chem}}.
\label{eq:wall_molecular_exchange}
\end{equation}
The sums include all outgoing products. Molecular energies comprise translational, rotational and vibrational contributions; $E_{\mathrm{chem}}$ is the separately prescribed chemical-energy deposit, positive for transfer to the wall and zero for non-reactive events. Summing the weighted impulses and energy transfers over impacts and dividing by area and time gives pressure, shear and heat flux. SPARTA's total wall-energy tally, $e_{\mathrm{tot}}$, already includes the chemical contribution \citep{SPARTA24Sep2025}. The oxygen and nitrogen deposits, $8.215\times10^{-19}$~J and $1.563\times10^{-18}$~J, are inputs to the solver reconstruction, not quantities supplied by the benchmark, and were not added again in post-processing.

The Maxwell sweep varies $\alpha_M$ and hence the fraction of events that accommodate internal energy. In the CLL sweep, only $(\alpha_n,\alpha_t)$ varies; rotational and vibrational accommodation remain unity, and the catalytic treatment is unchanged \citep{Naevdal2026DataRelease}. CLL--CLL comparisons therefore retain common internal-mode accommodation, whereas Maxwell--CLL differences in $C_H$ can include differences in the ensemble treatment of rotational and vibrational energy.

\subsection{DSMC numerical implementation}
\label{subsec:dsmc_numerical_implementation}

SPARTA's planar two-dimensional formulation advances positions in the $x$--$y$ plane while retaining all three velocity components, $(v_x,v_y,v_z)$. The required periodic setting in the unused $z$ direction does not represent a resolved finite-span slab. Uncut Cartesian cell volumes reduce to areas $\Delta x\Delta y$, and wall segments carry unit out-of-plane span. Forces are reported per unit span; surface tallies, normalised by segment length and time, retain the conventional area-normalised units of pressure, shear stress and heat flux under this convention \citep{SPARTA24Sep2025}.

Intermolecular collisions used SPARTA's variable-soft-sphere (VSS) implementation \citep{KouraMatsumoto1991VSS,SPARTA24Sep2025} with the archived five-species collision data \citep{Naevdal2026DataRelease}. The angular-scattering parameter was unity for every active species, recovering the isotropic variable-hard-sphere (VHS) angular law while retaining the species-dependent collision diameters, viscosity exponents, reference temperatures and relaxation parameters \citep{Bird1994,SPARTA24Sep2025}. O$_2$ and N$_2$ carry rotational and vibrational energy with variable relaxation during gas--gas collisions; the monatomic species carry translational energy only. Cell-volume particle weighting was active throughout the hierarchical grid.

\begin{figure}
    \centering
    \includegraphics[width=0.95\linewidth]{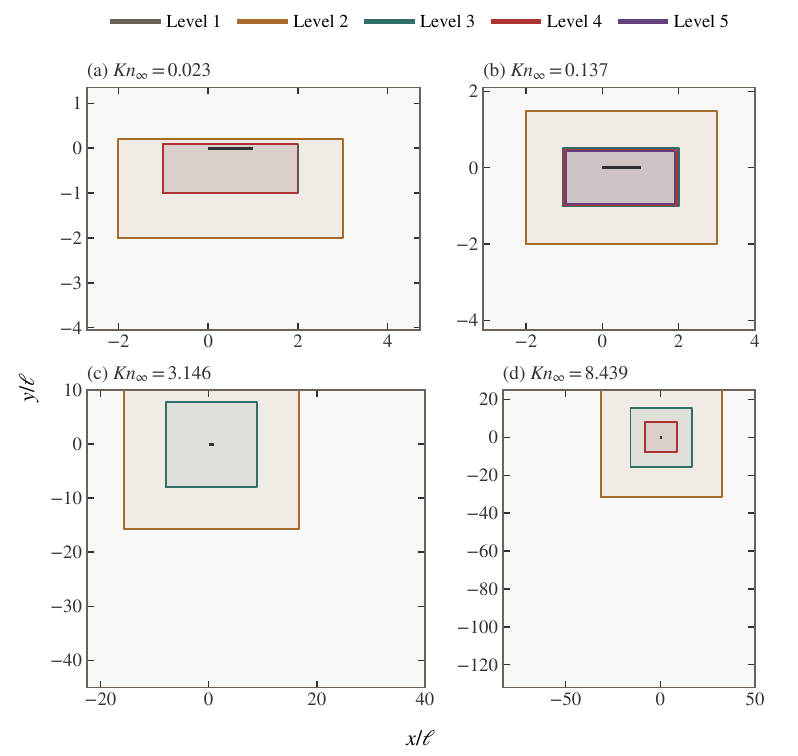}
    \caption{Computational domains and nested Cartesian refinement regions for (a) $Kn_\infty=0.023$, (b) $0.137$, (c) $3.146$ and (d) $8.439$. Outlines mark region boundaries, not individual cells. Level~1 spans each domain, and successive levels halve the cell spacing in both directions. The plate (black) lies entirely within the uniform finest-level region. Axis limits differ between panels.}
    \label{fig:mesh_overviews}
\end{figure}

The mesh varied between Knudsen conditions but was identical for every wall law within a condition. Each nested Cartesian hierarchy \citep{Plimpton2019PoF} covers the full domain at level~1 and halves the parent-cell spacing in both directions at each subsequent level. Rarefaction-based length scales provided initial domain and resolution estimates, which were then assessed through the numerical-control campaign below. Refinement surrounds the plate and its disturbance field; the entire plate lies within a uniform finest-level region. Figure~\ref{fig:mesh_overviews} shows the domains and refinement regions, and table~\ref{tab:production_controls} lists the production controls.

\begin{table}
\centering
\setlength{\tabcolsep}{2.7pt}
\renewcommand{\arraystretch}{1.08}
\begin{tabular*}{\textwidth}{@{\extracolsep{\fill}}
S[table-format=1.3]
c
S[table-format=1.0]
S[table-format=1.3e-1]
S[table-format=1.3e-1]
S[table-format=1.1e-1]
S[table-format=1.2e2]
@{}}
\toprule
{$Kn_\infty$}
& {$N_{x,1}\times N_{y,1}$}
& {$N_{\mathrm{lev}}$}
& {$\Delta x_{\min}$ [\si{\metre}]}
& {$\Delta y_{\min}$ [\si{\metre}]}
& {$\Delta t$ [\si{\second}]}
& {$W_p$} \\
\midrule
0.023 & $323\times235$ & 4 & 2.873e-3 & 2.872e-3 & 1.0e-7 & 1.00e19 \\
0.137 & $52\times47$   & 5 & 8.558e-3 & 8.445e-3 & 1.0e-6 & 5.00e16 \\
3.146 & $157\times138$ & 3 & 9.952e-2 & 9.964e-2 & 3.0e-6 & 7.41e15 \\
8.439 & $167\times197$ & 4 & 9.959e-2 & 9.962e-2 & 3.0e-6 & 3.70e15 \\
\bottomrule
\end{tabular*}
\caption{Production numerical controls, common to all wall laws at each $Kn_\infty$. $N_{x,1}\times N_{y,1}$ is the level-1 grid, $N_{\mathrm{lev}}$ the total number of mesh levels, and $\Delta x_{\min}$ and $\Delta y_{\min}$ the finest cell spacings. $W_p$ is the global reference particle-weight factor; local effective weights also include the cell-volume factor.}
\label{tab:production_controls}
\end{table}

Momentum and energy exchange were sampled on all 30 compression-side elements. SPARTA converted simulator-particle exchanges to physical surface fluxes using the effective particle weight, elapsed time and represented area \citep{SPARTA24Sep2025}. These records were accumulated in contiguous, non-overlapping physical-time blocks.

Profile resolution balances spatial detail against pointwise sampling precision. Benchmark comparisons retain all 30 elements at $Kn_\infty=0.023$ and $0.137$; at $3.146$ and $8.439$, each consecutive group of three elements is area-averaged within every retained block to give 10 comparison stations. The wall-model analysis uses this 10-station reduction at all four conditions, including the diffuse reference, so every wall-law profile has common spatial support. This aggregation reduces sampling variability at a resolution commensurate with the digitised curves. Integrated coefficients are calculated separately from all 30 native elements and are unaffected by profile aggregation.

Numerical sensitivity was assessed by perturbing one control at a time under fixed physical and wall-model conditions. Preliminary checks addressed the near-body envelope, buffer placement, far-field resolution and domain extent. With that provisional mesh fixed, the timestep, reference weight $W_p$ and retained averaging duration were examined. Spatial checks then proceeded from the near-body refinement layer through the buffer and far-field levels to the domain extent, retaining the accepted settings of all other controls. The primary monitors were $C_D$ and $C_L$ \citep{MyongKarchaniEjtehadi2019PoF,BurtBoyd2011CiCP}.

A control was adjusted further only when, for at least one $Q\in\{C_D,C_L\}$, a perturbation changed the retained-window mean by at least $1\%$ in absolute relative terms and the accepted and perturbed marginal 95\% intervals did not overlap. These intervals used the correlation-aware procedure in section~\ref{subsec:stationarity_uncertainty}. This conjunctive rule screens for material, sampling-resolved sensitivity; passing it is neither a formal equivalence test nor a $1\%$ bound on numerical error, and does not by itself establish convergence of local profiles or derived contrasts. The calculations were not tuned to the digitised benchmark \citep{OberkampfBarone2006JCP}. The refinement definitions, case registry, retained-window indices and block metadata are archived with the release \citep{Naevdal2026DataRelease}.

Movie~1 shows early filling in four separate diffuse-Maxwell visualisation reruns, one per Knudsen condition. Each started from an empty domain; the displayed sequence begins after injection has commenced and ends before the retained production sampling interval. These reruns contribute neither production means nor uncertainty estimates. Points represent simulator particles, whose effective weights vary with case and local cell area. Apparent occupancy changes at refinement interfaces need not be physical density discontinuities, and similar point concentrations need not indicate similar gas densities. Quantitative density comparisons use weighted grid tallies \citep{SPARTA24Sep2025,Naevdal2026DataRelease}.

\subsection{Response metrics}
\label{subsec:response_metrics}

\subsubsection{Surface coefficients and comparison metrics}

The molecular exchanges defined in section~\ref{subsubsec:wall_scattering_models} were reduced first to local surface moments and only then to integrated aerodynamic loads. On the compression side, $p_w$ denotes the positive wall-normal momentum flux, $\tau_w$ the wall-parallel shear stress in the plate-tangent direction, and $q_w$ the signed wall-energy flux, with $q_w>0$ denoting net transfer from the gas to the wall. The corresponding coefficients are
\begin{equation}
C_p(\xi)=\frac{p_w(\xi)}{q_\infty},
\qquad
C_f(\xi)=\frac{\tau_w(\xi)}{q_\infty},
\qquad
C_H(\xi)=\frac{q_w(\xi)}{q_\infty U_\infty},
\label{eq:surface_coefficients}
\end{equation}
where
\begin{equation}
q_\infty=\frac{1}{2}\rho_\infty U_\infty^2,
\qquad
\rho_\infty=n_\infty m_\infty,
\qquad
\xi=\frac{x}{\ell}.
\end{equation}
Under the common freestream normalisation, $C_p$, $C_f$ and $C_H$ represent wall-normal momentum, wall-parallel momentum and total gas-to-wall energy transfer, respectively \citep{Dogra1988FlatPlate}.

\PreprintBreak
\begin{compactmath}
Projecting the compression-side pressure and shear contributions onto the drag and lift directions and normalising the resulting forces per unit span by $q_\infty\ell$ gives \citep{Bird1994}:
\begin{equation}
C_D=\int_0^1\left[C_p(\xi)\sin\alpha+C_f(\xi)\cos\alpha\right]\,\mathrm{d}\xi,
\qquad
C_L=\int_0^1\left[C_p(\xi)\cos\alpha-C_f(\xi)\sin\alpha\right]\,\mathrm{d}\xi.
\label{eq:integrated_coefficients}
\end{equation}
Equation~\eqref{eq:integrated_coefficients} was evaluated in each retained block by area-weighted summation over all 30 native compression-side elements, without profile aggregation. The resulting blockwise coefficients were then averaged, and $L/D$ was formed as the ratio of retained means, $C_L/C_D$, not the mean of blockwise ratios.

Integrated benchmark discrepancies are reported as signed percentage differences relative to the digitised Dogra values \citep{Dogra1988FlatPlate},
\begin{equation}
\delta A=
100
\left(
\frac{A_{\mathrm{DSMC}}-A_{\mathrm{ref}}}
{A_{\mathrm{ref}}}
\right),
\qquad
A\in\left\{C_D,C_L,L/D\right\}.
\label{eq:benchmark_integrated_difference}
\end{equation}
For the wall-model sweep, the corresponding response relative to diffuse Maxwell reflection at the same $Kn_\infty$ is
\begin{equation}
\Delta A_m=
100
\left(
\frac{A_m-A_{\mathrm{DM}}}
{A_{\mathrm{DM}}}
\right),
\qquad
A\in\left\{C_D,C_L,L/D\right\},
\label{eq:gsi_integrated_response}
\end{equation}
where $m$ denotes the selected Maxwell or CLL model and $\mathrm{DM}$ denotes diffuse Maxwell reflection.

Surface-profile discrepancies were evaluated at the processed DSMC coordinates. Each digitised benchmark repeat was evaluated separately on the selected comparison grid; wall-model comparisons used the common 10-station grid at the same $Kn_\infty$. For $C\in\{C_p,C_f,C_H\}$ at ordered locations $\xi_1<\cdots<\xi_N$, the normalised root-mean-square and mean absolute errors are
\begin{equation}
\begin{aligned}
\mathrm{NRMSE}[C]
&=\left[
\frac{\sum_{i=1}^{N}w_i\left(C(\xi_i)-C_{\mathrm{ref}}(\xi_i)\right)^2}
{\sum_{i=1}^{N}w_i C_{\mathrm{ref}}(\xi_i)^2}
\right]^{1/2},\\
\mathrm{NMAE}[C]
&=\frac{\sum_{i=1}^{N}w_i\left|C(\xi_i)-C_{\mathrm{ref}}(\xi_i)\right|}
{\sum_{i=1}^{N}w_i\left|C_{\mathrm{ref}}(\xi_i)\right|}.
\end{aligned}
\label{eq:profile_error_metrics}
\end{equation}
For benchmark comparison, $C_{\mathrm{ref}}$ is the digitised benchmark profile and $w_i$ are normalised trapezoidal weights on the common grid. For wall-model comparison, $C_{\mathrm{ref}}$ is the diffuse Maxwell profile and $w_i=A_i/\sum_jA_j$, where $A_i$ is the grouped area represented by station $i$.

Each benchmark profile was digitised independently five times with WebPlotDigitizer \citep{WebPlotDigitizer_Rohatgi}. To select the comparison stations, we first averaged the streamwise coordinates of corresponding points across the five repeats; the first and last mean coordinates defined the reference range. At the two lower Knudsen numbers, only native DSMC stations within this range were retained. At the two higher values, all ten stations were retained, with any extension beyond the reference range limited to one tenth of the smaller median spacing of the DSMC and mean reference grids. Each original repeat was then evaluated separately at the selected stations using piecewise-linear interpolation, with its endpoint values held constant outside its own range. The resulting values were used to calculate the reference mean and pointwise Student-$t$ repeatability interval. This procedure retains repeat-to-repeat variation in coordinates and values but does not remove systematic digitisation error.

To retain the direction of the surface response without averaging unstable local percentages, the signed change is defined from the weighted mean coefficient. For model $m$ and its diffuse Maxwell reference,
\begin{equation}
\overline{C}_m
=
\sum_{i=1}^{N}w_i C_m(\xi_i),
\qquad
\overline{C}_{\mathrm{DM}}
=
\sum_{i=1}^{N}w_i C_{\mathrm{DM}}(\xi_i),
\end{equation}
and
\begin{equation}
\Delta\overline{C}_m
=
100
\frac{
\overline{C}_m-\overline{C}_{\mathrm{DM}}
}{
\left|\overline{C}_{\mathrm{DM}}\right|
}.
\label{eq:weighted_surface_change}
\end{equation}
This ratio of weighted means avoids the disproportionate influence of stations where the diffuse coefficient is small, which can arise when local percentages are averaged \citep{OberkampfBarone2006JCP}. Its sign gives the direction of the mean surface response; NRMSE and NMAE retain the unsigned profile discrepancy.

\end{compactmath}

\subsubsection{Common-incidence constraints}
\label{subsubsec:common_incidence_constraints}

At fixed incident-state flux, the Maxwell kernel in \eqref{eq:maxwell_scattering_kernel} makes each surface moment affine in $\alpha_M$. We test this implication by comparing the two intermediate production states with the chord joining the separately simulated diffuse and specular endpoints. For $C\in\{C_p,C_f,C_H\}$,
\begin{equation}
\overline C_{\mathrm{aff}}^{M}(\alpha_M)
=
\alpha_M\overline C_M(1)
+
(1-\alpha_M)\overline C_M(0),
\qquad
\mathcal D_C^{M}(\alpha_M)
=
100
\frac{
\overline C_M(\alpha_M)
-
\overline C_{\mathrm{aff}}^{M}(\alpha_M)
}{
\left|\overline C_M(1)\right|
},
\label{eq:maxwell_affine_defect}
\end{equation}
with $\alpha_M\in\{0.8,0.5\}$. The endpoint means $\overline C_M(1)$ and $\overline C_M(0)$ come from separate self-consistent simulations. In SPARTA, the prescribed reaction branch is selected before the Maxwell draw; its fixed probabilities and the subsequent specular--diffuse mixture preserve affinity at fixed incident flux. If the impact rate and flux-weighted incident-state distribution were invariant across the sweep, $\mathcal D_C^{M}$ would vanish. A resolved defect therefore diagnoses departure from that invariant-incidence condition, subject to numerical accuracy, rather than nonlinearity of the local Maxwell kernel.

A complementary constraint links the three CLL cases to diffuse Maxwell using total, rather than translational-only, heat transfer. For the prescribed probability-exchange chemistry, reaction selection precedes scattering and is independent of $(\alpha_n,\alpha_t)$ at a fixed complete incident state. A reaction may change the product mass before scattering, while full rotational and vibrational accommodation fixes the outgoing internal-energy expectation at the wall temperature. The CLL second moments therefore give, for one common complete incident flux $\mathcal F$,
\begin{equation}
Q_H(\alpha_n,\alpha_t;\mathcal F)
=\alpha_n\mathcal A_n[\mathcal F]
+\alpha_t\mathcal A_t[\mathcal F]
+\mathcal A_0[\mathcal F].
\label{eq:cll_total_energy_affinity}
\end{equation}
Here $Q_H$ is the mean total wall-energy flux and the three coefficients depend on the fixed incoming population and reaction probabilities. The intercept $\mathcal A_0$ retains the accommodation-independent contributions from internal energy, chemical energy and the difference between reactant and product masses; it is not set to zero. Under the registered stationary-wall and internal-mode treatment, diffuse Maxwell has the same conditional outgoing expectation as CLL $(1,1)$ \citep{SPARTA24Sep2025,Naevdal2026DataRelease}. Equation~\eqref{eq:cll_total_energy_affinity} is thus a property of this specified wall construction, not a general constraint on arbitrary reactive surfaces.

Let $D=\overline C_{H,\mathrm{DM}}$, $A=\overline C_{H,0.8,0.8}$, $B=\overline C_{H,0.8,0.2}$ and $C=\overline C_{H,0.2,0.8}$ denote the retained production area means. The combination $B+C-5A+3D$ cancels the normal, tangential and constant terms in \eqref{eq:cll_total_energy_affinity}. We therefore define the diffuse heat coefficient implied by the three CLL states and its departure from the simulated diffuse value as
\begin{equation}
D_{\mathrm{aff}}=\frac{5A-B-C}{3},
\qquad
T_H=D_{\mathrm{aff}}-D,
\qquad
\mathcal E_H=100\frac{T_H}{D}.
\label{eq:cll_heat_plane_defect}
\end{equation}
The linear residual $T_H$ tests the common-incidence constraint, while $\mathcal E_H$ expresses its magnitude relative to the positive diffuse heat coefficient. Both use total production heat, with the chemical contribution included once. A resolved residual is incompatible with common incidence under the stated implementation, subject to numerical accuracy; a zero residual would not establish identical incident distributions. The production component histories needed to assign this residual separately to translation, internal modes and chemistry are not available. The separately acquired component-rate means below are therefore not used as a production decomposition.

\Needspace{6\baselineskip}
For the two asymmetric CLL perturbations, direct responses relative to the symmetric case were defined as
\begin{equation}
\Delta_t^{\mathrm{CLL}}\overline C
=
100
\frac{
\overline C_{0.8,0.2}-\overline C_{0.8,0.8}
}{
\left|\overline C_{0.8,0.8}\right|
},
\qquad
\Delta_n^{\mathrm{CLL}}\overline C
=
100
\frac{
\overline C_{0.2,0.8}-\overline C_{0.8,0.8}
}{
\left|\overline C_{0.8,0.8}\right|
},
\label{eq:cll_pairwise_surface_responses}
\end{equation}
where the subscripts denote $(\alpha_n,\alpha_t)$. For an incident molecule $i$ of mass $m_i$ and velocity $\boldsymbol v_i$, \eqref{eq:cll_conditional_memory_moments} gives the non-reactive wall-tangential impulse
\begin{equation}
\left\langle\Delta p_{w,t}\mid\boldsymbol v_i,m_i\right\rangle
=
m_i\!\left(1-\sqrt{1-\alpha_t}\right)v_{i,t},
\qquad
\mathcal K_{f,t}^{\mathrm{fp}}
=
100\left[
\frac{1-\sqrt{1-0.2}}
     {1-\sqrt{1-0.8}}
-1
\right].
\label{eq:cll_tangential_impulse_reference}
\end{equation}
For any fixed non-reactive incident-state flux with positive reference shear, reducing $\alpha_t$ from $0.8$ to $0.2$ gives $\mathcal K_{f,t}^{\mathrm{fp}}\simeq-80.90\%$. Departures of $\Delta_t^{\mathrm{CLL}}\overline C_f$ can arise from changes in the incident population or impact rate and from the separately retained catalytic contribution. Finite-temperature normal-momentum and heat-transfer references are developed in section~\ref{subsubsec:finite_temperature_references}.

\subsection{Mechanism diagnostics and kernel references}
\label{subsec:selected_diagnostics}

A separate diagnostic campaign sampled the incident and reflected populations, the conditional wall transformation, and the associated density and surface-rate fields. These observables were acquired in separate simulations from those supplying the production coefficient records. Twenty-three fresh-start simulations yielded 15 non-reactive wall-event records, nine total-mixture density records and 15 compression-side rate records. The selected conditions, $Kn_\infty=0.023$, $0.137$ and $8.439$, span the benchmark range through its two endpoints and one intermediate state. Wall-event and surface-rate records cover the diffuse and specular Maxwell limits and CLL $(\alpha_n,\alpha_t)=(0.8,0.8),(0.8,0.2),(0.2,0.8)$; density records cover CLL only. No auxiliary acquisition was made at $Kn_\infty=3.146$ or for Maxwell $\alpha_M=0.8$ and $0.5$. Table~\ref{tab:selected_diagnostic_coverage} distinguishes record counts from run counts and identifies co-acquired products \citep{Naevdal2026DataRelease}.

The separate continuous wall-switch density campaign underlying movie~2 is described in section~\ref{subsec:movie_density_acquisition} and is excluded from every count in table~\ref{tab:selected_diagnostic_coverage}.

Each diagnostic simulation retained its case freestream, gas--gas collision model, catalytic treatment, timestep, particle-weighting scheme and near-body controls; only the output schedule and, where required, outer-domain extent differed by product. Processing used each acquisition's registered final interval, not the production window selected from coefficient histories. Endpoints, cadences, slice and averaging counts, and eligible event populations are archived \citep{Naevdal2026DataRelease}. Quantities evaluated from the same event record were processed jointly, retaining their covariance. No covariance was estimated between that record and the density or surface-rate aggregates, or between diagnostic and production acquisitions. None of these diagnostics enters the production coefficients or their uncertainty estimates.

\begin{table}
\centering

\begingroup
\small
\setlength{\tabcolsep}{4pt}
\renewcommand{\arraystretch}{1.15}

\setlength{\cmidrulewidth}{0.35pt}
\setlength{\lightrulewidth}{0.4pt}
\setlength{\aboverulesep}{2pt}
\setlength{\belowrulesep}{3pt}
\setlength{\cmidrulekern}{3pt}

\begin{tabularx}{\linewidth}
{@{}c c c c c @{\hspace{10pt}}>{\raggedright\arraybackslash}X@{}}
\toprule

& & \multicolumn{3}{c}{Diagnostic records} & \\
\cmidrule(lr){3-5}

$Kn_\infty$
& Runs
& Events
& Density
& Surface rates
& Co-acquired records \\
\midrule

0.023
& 10
& 5
& 3
& 5
& Density and surface rates (CLL) \\
\addlinespace[3pt]

0.137
& 8
& 5
& 3
& 5
& Events and surface rates (all five laws) \\
\addlinespace[3pt]

8.439
& 5
& 5
& 3
& 5
& All three (CLL); events and surface rates (Maxwell) \\

\cmidrule(r){1-5}
Total
& 23
& 15
& 9
& 15
& \\

\bottomrule
\end{tabularx}
\par
\endgroup

\caption{Run and record counts for the selected diagnostic campaign. Runs are fresh-start simulations. Co-acquired records share a run for a given wall law; other records were acquired separately. Event and surface-rate records cover the diffuse and specular Maxwell limits and the three CLL laws; density records cover CLL only. Event records contain non-reactive impacts from two compression-side patches. Production simulations and the movie~2 campaign are excluded.}
\label{tab:selected_diagnostic_coverage}
\end{table}

\PreprintBreak
\subsubsection{Wall-event acquisition and incident moments}
\label{subsubsec:wall_event_acquisition}
Non-reactive wall-collision events were acquired from complete compression-side elements selected by their centres around the nominal leading and trailing windows
\begin{equation}
0.075\leq x/\ell\leq0.125,
\qquad
0.875\leq x/\ell\leq0.925,
\label{eq:selected_event_windows}
\end{equation}
and were subsequently trimmed to these inclusive bounds using the recorded collision coordinate. Species identity and all three pre- and post-collision translational velocity components were retained. The event archive contains no persistent particle identity, prior-impact count or gas-collision history and therefore characterises the local incident and reflected populations without identifying first impacts, return impacts or molecular lineage. Under cell-volume weighting, the physical-molecule weight represented by event $i$ was reconstructed as
\begin{equation}
\omega_i=W_pA_{\mathrm{parent},i},
\label{eq:selected_event_weight}
\end{equation}
where $W_p$ is the reference particle-weight factor registered for the acquisition run and $A_{\mathrm{parent},i}$ is the uncut Cartesian area of the cell containing the incident particle at the start of the timestep in which the wall collision occurs. The area is reconstructed from that run's registered event-acquisition mesh and is not the clipped gas-only area of a cut cell. Within each acquired Knudsen family, the relevant cells share a verified common nominal area, up to floating-point coordinate roundoff. The resulting common event weight cancels from the normalisation of the reported angular moments, making physical-event and equal-event weighting algebraically equivalent \citep{SPARTA24Sep2025,Naevdal2026DataRelease}.

The collision tally records one movement timestep at each ten-timestep output interval; the intervening steps are not accumulated. Initialisation-only dumps before the first sampled movement step were excluded from the statistical sequence, while genuinely observed zero-collision timesteps were retained. The observed sequences contain $1000$, $2998$ and $33\,333$ slices at $Kn_\infty=0.023$, $0.137$ and $8.439$, respectively. Thus the observation exposure is $T_{\mathrm{obs}}=N_{\mathrm{obs}}\Delta t$, not the elapsed time spanning the output sequence. Each nominal patch has area $0.05~\mathrm{m}^2$ under the unit-span convention. These definitions determine absolute event-based fluxes; normalised angular and energy moments depend on the recorded events themselves \citep{Naevdal2026DataRelease}.

For an eligible reflected event, the projected in-plane leaving angle was
\begin{equation}
\theta_i
=
\operatorname{atan2}\!\left(-v^{\mathrm{post}}_{y,i},v^{\mathrm{post}}_{x,i}\right)
\frac{180^\circ}{\upi}
\in[0^\circ,180^\circ].
\label{eq:selected_event_angle}
\end{equation}
The empirical distributions use 60 non-periodic bins of width $3^\circ$. Quantitative moments are evaluated from the events, not from the filtered curves used for display. Let $s\in\{\mathrm{L},\mathrm{T}\}$ denote the leading and trailing stations and, for registered event slice $b$, define $N_{s,b}$, $S_{1,s,b}=\sum_{i\in(s,b)}\theta_i$ and $S_{2,s,b}=\sum_{i\in(s,b)}\theta_i^2$. The corresponding joint leading--trailing sufficient-statistic vector is
\begin{equation}
\boldsymbol z_b^{(\theta)}=
\begin{bmatrix}
N_{\mathrm L,b}&S_{1,\mathrm L,b}&S_{2,\mathrm L,b}&
N_{\mathrm T,b}&S_{1,\mathrm T,b}&S_{2,\mathrm T,b}
\end{bmatrix}^{\mathsf T},
\label{eq:selected_event_slice_vector}
\end{equation}
with the corresponding station subvector set to zero whenever an observed slice contains no eligible events at that station. Writing $N_s=\sum_bN_{s,b}$, $S_{1,s}=\sum_bS_{1,s,b}$ and $S_{2,s}=\sum_bS_{2,s,b}$, the station variance, pooled leading--trailing mean and trailing-to-leading variance ratio are
\begin{equation}
V_s=
\frac{S_{2,s}-S_{1,s}^{2}/N_s}{N_s-1},
\qquad
\overline\theta_{\mathrm{LT}}=
\frac{S_{1,\mathrm L}+S_{1,\mathrm T}}
{N_{\mathrm L}+N_{\mathrm T}},
\qquad
R_V=\frac{V_{\mathrm T}}{V_{\mathrm L}}.
\label{eq:selected_event_moments}
\end{equation}
A heteroskedasticity- and autocorrelation-consistent (HAC) covariance estimate was formed from the complete observed slice sequence $\{\boldsymbol z_b^{(\theta)}\}_b$, using the Bartlett window and lag rule in \eqref{eq:hac_centering_and_lag}--\eqref{eq:hac_covariance}. Analytic first-order multivariate linearisation (the delta method) propagates the joint covariance to $\overline\theta_{\mathrm{LT}}$, $V_s$ and $R_V$, with two-sided normal-approximation half-widths of $1.96$ standard errors \citep{JCGM1002008,Oehlert1992DeltaMethod}. Collision rows were not treated as independent temporal replicates.

The component-energy content of the incident population was measured by the ratio of pooled tangential to normal excess-energy sums:
\begin{equation}
c_{w,i}^{2}=\frac{2k_BT_w}{m_i},
\qquad
\mathcal R_{t/n}=
\frac{
\displaystyle\sum_{i\in\mathcal P_{\mathrm{LT}}}
\omega_i m_i\left[(v^{\mathrm{pre}}_{x,i})^2+(v^{\mathrm{pre}}_{z,i})^2-c_{w,i}^{2}\right]
}{
\displaystyle\sum_{i\in\mathcal P_{\mathrm{LT}}}
\omega_i m_i\left[(v^{\mathrm{pre}}_{y,i})^2-c_{w,i}^{2}\right]
}.
\label{eq:incident_component_energy_ratio}
\end{equation}
Here $\mathcal P_{\mathrm{LT}}$ is the union of the two nominal windows. Primitive sums are pooled before division; the common factor $1/2$ cancels, while molecular masses and event weights remain inside the sums. Negative excess-energy contributions are retained. Thus $\mathcal R_{t/n}$ describes the translational content of the acquired two-patch, non-reactive population, not total energy transfer over the full compression side. Acquisition uncertainty was propagated from the joint slice histories of the tangential and normal sums using the same Bartlett--HAC estimator. The primary delta-method interval was reported only when the 95\% interval for the pooled normal denominator excluded zero; the additional Fieller check is specified in section~\ref{subsubsec:ratio_confidence_sets}.

\subsubsection{Cross-kernel evaluation of incident populations}
\label{subsubsec:cross_kernel_evaluation}

To distinguish changes in the wall law from changes in incidence, we evaluate each scattering kernel on the incident records acquired under all five wall laws. Let $X_i=(\boldsymbol v_{i,\mathrm{pre}},\sigma_i,m_i)$ denote the incident velocity, species and mass, and $h_a^{(\nu)}(X)$ the conditional mean transfer to the wall per collision under law $a$. The index $\nu\in\{N,T,H_{\mathrm{tr}}\}$ labels normal momentum, tangential momentum and translational energy. With $u=v_y^{\mathrm{pre}}>0$, $v=v_x^{\mathrm{pre}}$ and $w=v_z^{\mathrm{pre}}$, the non-reactive CLL expressions are
\begin{equation}
\begin{aligned}
h_a^{(N)}(X)&=m\left[u+\mu_{r,n}(u;\alpha_{n,a})\right],\\
h_a^{(T)}(X)&=m\left(1-\sqrt{1-\alpha_{t,a}}\right)v,\\
h_a^{(H_{\mathrm{tr}})}(X)&=\alpha_{n,a}\left(\frac{mu^2}{2}-k_BT_w\right)
+\alpha_{t,a}\left[\frac{m(v^2+w^2)}{2}-k_BT_w\right].
\end{aligned}
\label{eq:cross_kernel_conditional_moments}
\end{equation}
Here $\mu_{r,n}$ is the species-specific mean reflected normal speed given by \eqref{eq:cll_conditional_normal_speed_mean}. The diffuse and specular translational limits correspond to $(\alpha_n,\alpha_t)=(1,1)$ and $(0,0)$, with $\mu_{r,n}=\sqrt{\upi k_BT_w/(2m)}$ and $u$, respectively. These are per-collision impulses and energies, evaluated analytically without further reflected-velocity sampling; signed tangential velocities and excess energies are retained.

Let $\mathcal F_{b,P}$ denote the non-reactive incident-flux measure on patch $P$ in the flow obtained under law $b$. The corresponding moment and its event-based estimator are
\begin{equation}
\begin{aligned}
Q_{ab,P}^{(\nu)}&=\int h_a^{(\nu)}(X)\,\mathrm d\mathcal F_{b,P}(X),\\
\widehat Q_{ab,P}^{(\nu)}&=
\frac{1}{A_PT_{\mathrm{obs},b}}
\sum_{i\in(b,P)}\omega_i h_a^{(\nu)}(X_i).
\end{aligned}
\label{eq:cross_kernel_moments}
\end{equation}
The total measure is the non-reactive impact rate per unit area, so $Q_{ab,P}^{(\nu)}$ retains both the encounter rate and the distribution of incident states. Recorded impacts already sample this flux; no additional factor of $u$ enters the estimator. Rows identify applied kernels and columns identify acquired populations. Off-diagonal entries therefore describe alternative wall laws acting on unchanged incidence, rather than new self-consistent flow solutions.

Normalisation by $q_\infty$ for momentum and by $q_\infty U_\infty$ for energy gives dimensionless patch-mean coefficients. Every term in a contrast uses the same support and normalisation. Leading and trailing patches are analysed separately; pooled fluxes combine weighted totals over the summed area--time exposure. All comparisons concern the selected non-reactive flux on these patches, distinct from the full-side, all-event production loads and total heat.

For either asymmetric CLL law, let $0$ denote the symmetric reference and $1$ the perturbed case. Hereafter, $Q_{ab}$ denotes the finite-record estimate in \eqref{eq:cross_kernel_moments}, with patch and moment indices suppressed. The kernel change can be evaluated on either population, and the population change under either kernel. Averaging the two orders gives
\begin{equation}
\begin{aligned}
\Delta Q_{\mathrm{ker}}&=\tfrac12\left[(Q_{10}-Q_{00})+(Q_{11}-Q_{01})\right],\\
\Delta Q_{\mathrm{pop}}&=\tfrac12\left[(Q_{01}-Q_{00})+(Q_{11}-Q_{10})\right],\\
\Delta Q&=Q_{11}-Q_{00}=\Delta Q_{\mathrm{ker}}+\Delta Q_{\mathrm{pop}}.
\end{aligned}
\label{eq:kernel_population_allocation}
\end{equation}
This symmetric allocation is exact but is not a unique causal partition. The contributions can oppose one another and exceed the net change in magnitude. Percentage contributions are normalised by the common non-zero $|Q_{00}|$, rather than by the net contrast. Because changing only $\alpha_t$ leaves $h^{(N)}$ unchanged at every incident state, the kernel contribution to pressure vanishes identically; changing only $\alpha_n$ gives the corresponding result for shear. Any conditional change in these complementary moments is therefore assigned to the incident population.

A response can depart from its freestream reference because baseline incidence is already modified, because the wall perturbation changes it further, or both. For non-zero reference moments, define $r_{\mathrm{self}}=(Q_{11}-Q_{00})/|Q_{00}|$ and the response at fixed baseline incidence $r_{\mathrm{fp},0}=(Q_{10}-Q_{00})/|Q_{00}|$. Let $r_{\mathrm{fp},\infty}=(Q_{1,\infty}-Q_{0,\infty})/|Q_{0,\infty}|$ denote the same kernel comparison on a freestream flux matched to the non-reactive selection. Then
\begin{equation}
r_{\mathrm{self}}-r_{\mathrm{fp},\infty}
=\left(r_{\mathrm{fp},0}-r_{\mathrm{fp},\infty}\right)
+\frac{Q_{11}-Q_{10}}{|Q_{00}|}.
\label{eq:reference_population_split}
\end{equation}
The first term compares fixed-incidence responses on the baseline and freestream populations. The second measures the population change under the perturbed kernel, so this ordered split differs from the symmetric allocation in \eqref{eq:kernel_population_allocation}. On the incoming half-space $\boldsymbol v\bcdot\boldsymbol n<0$, the matched freestream measure for species $s$ is $p_{\mathrm{nr},s}|\boldsymbol v\bcdot\boldsymbol n|f_{\infty,s}(\boldsymbol v)\,\mathrm d\boldsymbol v$, where $p_{\mathrm{nr},s}$ is the prescribed probability of no surface reaction. It is evaluated by replacing $X_s$ with $X_sp_{\mathrm{nr},s}$ in the finite-temperature sums of section~\ref{subsubsec:finite_temperature_references}, without renormalisation. This removes the known selection mismatch from the comparison; the unfiltered references below remain unchanged. These comparisons concern population moments, not individual impact histories.

For a finite sample, conditional and realised wall exchanges need not coincide. Let $Q_b^{\mathrm{obs}}$ denote the non-reactive exchange flux calculated from the recorded pre- and post-collision velocities. The diagonal estimate $Q_{bb}$ instead uses the conditional mean transfer under law $b$ on the incident population acquired under that law. Both estimates use the same events, weights and area--time normalisation. The sampling remainder satisfies
\begin{equation}
\epsilon_b=Q_b^{\mathrm{obs}}-Q_{bb},
\qquad
Q_1^{\mathrm{obs}}-Q_0^{\mathrm{obs}}
=\Delta Q_{\mathrm{ker}}+\Delta Q_{\mathrm{pop}}+\epsilon_1-\epsilon_0.
\label{eq:conditional_observed_closure}
\end{equation}
The remainder accounts for finite-sample variability in wall scattering and is retained separately from the population contribution. Joint timestep vectors contain both patches, all applied-kernel moments and the recorded exchanges. Their Bartlett--HAC covariance is propagated through the allocations and normalised contrasts, preserving dependence between quantities evaluated on the same records. Because wall scattering also affects subsequent incidence, conditional averaging need not reduce the long-run sampling variance. Ordered allocations, closure residuals and full moment matrices are included in the data release \citep{Naevdal2026DataRelease}.

\PreprintBreak
\subsubsection{Conditional angular-variance decomposition}
\label{subsubsec:conditional_angular_variance}

Angular width poses a different question from mean exchange: how much leaving-angle variance arises at fixed incidence, and how much from variation of the conditional mean across incident states? To answer it, the exact translational SPARTA kernel was conditionally resampled at each acquired state $X_i$. For $M$ reflected draws at state $i$, let $\mu_i$ and $s_i^2$ denote the sample mean and unbiased sample variance of the conditional leaving-angle distribution. With
\[
W=\sum_i\omega_i,
\qquad
W_2=\sum_i\omega_i^2,
\qquad
D_\omega=W-\frac{W_2}{W},
\qquad
\overline\mu=\frac{\sum_i\omega_i\mu_i}{W},
\]
the decomposition is
\begin{equation}
\begin{aligned}
V_{\mathrm{ker}}
&=\frac{\sum_i\omega_i s_i^2}{W},\\
V_{\mathrm{inc}}
&=\frac{\sum_i\omega_i(\mu_i-\overline\mu)^2}{D_\omega}
-
\frac{
\sum_i\omega_i s_i^2-W^{-1}\sum_i\omega_i^2s_i^2
}{M D_\omega},\\
V_{\mathrm{dec}}
&=V_{\mathrm{ker}}+V_{\mathrm{inc}}.
\end{aligned}
\label{eq:outgoing_angle_variance}
\end{equation}
Here $V_{\mathrm{ker}}$ estimates the weighted mean conditional variance $\operatorname{Var}(\theta_r\mid X_i)$, whereas $V_{\mathrm{inc}}$ estimates the finite-weight dispersion of the conditional means $\mathbb E(\theta_r\mid X_i)$. The subtracted term removes the finite-$M$ Monte Carlo contribution to that dispersion. This correction is conditional on the acquired incident states; it does not establish unbiasedness with respect to correlated DSMC acquisition. Corrected estimates of $V_{\mathrm{inc}}$ may be negative and were retained without truncation.

Eight independently and reproducibly seeded Monte Carlo streams were evaluated. The first common draw count satisfying the archived convergence criterion for every population and component was $M=128$ \citep{Naevdal2026DataRelease}. The associated standard errors describe conditional resampling error only, not acquisition uncertainty. Each wall law is applied to its own acquired incident population: \eqref{eq:outgoing_angle_variance} is therefore a within-model variance decomposition, not a common-input causal separation of wall and flow effects.

For the isotropic diffuse wall, the projected-angle density is $p_D(\vartheta)=\tfrac12\sin\vartheta$ on $0\leq\vartheta\leq\upi$, with $\vartheta$ in radians. Its exact conditional variance is $V_D=(180/\upi)^2(\upi^2/4-2)\simeq1534.3873~\mathrm{deg}^2$, independent of incident state, species mass and wall temperature. This analytical reference is distinct from the finite-draw estimate and its conditional Monte Carlo error.

\subsubsection{Density-field morphology}
For each CLL model, total-mixture number density was time-averaged over the registered final interval in every resolved gas cell of the adaptive Cartesian mesh. These native cell averages define the density fields used for quantitative comparison. The two finite CLL accommodation contrasts were defined as
\begin{equation}
\Delta_t(\boldsymbol x)=
\frac{n_{0.8,0.8}(\boldsymbol x)-n_{0.8,0.2}(\boldsymbol x)}{n_\infty},
\qquad
\Delta_n(\boldsymbol x)=
\frac{n_{0.8,0.8}(\boldsymbol x)-n_{0.2,0.8}(\boldsymbol x)}{n_\infty}.
\label{eq:density_contrasts}
\end{equation}
Here $\Delta_t$ and $\Delta_n$ are the finite contrasts obtained by reducing $\alpha_t$ and $\alpha_n$, respectively, from $0.8$ to $0.2$; positive values denote greater local density under CLL $(0.8,0.8)$ than under the corresponding reduced-accommodation model. Reference and comparison fields were subtracted only after exact matching of resolved cell identities and physical bounds, including reconstruction of the upper and lower gas subcells cut by the thin plate. All density integrals below are evaluated over this finite registered diagnostic region and are not whole-domain mass-balance measures. The area $A_j^{(c)}$ and centroid $\boldsymbol x_j$ refer to the gas-only footprint of native cell $j$, excluding the solid plate even in connected plate-tip cells; any rectangular subdivision for rasterisation retains the cell's single density observation. With contrast $\Delta_j$, the dimensionless area-integrated response magnitude, signed-lobe centroids and directed dipole are
\begin{equation}
\begin{gathered}
\mathcal L_1(\Delta)
=
\frac{1}{\ell^2}
\sum_jA_j^{(c)}|\Delta_j|,
\qquad
W_{\pm}=\sum_jA_j^{(c)}[\pm\Delta_j]_+,\\
\boldsymbol c_{\pm}=
\frac{\sum_jA_j^{(c)}[\pm\Delta_j]_+\boldsymbol x_j}{W_{\pm}},
\qquad
\boldsymbol d=\boldsymbol c_+-\boldsymbol c_-,
\end{gathered}
\label{eq:density_morphology}
\end{equation}
where $[a]_+=\max(a,0)$. Because $\boldsymbol d=\boldsymbol c_+-\boldsymbol c_-$, each directed dipole points from the negative to the positive contrast lobe. The smaller angular separation between the two directed response dipoles and the $|\Delta|$-weighted wall-normal distance from the compression-face line are
\begin{equation}
\begin{aligned}
\Delta\varphi
&=
\operatorname{atan2}\!\left(
|d_{t,x}d_{n,y}-d_{t,y}d_{n,x}|,
\boldsymbol d_t\bcdot\boldsymbol d_n
\right)\in[0,\upi],\\
\frac{D_w}{\ell}
&=
\frac{\sum_jA_j^{(c)}|\Delta_j|\,|y_j-y_c|}
{\ell\sum_jA_j^{(c)}|\Delta_j|},
\qquad y_c=-5.0\times10^{-5}\ \mathrm m.
\end{aligned}
\label{eq:density_dipole_and_distance}
\end{equation}
Here $D_w/\ell$ is the $|\Delta|$-weighted wall-normal distance of the native-cell centroids from the compression-face line $y=y_c$; $\Delta\varphi$ was converted to degrees for reporting.

Sensitivity to the finite integration support was assessed by moving each boundary inward by $k=0,\ldots,4$ multiples of its side-specific median outer-cell width. At each $k$, whole cells with centroids on or within the contracted bounds were retained and the metrics recomputed. The resulting range describes sensitivity to cropping the sampled field; it is neither a confidence interval nor a test of the computational outer-boundary placement. Exact-overlap rasterisation, mask-normalised Gaussian filtering, bilinear rendering and symmetric inverse-hyperbolic-sine colour scaling were used only for display. Quantitative morphology was evaluated on the unregridded, unsmoothed native fields \citep{Naevdal2026DataRelease}.

\subsubsection{Surface-rate factorisation}

Each compression-side rate record contains final-window elementwise aggregates over all wall encounters, including reactive events, on the 30 lower-face elements. On any sampled surface support, the signed wall-energy flux factorises into the physical incident-number flux and mean energy transfer per incident molecule,
\begin{equation}
q_w
=
\frac{1}{A_s\Delta t_s}
\sum_{h=1}^{N_{\mathrm{hit}}}
\omega_h\Delta E_{w,h}
=
\Gamma_{\mathrm{inc}}
\left\langle\Delta E_w\right\rangle_{\mathrm{hit}}.
\label{eq:surface_energy_balance}
\end{equation}
Here $A_s$ and $\Delta t_s$ are the sampled area and accumulation interval, $N_{\mathrm{hit}}$ is the number of simulator-particle wall encounters accumulated on that support, $h$ indexes those encounters and $\omega_h$ is the effective number of physical molecules represented by encounter $h$. Hence
\[
\Gamma_{\mathrm{inc}}
=
\frac{1}{A_s\Delta t_s}
\sum_h\omega_h,
\qquad
\left\langle\Delta E_w\right\rangle_{\mathrm{hit}}
=
\frac{\sum_h\omega_h\Delta E_{w,h}}
{\sum_h\omega_h}.
\]
The encounter energy $\Delta E_{w,h}$ is defined in \eqref{eq:wall_molecular_exchange}. Complete-side rates were obtained from SPARTA's physical surface tallies, not reconstructed from the local non-reactive event archive. Let $j$ index the 30 lower-face elements, $m$ the wall model, $\ell_j=\ell/30$ the element length (surface area per unit span), and $k\in\{\mathrm{tr},\mathrm{rot},\mathrm{vib},\mathrm{chem},\mathrm{tot}\}$ the energy component. The incident-number flux $\Phi_{N,jm}$ and signed component energy flux $q''_{k,jm}$ were integrated as
\begin{equation}
\dot N'_m=\sum_{j=1}^{30}\ell_j\Phi_{N,jm},
\qquad
\dot E'_{k,m}=\sum_{j=1}^{30}\ell_jq''_{k,jm},
\qquad
\varepsilon_{k,m}=\frac{\dot E'_{k,m}}{\dot N'_m}.
\label{eq:selected_surface_rates}
\end{equation}
The primes denote rates per unit span. Positive energy denotes transfer from gas to wall. The energy per incident molecule $\varepsilon_{k,m}$ is a ratio of separately integrated physical rates, not an average of elementwise ratios. For the total-energy quantities used in section~\ref{SUBSEC:pathway_diagnostics},
\[
\dot N'_{\mathrm{hit},m}\equiv\dot N'_m,
\qquad
\varepsilon_{\mathrm{hit},m}\equiv\varepsilon_{\mathrm{tot},m},
\qquad
\dot E'_{\mathrm{wall},m}\equiv\dot E'_{\mathrm{tot},m}.
\]
To localise where the complete-side incident-rate response developed, the 30 lower-face elements were ordered from the leading to the trailing edge. For wall model $m$ relative to the diffuse Maxwell reference at the same $Kn_\infty$, the cumulative complete-side-normalised contrast at the downstream edge $x_k^+$ of element $k$ was
\begin{equation}
\begin{aligned}
\mathcal D_m(\xi_k)
&=
\frac{\displaystyle\int_0^{x_k^+}
\left[
\Gamma_{\mathrm{inc},m}(x)
-
\Gamma_{\mathrm{inc},\mathrm{DM}}(x)
\right]\,\mathrm{d}x}
{\displaystyle\int_0^\ell
\Gamma_{\mathrm{inc},\mathrm{DM}}(x)\,\mathrm{d}x}
\\
&=
\frac{\displaystyle\sum_{j=1}^{k}\ell_j
\left(
\Phi_{N,jm}-\Phi_{N,j\mathrm{DM}}
\right)}
{\displaystyle\sum_{j=1}^{30}\ell_j
\Phi_{N,j\mathrm{DM}}},
\qquad
\xi_k=\frac{x_k^+}{\ell}.
\end{aligned}
\label{eq:cumulative_impact_rate_contrast}
\end{equation}
The curve includes the exact origin, $\mathcal D_m(0)=0$, and its endpoint satisfies
\[
\mathcal D_m(1)
=
\frac{\dot N'_{\mathrm{hit},m}}
{\dot N'_{\mathrm{hit},\mathrm{DM}}}
-1.
\]
The fixed denominator therefore makes $\mathcal D_m$ a streamwise decomposition of the complete-side response, not a response renormalised by the diffuse flux accumulated up to $x_k^+$. Accumulation locations were evaluated by treating each native element mean as piecewise constant over its represented interval.

Because $e_{\mathrm{tot}}$ (\texttt{etot}) already includes the surface-reaction energy, the separately retained chemical component was used for attribution and closure only, not added to the total \citep{SPARTA24Sep2025,Naevdal2026DataRelease}.

For a target wall model $m$ and a specified baseline $b$, with diffuse Maxwell used for diffuse-relative responses and CLL $(0.8,0.8)$ used for the asymmetric CLL attribution and component fractions, define
\begin{equation}
r_N=\frac{\dot N'_m}{\dot N'_b}-1,
\qquad
r_\varepsilon=
\frac{\varepsilon_{\mathrm{tot},m}}
{\varepsilon_{\mathrm{tot},b}}-1,
\qquad
r_E=
\frac{\dot E'_{\mathrm{tot},m}}
{\dot E'_{\mathrm{tot},b}}-1.
\end{equation}
The total-rate response factorises exactly,
\begin{equation}
1+r_E=(1+r_N)(1+r_\varepsilon),
\qquad
r_E=r_N+r_\varepsilon+r_Nr_\varepsilon.
\label{eq:wall_energy_factorisation}
\end{equation}
Averaging the two possible sequential allocations gives an order-symmetric two-factor attribution, under which the interaction term is divided equally,
\begin{equation}
\phi_N=r_N+\frac12r_Nr_\varepsilon,
\qquad
\phi_\varepsilon=r_\varepsilon+\frac12r_Nr_\varepsilon,
\qquad
\phi_N+\phi_\varepsilon=r_E.
\label{eq:wall_energy_symmetric_allocation}
\end{equation}
For $r_E\neq0$, the shares $\phi_N/r_E$ and $\phi_\varepsilon/r_E$ are signed algebraic allocations, not independently identified causal effects. They may be negative or exceed unity when the factors oppose one another and are not clipped to $[0,1]$. For componentwise contrasts
\[
\Delta\dot E'_k
=
\dot E'_{k,m}-\dot E'_{k,b},
\qquad
k\in\{\mathrm{tr},\mathrm{rot},\mathrm{vib},\mathrm{chem}\},
\]
translational dominance was reported using both
\begin{equation}
f_{\mathrm{tr,signed}}
=
\frac{\Delta\dot E'_{\mathrm{tr}}}
{\Delta\dot E'_{\mathrm{tot}}},
\qquad
f_{\mathrm{tr,abs}}
=
\frac{
\left|\Delta\dot E'_{\mathrm{tr}}\right|
}{
\displaystyle\sum_k
\left|\Delta\dot E'_k\right|
}.
\label{eq:wall_energy_translational_fractions}
\end{equation}
The signed fraction preserves algebraic cancellation and may therefore lie outside $[0,1]$, whereas the absolute-component fraction measures dominance without cancellation and lies in $[0,1]$. The four component contrasts were checked against the direct total-rate contrast, with $e_{\mathrm{tot}}$ retained as the authoritative total. Because each surface-rate record contains a single final-window elementwise aggregate rather than a retained temporal history, these quantities are descriptive point estimates without temporal confidence intervals \citep{Naevdal2026DataRelease}.

\subsubsection{Finite-temperature kernel references}
\label{subsubsec:finite_temperature_references}

The CLL second moments link the incident excess-energy partition to the finite heat-transfer response. From \eqref{eq:cll_kernel_factorization}--\eqref{eq:cll_tangential_kernel}, the conditional mean tangential and normal translational-energy transfers at fixed non-reactive incident state $X_i$ are
\begin{equation}
\begin{aligned}
\left\langle \Delta E_{w,t}^{\mathrm{tr}}\mid X_i\right\rangle
&=
\frac{\alpha_t m_i}{2}
\left(
v_{i,t_1}^{2}+v_{i,t_2}^{2}-c_{w,i}^{2}
\right),\\
\left\langle \Delta E_{w,n}^{\mathrm{tr}}\mid X_i\right\rangle
&=
\frac{\alpha_n m_i}{2}
\left(
v_{i,n}^{2}-c_{w,i}^{2}
\right).
\end{aligned}
\label{eq:cll_conditional_translational_transfer}
\end{equation}
Reducing either coefficient from $0.8$ to $0.2$ gives a reference-minus-perturbed contrast of $0.6m_i/2$ times the corresponding excess-energy term. At fixed incidence, the ratio of the weighted tangential and normal contrasts is therefore exactly $\mathcal R_{t/n}$ in \eqref{eq:incident_component_energy_ratio}: the common factor $0.6/2$ cancels, but each molecular mass remains within the sum. The event-based comparison uses the pooled nominal-window incident population from CLL $(0.8,0.8)$.

Direct integration of each species' drifting Maxwellian over the incident half-space gives an analytic finite-temperature mixture reference. Let
\begin{equation}
U_n=U_\infty\sin\alpha,
\qquad
U_t=U_\infty\cos\alpha,
\qquad
a_s=\left(\frac{k_BT_\infty}{m_s}\right)^{1/2},
\qquad
\eta_s=\frac{U_n}{a_s},
\end{equation}
and let $\varphi$ and $\Phi$ denote the standard normal density and cumulative-distribution functions. The first and third incident half-space moments for species $s$ are
\begin{equation}
\begin{aligned}
J_{1,s}
&=U_n\Phi(\eta_s)+a_s\varphi(\eta_s),\\
J_{3,s}
&=(U_n^3+3U_na_s^2)\Phi(\eta_s)
 +(U_n^2a_s+2a_s^3)\varphi(\eta_s),
\end{aligned}
\end{equation}
so that
\[
\left\langle v_n^2\right\rangle_{\mathrm{hit},s}
=
\frac{J_{3,s}}{J_{1,s}},
\qquad
\left\langle v_{t_1}^2+v_{t_2}^2\right\rangle_{\mathrm{hit},s}
=
U_t^2+2a_s^2.
\]
For the registered case-specific mole fractions $X_s$ \citep{Naevdal2026DataRelease}, the species-resolved, incident-flux-weighted mixture reference is
\begin{equation}
\mathcal R_{\mathrm{mix}}=
\frac{
\displaystyle\sum_sX_sJ_{1,s}m_s
\left(U_t^2+2a_s^2-c_{w,s}^2\right)
}{
\displaystyle\sum_sX_sJ_{1,s}m_s
\left(J_{3,s}/J_{1,s}-c_{w,s}^2\right)
},
\qquad
c_{w,s}^2=\frac{2k_BT_w}{m_s}.
\label{eq:finite_temperature_mixture_ratio}
\end{equation}
As $T_\infty,T_w\rightarrow0$ at fixed $U_\infty$ and $\alpha$, $\mathcal R_{\mathrm{mix}}$ approaches $U_t^2/U_n^2=\cot^2\alpha$, the directed hyperthermal limit.

Unlike tangential impulse and componentwise translational-energy exchange, the first reflected-normal-speed moment does not admit a distribution-independent multiplicative accommodation factor \citep{Cercignani1971KineticModels,Lord1991ExtensionsCLL,ChenEtAl2023Physisorption}. The normal-momentum reference is therefore obtained by integrating the conditional mean over the prescribed incident distribution. For inward normal speed $u>0$ and species $s$, define
\begin{equation}
\begin{aligned}
\nu_{n,s}(u,\alpha_n)
&=
\sqrt{1-\alpha_n}\,u,
\qquad
\sigma_{n,s}^{2}(\alpha_n)
=
\frac{\alpha_n k_BT_w}{m_s},
\\
\chi_s(u,\alpha_n)
&=
\frac{\nu_{n,s}^{2}(u,\alpha_n)}
     {4\sigma_{n,s}^{2}(\alpha_n)}
=
\frac{(1-\alpha_n)u^2}
     {4\sigma_{n,s}^{2}(\alpha_n)}.
\end{aligned}
\end{equation}
The CLL normal kernel is the radial law of Lord's displaced two-dimensional Gaussian construction \citep{Lord1991ExtensionsCLL}. Thus $Y=v_{r,n}/\sigma_{n,s}$ has Park's $N=2$ generalised Rayleigh distribution, or Rice distribution, with noncentrality $y_0=\nu_{n,s}/\sigma_{n,s}$. Setting the moment order to $a=1$ in \citet{Park1961GeneralizedRayleighMoments} and writing the result in modified-Bessel form gives
\begin{equation}
\mu_{r,n,s}(u;\alpha_n)
=
\sigma_{n,s}\sqrt{\frac{\upi}{2}}\,
\mathrm{e}^{-\chi_s}
\left[
\left(1+2\chi_s\right)\mathrm{I}_0(\chi_s)
+
2\chi_s \mathrm{I}_1(\chi_s)
\right],
\label{eq:cll_conditional_normal_speed_mean}
\end{equation}
where $\mathrm{I}_1$ is the modified Bessel function of the first kind of order one. Both modified-Bessel functions are evaluated in exponentially scaled form to avoid overflow. At $\alpha_n=0$ the limiting mean is $u$, while at $\alpha_n=1$ it is $\sqrt{\upi k_BT_w/(2m_s)}$. With
\[
g_s(u)
=
\frac{1}{a_s}
\varphi\!\left(\frac{u-U_n}{a_s}\right),
\qquad u>0,
\]
denoting the freestream normal-velocity density on the incident half-space, flux-weighting $m_s[u+\mu_{r,n,s}(u;\alpha_n)]$ gives the wall-normal momentum flux per unit freestream number density,
\begin{equation}
\begin{aligned}
\Pi_n(\alpha_n)
&=
\sum_s X_s m_s
\int_0^\infty
u\left[
u+\mu_{r,n,s}(u;\alpha_n)
\right]g_s(u)\,\mathrm du,
\\
\mathcal K_{p,n}^{\mathrm{mix}}
&=
100\left[
\frac{\Pi_n(0.2)}{\Pi_n(0.8)}-1
\right].
\end{aligned}
\label{eq:cll_normal_impulse_reference}
\end{equation}
The common factor $n_\infty$ cancels in the ratio. In the directed cold limit, $\mu_{r,n,s}\rightarrow\sqrt{1-\alpha_n}\,u$ and
\[
\mathcal K_{p,n}^{\mathrm{mix}}
\rightarrow
100\left[
\frac{1+\sqrt{1-0.2}}
     {1+\sqrt{1-0.8}}
-1
\right]\simeq+30.90\%.
\]

The same half-space integration supplies finite-temperature collisionless force references,
\begin{equation}
C_N^{\mathrm{FM}}=\frac{n_\infty\Pi_n(\alpha_n)}{q_\infty},
\qquad
C_T^{\mathrm{FM}}=\frac{n_\infty U_t}{q_\infty}
\left(1-\sqrt{1-\alpha_t}\right)
\sum_sX_sm_sJ_{1,s}.
\label{eq:finite_temperature_force_coefficients}
\end{equation}
They are projected into drag and lift with \eqref{eq:integrated_coefficients}, retaining the prescribed $m_\infty$ in $q_\infty$ rather than substituting a differently rounded species sum. These are unfiltered, non-reactive reference coefficients. They support the cold-beam symmetry comparison in section~\ref{SUBSEC:aerodyn_resp}, but their difference from the reactive production loads is not attributed solely to intermolecular collisions.

Finally, substituting the two asymmetric CLL parameter pairs into \eqref{eq:cll_conditional_translational_transfer} and normalising by CLL $(0.8,0.8)$ yields the corresponding non-reactive fixed-population translational heat-transfer references. In terms of $\mathcal R_{\mathrm{mix}}$,
\begin{equation}
\begin{aligned}
\mathcal K_{H,t}^{\mathrm{mix}}
&=
100\left[
\frac{0.2\mathcal R_{\mathrm{mix}}+0.8}
     {0.8(\mathcal R_{\mathrm{mix}}+1)}
-1
\right],\\
\mathcal K_{H,n}^{\mathrm{mix}}
&=
100\left[
\frac{0.8\mathcal R_{\mathrm{mix}}+0.2}
     {0.8(\mathcal R_{\mathrm{mix}}+1)}
-1
\right].
\end{aligned}
\label{eq:cll_mixture_heat_references}
\end{equation}
These are non-reactive translational fixed-population references for the common collisionless freestream incident half-space. The production $C_H$ responses contain total energy transfer and can differ through changes in the self-consistent incident population, impact rate, internal modes or catalytic exchange.

The corresponding production heat-response ratio uses the area-weighted retained means $A$, $B$ and $C$ defined for \eqref{eq:cll_heat_plane_defect}. Its numerator $A-B$ is the low-$\alpha_t$ contrast and its denominator $A-C$ the low-$\alpha_n$ contrast,
\begin{equation}
R_{\mathrm{prod}}=\frac{A-B}{A-C},
\qquad
\bnabla_{(A,B,C)}R_{\mathrm{prod}}
=
\begin{bmatrix}
(B-C)/(A-C)^2\\
-1/(A-C)\\
(A-B)/(A-C)^2
\end{bmatrix}.
\label{eq:production_heat_response_ratio}
\end{equation}
Here $\bnabla_{(A,B,C)}R_{\mathrm{prod}}$ is the gradient of the ratio with respect to $(A,B,C)$. Let $\widehat V_A$, $\widehat V_B$ and $\widehat V_C$ denote the three separate production Bartlett--HAC variance estimates. Under the adopted zero cross-run-covariance approximation,
\[
\widehat{\matSigma}_{ABC}
=
\operatorname{diag}
\left(
\widehat V_A,\widehat V_B,\widehat V_C
\right),
\qquad
\widehat{\operatorname{Var}}(R_{\mathrm{prod}})
=
\bnabla R_{\mathrm{prod}}^{\mathsf T}
\widehat{\matSigma}_{ABC}
\bnabla R_{\mathrm{prod}}.
\]
Although the primitive cross-run covariances are set to zero, the shared symmetric-reference estimate remains in both finite differences and therefore gives
\[
\widehat{\operatorname{Cov}}(A-B,A-C)=\widehat V_A.
\]
The primary delta-method interval was reported only when the 95\% interval for $A-C$ excluded zero; the Fieller comparison uses the same shared-reference covariance. The analytic mixture reference, two-patch event estimate and production ratio describe different populations and energy contents, with different uncertainty sources. Their comparison tests consistency for this configuration; it is not a universal equality, and no joint interval was constructed across them.

Uncertainty treatment follows the sampling structure: retained-block HAC for production coefficients, observed-timestep HAC for jointly evaluated event moments, and conditional Monte Carlo standard errors for angular resampling. Density morphology carries deterministic support-sensitivity ranges, while final-window surface rates are point estimates without temporal intervals. Cells and surface elements are quadrature units, not independent replicates. Joint event-record covariance is retained, but no covariance is inferred between the event, density, surface-rate and production acquisitions. Raw products, estimators, processed tables and validation records are archived with the data release \citep{Naevdal2026DataRelease}.

\subsection{Auxiliary time-resolved density campaign}
\label{subsec:movie_density_acquisition}

Movie~2 uses a separate density-only campaign at $Kn_\infty=0.023$ and $8.439$, excluded from both production and selected-diagnostic statistics. The former retains the registered upper and downstream density-domain extension; the latter uses the unchanged domain. For case $c$, source bin $b$, native gas cell $i$ and seed realisation $m$, let $n_{c,b,i}^{(m)}$ be SPARTA's physical total-mixture number-density estimate. The source-bin ensemble mean and its normalised value are
\begin{equation}
\overline n_{c,b,i}
=
\frac{1}{M}\sum_{m=1}^{M}n_{c,b,i}^{(m)},
\qquad
n^*_{c,b,i}
=
\frac{\overline n_{c,b,i}}{n_{\infty,c}},
\qquad M=32.
\label{eq:movie_density_ensemble}
\end{equation}
Only $n^*$ is visualised. The across-realisation sample variance, standard deviation and standard error are retained with the separate movie~2 processing records, but no pixelwise uncertainty or significance field is shown.

Each realisation began from an empty domain and was advanced by one uninterrupted eight-rank SPARTA process through
\[
(\alpha_n,\alpha_t):
(0.8,0.8)\rightarrow(0.8,0.2)\rightarrow(0.8,0.8)
\rightarrow(0.2,0.8)\rightarrow(0.8,0.8).
\]
At each internal switch, only the active plate collision assignment changed; the particles, mesh, inflow, gas-collision state, chemistry and random-number state remained continuous, and no restart was read. The first state therefore includes empty-domain filling and flow development, whereas the later states inherit the complete preceding trajectory. The repeated $(0.8,0.8)$ states are return states within that trajectory rather than independent replicate blocks.

Each wall state comprised 608 non-overlapping source bins of duration $16.5~\si{\micro\second}$ at $Kn_\infty=0.023$ and $129~\si{\micro\second}$ at $8.439$, giving state durations of $10.032$ and $78.432~\mathrm{ms}$. Within each state, displayed frame $j=0,\ldots,149$ averages source bins $4j+4,\ldots,4j+7$, using zero-based indices. The first and last four source bins are retained as guards for matched-centre temporal-window checks; the intervening 600 bins yield 150 displayed frames per state and 750 in total. The four-bin exposures are $66$ and $516~\si{\micro\second}$, respectively, and no displayed window crosses a wall switch. No temporal interpolation, frame duplication, cumulative averaging or optical-flow reconstruction was used.

Native cut-cell densities were projected by exact gas-area overlap with raster pixels, retaining separate gas components on either side of the finite plate. The unsmoothed normalised raster was retained, and a plate-aware Gaussian filter with $\sigma=2$ scientific-raster pixels at $Kn_\infty=0.023$ and $\sigma=6$ at $8.439$ was applied only for display. The case-specific colour scales were fixed throughout the movie, with ranges $[0,36]$ and $[0,3.7]$ and the neutral colour at $n^*=1$.

The panels share wall state and frame index, not dimensional time. Each clock gives the midpoint of the included movement-step times, measured since the current state began: since the start in the first state and since switching thereafter. State durations are prescribed design horizons, not measured relaxation times \citep{SPARTA24Sep2025,Naevdal2026DataRelease}.

\subsection{Stationarity and uncertainty quantification}
\label{subsec:stationarity_uncertainty}

\subsubsection{Retained-window selection}

Transient removal and uncertainty estimation use a common blockwise framework for correlated DSMC output \citep{MyongKarchaniEjtehadi2019PoF}. Candidate windows were assessed from integrated-load histories \citep{BurtBoyd2011CiCP}, with the seven wall-model cases in each Knudsen family aligned on their earliest common archived output. At candidate boundary $k$, the largest even-length prefix of the candidate-to-final sequence was divided into equal early ($r=1$) and late ($r=2$) portions. A Bartlett--HAC estimate was formed separately for each portion of $\boldsymbol z_b=(C_{D,b},C_{L,b})^{\mathsf T}$ \citep{NeweyWest1987HAC}, giving
\begin{equation}
\mathrm{SE}^{(k)}_{r,Q}
=
\left[\widehat{\matSigma}^{(k)}_{r,\mathrm{HAC}}\right]_{QQ}^{1/2},
\qquad
\widehat{\matSigma}^{(k)}_{r,\mathrm{HAC}}
=
\frac{\widehat{\matOmega}^{(k)}_r}{B_r^{(k)}},
\qquad Q\in\{C_D,C_L\}.
\label{eq:stationarity_half_hac_se}
\end{equation}
Here $\mathrm{SE}$ denotes a standard error; $B_r^{(k)}$, $\mu_{r,Q}^{(k)}$ and $\widehat{\matOmega}_r^{(k)}$ are the block count, component mean and $2\times2$ long-run covariance estimate for portion $r$. The lag rule and Bartlett weights below are applied with $B=B_r^{(k)}$. A candidate passed the primary drift screen only when
\begin{equation}
\left|\mu^{(k)}_{2,Q}-\mu^{(k)}_{1,Q}\right|
\leq
\kappa
\left[\left(\mathrm{SE}^{(k)}_{1,Q}\right)^2+\left(\mathrm{SE}^{(k)}_{2,Q}\right)^2\right]^{1/2},
\qquad Q\in\{C_D,C_L\},
\label{eq:stationarity_gate}
\end{equation}
with $\kappa=2$ and at least three blocks per portion. This is a prespecified engineering tolerance, not a critical value with calibrated confidence coverage. The primary screen combines half-window variances in quadrature; the direct contrast below checks dependence across their boundary. The first candidate passing for both coefficients in all seven cases defines the primary-selected family boundary. This boundary is then applied to $C_H$ and local surface histories, which do not enter the gate. At that boundary, each component is also fitted to
\begin{equation}
z_{b,Q}
=
\beta_{0,Q}
+
\beta_{1,Q}\mathbf{1}\!\left\{b\in\mathcal B_{\mathrm{late}}\right\}
+
\varepsilon_{b,Q},
\label{eq:stationarity_direct_hac_contrast}
\end{equation}
where $\mathbf1\{\cdot\}$ selects the late block set $\mathcal B_{\mathrm{late}}$, $\beta_{0,Q}$ is the early mean, $\beta_{1,Q}$ the late-minus-early contrast and $\varepsilon_{b,Q}$ the residual. A Bartlett--HAC sandwich covariance over the complete usable sequence accounts for dependence across the early--late boundary. The primary-selected boundary was retained only if $|\widehat\beta_{1,Q}|\leq2\,\mathrm{SE}_{\mathrm{HAC}}(\widehat\beta_{1,Q})$ for both coefficients in all seven cases. Exactly 25 later common candidates supplied a persistence audit; their archived pass states did not select or revise the accepted boundary.

A common retained window for all wall laws at each $Kn_\infty$ prevents model-dependent truncation and averaging duration from confounding the comparison. The four windows contain $1300$, $2999$, $49\,999$ and $55\,971$ blocks, spanning $0.01300$, $0.02999$, $0.6000$ and $0.6717~\mathrm{s}$ in increasing Knudsen order. Blocks span $1.0\times10^{-5}~\mathrm{s}$ in the two lower-$Kn_\infty$ families and $1.2\times10^{-5}~\mathrm{s}$ in the two higher-$Kn_\infty$ families. For odd-length candidate windows, both stationarity checks use the largest even-length prefix; final means and uncertainties use all retained blocks. Search records, selected ranges and all gate, contrast and persistence diagnostics are archived \citep{Naevdal2026DataRelease}. The first common archived candidate passed at $Kn_\infty=0.023$, $0.137$ and $3.146$; $8.439$ required a later boundary. Selection therefore identifies the earliest passing boundary on the archived output grid, not an earlier physical onset outside that record.

\PreprintBreak
\subsubsection{Correlation-aware uncertainty}

The retained non-overlapping block means, rather than the underlying tally samples, form the elementary observations for uncertainty estimation. For the integrated coefficients, block $b$ provides the paired vector
\begin{equation}
\boldsymbol{z}_b=
\begin{bmatrix}
C_{D,b}\\
C_{L,b}
\end{bmatrix},
\qquad
\overline{\boldsymbol{z}}
=
\frac{1}{B}
\sum_{b=1}^{B}\boldsymbol{z}_b,
\label{eq:integrated_block_vector}
\end{equation}
where $B$ is the number of retained blocks. Residual dependence between successive block estimates was accounted for with the same Bartlett--HAC estimator. The centred sequence and deterministic sample-size-dependent lag cutoff are
\begin{equation}
\boldsymbol{u}_b=\boldsymbol{z}_b-\overline{\boldsymbol{z}},
\qquad
L_H=
\min\!\left\{
B-1,\,
\left\lfloor
4\left(\frac{B}{100}\right)^{2/9}
\right\rfloor
\right\}.
\label{eq:hac_centering_and_lag}
\end{equation}
The power-law term in \eqref{eq:hac_centering_and_lag} is one of the pilot lag-selection formulas examined by Newey and West \citep{NeweyWest1994LagSelection}. It is used here directly as the final deterministic Bartlett truncation lag, with the outer minimum clamping it to $B-1$; the subsequent data-dependent plug-in bandwidth step is not applied. The uncertainty audit includes grouped batch means and a $5000$-replicate circular block bootstrap \citep{Schmeiser1982BatchMeans,PolitisRomano1992CircularBootstrap,Naevdal2026DataRelease}. The production and event contrasts were evaluated at the primary lag and at half, double and quadruple that lag, with integer truncation and clamping to the available sequence. These sensitivity calculations neither revise the retained windows nor select a different primary bandwidth. For integer lags $h=0,1,\ldots,L_H$, the lag-$h$ covariance, long-run covariance and covariance of the retained mean are
\begin{align}
\widehat{\matGamma}_h
&=
\frac{1}{B}
\sum_{b=h+1}^{B}
\boldsymbol{u}_b\boldsymbol{u}_{b-h}^{\mathsf T},
\\
\widehat{\matOmega}
&=
\widehat{\matGamma}_0
+
\sum_{h=1}^{L_H}
\left(
1-\frac{h}{L_H+1}
\right)
\left(
\widehat{\matGamma}_h+
\widehat{\matGamma}_h^{\mathsf T}
\right),
\\
\widehat{\matSigma}_{\overline{\boldsymbol{z}}}
&=
\frac{\widehat{\matOmega}}{B}.
\label{eq:hac_covariance}
\end{align}
The standard errors of $C_D$ and $C_L$ are obtained from the diagonal entries of $\widehat{\matSigma}_{\overline{\boldsymbol{z}}}$. The off-diagonal entries are retained because $C_D$ and $C_L$ are assembled from the same blockwise pressure and shear contributions and are therefore generally correlated. That within-case covariance is required when propagating uncertainty into $L/D$.

The lift-to-drag uncertainty was propagated from the full paired covariance by the delta method \citep{JCGM1002008}. For
\begin{equation}
g(C_D,C_L)=\frac{C_L}{C_D},
\qquad
\bnabla g\big|_{\overline{\boldsymbol{z}}}
=
\begin{bmatrix}
-\overline{C}_L/\overline{C}_D^{\,2}\\
1/\overline{C}_D
\end{bmatrix},
\end{equation}
the variance estimate is
\begin{equation}
\mathrm{SE}_{L/D}^{2}
=
\bnabla g^{\mathsf T}
\widehat{\matSigma}_{\overline{\boldsymbol{z}}}
\bnabla g.
\label{eq:ld_uncertainty}
\end{equation}
This covariance-preserving construction avoids treating the drag and lift estimates as independent quantities.

Each profile's retained block-by-station matrix was supplied to a multistation Bartlett--HAC estimator after the blockwise spatial reduction in section~\ref{subsec:dsmc_numerical_implementation}. Benchmark matrices retain 30 stations at the two lower Knudsen numbers and 10 at the two higher values; wall-model comparisons use 10 throughout. Diagonal entries of the mean covariance give the pointwise 95\% intervals. Off-diagonal cross-station entries were not used to construct simultaneous bands or stationwise model--reference response intervals.

\PreprintBreak
\subsubsection{Derived responses and shared references}

All wall-model cases used the same configured seed, but wall-law-dependent reflections caused trajectories, collision histories and subsequent random-number consumption to diverge. The cases were not deliberately synchronised as a common-random-number comparison, and divergence alone does not establish independence. Cross-run covariance was not estimated and was set to zero as an explicit approximation. For separately simulated retained means $A_m$ and $A_n$, with marginal Bartlett--HAC variances $\widehat V_m$ and $\widehat V_n$,
\begin{equation}
\widehat{\operatorname{Var}}(A_m-A_n)
=
\widehat V_m+\widehat V_n.
\label{eq:separate_case_difference_variance}
\end{equation}

For wall-law case $q\in\{\mathrm{ref},t,n\}$, where $\mathrm{ref}$ denotes CLL $(0.8,0.8)$, $t$ denotes CLL $(0.8,0.2)$ and $n$ denotes CLL $(0.2,0.8)$, let $\boldsymbol x_q$ denote the retained plate-fixed force-coefficient vector. Because the transformation implied by \eqref{eq:integrated_coefficients} is symmetric and orthogonal, the same matrix maps the retained drag--lift mean and covariance back to plate-fixed coordinates:
\begin{equation}
\boldsymbol x_q
=
\begin{bmatrix}
C_{N,q}\\
C_{T,q}
\end{bmatrix}
=
\mathsfbi{P}
\begin{bmatrix}
C_{D,q}\\
C_{L,q}
\end{bmatrix},
\qquad
\mathsfbi{P}=
\begin{bmatrix}
\sin\alpha & \cos\alpha\\
\cos\alpha & -\sin\alpha
\end{bmatrix},
\qquad
\widehat{\matSigma}^{\mathrm{NT}}_q
=
\mathsfbi{P}
\widehat{\matSigma}^{\mathrm{DL}}_q
\mathsfbi{P}^{\mathsf T}.
\label{eq:plate_fixed_response_covariance}
\end{equation}
Here $C_{N,q}$ and $C_{T,q}$ are the integrated plate-normal and plate-tangential coefficients, and $\widehat{\matSigma}^{\mathrm{DL}}_q$ is the full within-case Bartlett--HAC covariance of the retained $(C_D,C_L)$ mean. Define the low-$\alpha_t$ and low-$\alpha_n$ response vectors as
\[
\boldsymbol{\delta}_t
=
\boldsymbol x_t-\boldsymbol x_{\mathrm{ref}},
\qquad
\boldsymbol{\delta}_n
=
\boldsymbol x_n-\boldsymbol x_{\mathrm{ref}}.
\]
Under the adopted zero cross-run covariance,
\begin{equation}
\widehat{\operatorname{Cov}}\!\left(
\begin{bmatrix}
\boldsymbol{\delta}_t\\
\boldsymbol{\delta}_n
\end{bmatrix}
\right)
=
\begin{bmatrix}
\widehat{\matSigma}^{\mathrm{NT}}_t+
\widehat{\matSigma}^{\mathrm{NT}}_{\mathrm{ref}}
&
\widehat{\matSigma}^{\mathrm{NT}}_{\mathrm{ref}}\\
\widehat{\matSigma}^{\mathrm{NT}}_{\mathrm{ref}}
&
\widehat{\matSigma}^{\mathrm{NT}}_n+
\widehat{\matSigma}^{\mathrm{NT}}_{\mathrm{ref}}
\end{bmatrix}.
\label{eq:cll_response_vector_covariance}
\end{equation}
The off-diagonal blocks arise because both response vectors subtract the same estimated CLL $(0.8,0.8)$ reference; no additional cross-run covariance is introduced. The complementary-component fractions and departure from orthogonality were
\begin{equation}
f_t=
\frac{|\delta_{t,N}|}{\|\boldsymbol{\delta}_t\|_2},
\qquad
f_n=
\frac{|\delta_{n,T}|}{\|\boldsymbol{\delta}_n\|_2},
\qquad
\psi=
\cos^{-1}
\left(
\frac{
\boldsymbol{\delta}_t^{\mathsf T}\boldsymbol{\delta}_n
}{
\|\boldsymbol{\delta}_t\|_2\|\boldsymbol{\delta}_n\|_2
}
\right),
\qquad
d_\perp=
\left|\psi-\frac{\upi}{2}\right|.
\label{eq:cll_response_selectivity_metrics}
\end{equation}
Here $\delta_{t,N}$ is the plate-normal component of $\boldsymbol{\delta}_t$, $\delta_{n,T}$ is the plate-tangential component of $\boldsymbol{\delta}_n$, and $\|\cdot\|_2$ denotes the Euclidean norm.

Analytic delta-method Jacobians were applied to the joint covariance in \eqref{eq:cll_response_vector_covariance}, with $\psi$ and $d_\perp$ converted to degrees. For $g\in\{f_t,f_n,d_\perp\}$, the reported upper endpoint is $\widehat g+1.96\,\mathrm{SE}(\widehat g)$. At non-zero signed estimates, this is also the upper endpoint obtained by mapping a symmetric normal-approximation interval for the corresponding signed quantity, $\delta_{t,N}/\|\boldsymbol\delta_t\|_2$, $\delta_{n,T}/\|\boldsymbol\delta_n\|_2$ or $\psi-\upi/2$, through the absolute-value transformation. This interpretation does not require the non-negative metric itself to have a symmetric sampling distribution; it retains the first-order approximation for the signed quantity.

For the weighted surface response defined in \eqref{eq:weighted_surface_change}, $\overline C_{m,b}=\sum_i w_i C_{m,b}(\xi_i)$ was formed within every retained block and the scalar Bartlett--HAC estimator was applied to the resulting block history. Let $\widehat V_m$ and $\widehat V_{\mathrm{DM}}$ denote the corresponding Bartlett--HAC estimates of $\mathrm{Var}(\widehat{\overline C}_m)$ and $\mathrm{Var}(\widehat{\overline C}_{\mathrm{DM}})$, respectively. Under this zero-cross-covariance approximation, first-order propagation gives
\begin{equation}
\mathrm{SE}_{\Delta\overline C_m}^{2}
=
\left(
\frac{100}{|\overline C_{\mathrm{DM}}|}
\right)^2 \widehat V_m
+
\left(
\frac{-100\,\operatorname{sgn}(\overline C_{\mathrm{DM}})\overline C_m}
{|\overline C_{\mathrm{DM}}|^2}
\right)^2 \widehat V_{\mathrm{DM}}.
\label{eq:weighted_surface_response_uncertainty}
\end{equation}
Here $\operatorname{sgn}(\cdot)$ is the sign function. Forming the weighted scalar before HAC estimation includes contemporaneous and lagged cross-station covariance in its long-run variance; it does not propagate the separate multistation covariance estimate used for pointwise intervals.

Endpoint-affine defects and direct CLL responses were formed from primitive retained weighted means before propagation. The Jacobian for $\mathcal D_C^M$ includes the intermediate state, both Maxwell endpoints and the repeated diffuse mean in the numerator and denominator. CLL responses use \eqref{eq:weighted_surface_response_uncertainty} with the symmetric CLL reference. Cross-run covariances are zero by assumption, while shared-reference dependence is retained within each derived quantity. Kernel references in \eqref{eq:cll_tangential_impulse_reference}, \eqref{eq:cll_normal_impulse_reference} and~\eqref{eq:cll_mixture_heat_references} are deterministic conditional on the registered inputs, so comparison intervals include production sampling uncertainty only. Analytic Jacobians were checked against centred numerical differentiation at unrounded primitive values.

The total-energy residual in \eqref{eq:cll_heat_plane_defect} is linear in the primitive means $\boldsymbol z_H=(D,A,B,C)^{\mathsf T}$. With their mean covariance $\widehat{\matSigma}_H$,
\begin{equation}
\widehat{\operatorname{Var}}(T_H)
=\boldsymbol c_H^{\mathsf T}\widehat{\matSigma}_H\boldsymbol c_H,
\qquad
\boldsymbol c_H=\begin{bmatrix}-1&5/3&-1/3&-1/3\end{bmatrix}^{\mathsf T}.
\label{eq:cll_heat_plane_uncertainty}
\end{equation}
Under zero cross-run covariance this reduces to $\widehat V_D+(25\widehat V_A+\widehat V_B+\widehat V_C)/9$. The percentage $\mathcal E_H$ is propagated jointly, retaining $D$ in both the residual and its denominator. For the cross-kernel comparisons, the mean covariance contains all kernels, moments, recorded exchanges and both patches from each acquired run before any contrast or common-reference normalisation is formed. Thus the covariance between $\Delta Q_{\mathrm{ker}}$, $\Delta Q_{\mathrm{pop}}$ and the observed--conditional remainder is retained rather than treating their separate error bars as independent.

For the integrated wall-model response $\Delta A_m$ defined in \eqref{eq:gsi_integrated_response}, the marginal standard errors of $A_m$ and $A_{\mathrm{DM}}$ were evaluated from their respective retained block histories using Bartlett--HAC estimation. The same zero-cross-covariance approximation was used for the model--reference pair. For each case, the $L/D$ standard error retained the within-case $C_D$--$C_L$ covariance through \eqref{eq:ld_uncertainty}. First-order delta-method propagation for the two separately simulated cases then gives
\begin{equation}
\mathrm{SE}_{\Delta A_m}^{2}
=
\left(
\frac{100}{A_{\mathrm{DM}}}
\right)^2
\mathrm{SE}_{A_m}^{2}
+
\left(
\frac{100A_m}{A_{\mathrm{DM}}^{\,2}}
\right)^2
\mathrm{SE}_{A_{\mathrm{DM}}}^{2}.
\label{eq:integrated_response_uncertainty}
\end{equation}

\subsubsection{Sensitivity to cross-run covariance}
\label{subsubsec:cross_run_covariance_bound}

The unmeasured covariance between runs is assessed separately from the primary zero-covariance calculation. For a scalar statistic $g$ formed from $r$ distinct run estimates $\boldsymbol x_q$, let $\boldsymbol a_q=\bnabla_{\boldsymbol x_q}g$ at the unrounded means and $s_q^2=\boldsymbol a_q^{\mathsf T}\widehat{\matSigma}_q\boldsymbol a_q$. Holding each within-run covariance block fixed, the maximal linearised standard error over admissible joint covariance matrices is
\begin{equation}
\mathrm{SE}_{\max}(g)=\sum_{q=1}^{r}s_q,
\qquad
\mathrm{SE}_{0}(g)=\left(\sum_{q=1}^{r}s_q^2\right)^{1/2},
\qquad
\mathrm{SE}_{\max}(g)\leq\sqrt r\,\mathrm{SE}_{0}(g).
\label{eq:cross_run_covariance_bound}
\end{equation}
Cauchy--Schwarz bounds the covariance between the two scalar run contributions by $s_qs_{q'}$; summing the bounds gives \eqref{eq:cross_run_covariance_bound}. All occurrences of one reference run are combined in its gradient before calculating $s_q$, so shared-reference dependence is not counted twice. The bound was evaluated for the principal force, heat-residual and cross-kernel contrasts at the same four lag choices. For an absolute metric $|h|$, its upper endpoint is $|\widehat h|+1.96\,\mathrm{SE}_{\max}(h)$, using the corresponding signed differentiable quantity and non-zero response-vector norms. This is a sensitivity bound under the estimated marginal HAC covariances and first-order normal approximation, not a simultaneous interval or a bound on discretisation and model-form error.

\PreprintBreak
\subsubsection{Ratio confidence sets}
\label{subsubsec:ratio_confidence_sets}

The primary delta-method intervals for the production heat-response ratio and the acquired incident-energy ratios were checked using Fieller confidence sets \citep{Fieller1954IntervalEstimation}. For an estimated numerator $\widehat X$ and denominator $\widehat Y$, with their joint HAC covariance, the normal-approximation set is
\begin{equation}
\mathcal C_R=\left\{r\in\mathbb R:
(\widehat X-r\widehat Y)^2
\leq1.96^2\left[
\widehat V_X-2r\widehat C_{XY}+r^2\widehat V_Y
\right]\right\}.
\label{eq:fieller_confidence_set}
\end{equation}
For $R_{\mathrm{prod}}$, $X=A-B$, $Y=A-C$ and $\widehat C_{XY}=\widehat V_A$ under the primary cross-run approximation. For $\mathcal R_{t/n}$, the numerator, denominator and covariance come from the jointly acquired excess-energy sums. The quadratic inequality is evaluated without imposing a bounded-interval branch or discarding a denominator that is unresolved from zero. These sets are sensitivity checks using the same retained records and lag choices; they inherit the asymptotic covariance approximation rather than providing exact finite-sample coverage.

\subsubsection{Confidence intervals and benchmark comparison}

For quantities whose reported uncertainty arises solely from DSMC sampling, the primary two-sided 95\% normal-approximation confidence intervals use the half-width
\begin{equation}
U_{95}=1.96\,\mathrm{SE}.
\label{eq:confidence_halfwidth}
\end{equation}
An individual wall-model response was classified as resolved from zero when its 95\% interval excluded zero. No familywise-error or false-discovery-rate adjustment was applied; these are comparison-wise statements, not simultaneous coverage guarantees. Benchmark profiles use five separately digitised and interpolated curves, and integrated references use five separately digitised scalar values. Let $\mathrm{SE}_{\mathrm{DSMC}}$ be the present mean's Bartlett--HAC standard error and $s_{\mathrm{dig}}$ the repeat sample standard deviation, giving $\mathrm{SE}_{\mathrm{dig}}=s_{\mathrm{dig}}/\sqrt5$. Assuming independence of present-DSMC sampling and digitisation repeatability, the combined comparison half-width uses the Welch--Satterthwaite approximation,
\begin{equation}
U_{\mathrm{comb}}
=
t_{0.975,\nu_{\mathrm{eff}}}
\left(\mathrm{SE}_{\mathrm{DSMC}}^2+\mathrm{SE}_{\mathrm{dig}}^2\right)^{1/2},
\qquad
\nu_{\mathrm{eff}}
=
\frac{\left(\mathrm{SE}_{\mathrm{DSMC}}^2+\mathrm{SE}_{\mathrm{dig}}^2\right)^2}
{\mathrm{SE}_{\mathrm{dig}}^4/4},
\label{eq:benchmark_combined_uncertainty}
\end{equation}
where $t_{0.975,\nu_{\mathrm{eff}}}$ is the 97.5th percentile of Student's $t$ distribution. The coverage-factor calculation assigns effectively infinite degrees of freedom to the DSMC contribution and four to the five digitisation repeats \citep{JCGM1002008}. With $D_A=A_{\mathrm{DSMC}}-A_{\mathrm{ref}}$, positive differences place the present mean above the digitised reference. The ratio $|D_A|/U_{\mathrm{comb}}$ expresses discrepancy in units of the combined half-width; it is a descriptive scale, not a validation acceptance test.

These intervals quantify present-DSMC sampling and repeat-digitisation variability conditional on the adopted models. They exclude uncertainty in the original benchmark, systematic digitisation error and systematic effects of GSI model form, freestream reconstruction, collision data, catalytic treatment and numerical implementation. They are therefore not total predictive-uncertainty bounds \citep{OberkampfBarone2006JCP}.

The same family-selected retained windows, blockwise integrations and uncertainty estimators were applied to the reproduced diffuse Maxwell baseline and to every Maxwell and CLL sensitivity case. Comparisons within each $Kn_\infty$ family therefore share common temporal support and a common statistical construction, while residual DSMC sampling variability is estimated separately for each simulation.
\renewcommand{\floatpagefraction}{0.90}
\section{Results and discussion}
\label{SEC:res}

\subsection{Diffuse Maxwell benchmark reproduction}
\label{SUBSEC:baseline}

\begin{figure}
    \centering
    \includegraphics[width=\textwidth]{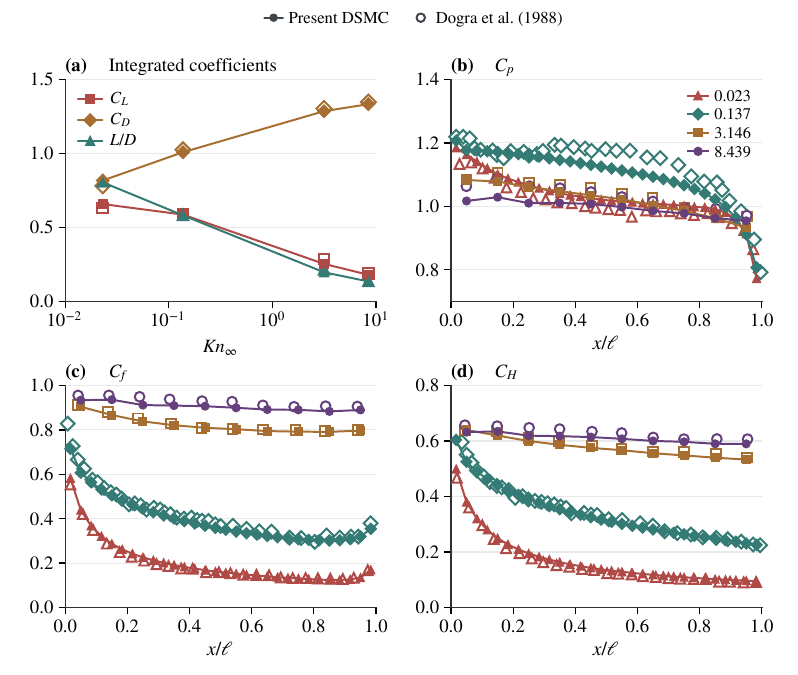}
    \caption{Compression-side comparison with the diffuse Maxwell DSMC benchmark of \citet{Dogra1988FlatPlate}: (a) integrated $C_D$, $C_L$ and $L/D$; (b) pressure coefficient $C_p$; (c) skin-friction coefficient $C_f$; and (d) heat-transfer coefficient $C_H$. Filled markers denote present DSMC means; open markers denote means of five benchmark digitisations. The $Kn_\infty$ legend in (b) applies to (b--d). Lines are guides to the eye. Pointwise 95\% confidence intervals for the surface profiles are tabulated in the data release and omitted here.}
    \label{fig:dogra_validation}
\end{figure}

The diffuse Maxwell branch, $\alpha_M=1.0$, provides the common reference for the wall-model comparisons. Figure~\ref{fig:dogra_validation} and tables~\ref{tab:baseline_integrated_metrics}--\ref{tab:baseline_surface_metrics} compare this branch with the published benchmark at all four conditions. The comparison assesses the reconstruction described in section~\ref{SEC:meth}; it does not calibrate the accommodation parameters. For this comparison, present profiles retain 30 native elements at the two lower $Kn_\infty$ values and use three-element blockwise aggregation at the two higher values. Benchmark metrics use 28--30 native comparison locations at the lower conditions and all ten stations at the higher conditions, with repeat-wise reference evaluation as specified in section~\ref{subsec:response_metrics}.

The diffuse branch reproduces the benchmark trend: $C_D$ increases along the Knudsen sequence, while $C_L$ and $L/D$ decrease. Drag discrepancies remain below $4.65\%$ at all four conditions and below $2\%$ at the two highest. The largest relative lift discrepancy is $-11.66\%$ at $Kn_\infty=3.146$, corresponding to an absolute coefficient difference of $-0.0335$. The smaller lift at high rarefaction makes this relative comparison particularly sensitive to differences in the opposing pressure and shear contributions.

The surface profiles reproduce the complementary benchmark behaviour: pressure varies comparatively little across the conditions, whereas skin friction and heat transfer change substantially \citep{Dogra1988FlatPlate}. Every weighted NRMSE and NMAE is below $0.057$. The closest profile agreement occurs at $Kn_\infty=3.146$, where all six metrics are at or below $0.0191$; at $8.439$, every NRMSE is below $0.040$. Pointwise 95\% intervals are provided in the released surface tables.

Small profile discrepancies do not imply statistical equivalence. Only the $C_L$ and $L/D$ comparisons at $Kn_\infty=0.137$ fall within the combined sampling and repeat-digitisation scale among the 12 integrated comparisons. That scale excludes uncertainty in the original calculation, systematic digitisation error and systematic differences between implementations. The benchmark therefore documents agreement in trends and profile shape, together with finite offsets; it is not an experimental validation, a material calibration or an independent bound on numerical error.

The subsequent comparisons ask a different question: how does changing the wall law alter the solution generated by this fixed formulation? Each contrast uses a reference at the same freestream state and under the same numerical and statistical treatment. It therefore measures the computed response to the scattering-law change, including the accompanying change in the incident population, rather than a correction to the benchmark.

\begin{table}
\centering
\setlength{\tabcolsep}{2.8pt}
\renewcommand{\arraystretch}{1.08}
\begin{tabular*}{\textwidth}{@{\extracolsep{\fill}}
S[table-format=1.3]
c
S[table-format=1.4]
S[table-format=1.4]
S[table-format=+2.2]
S[table-format=1.4]
S[table-format=2.2]
@{}}
\toprule
{$Kn_\infty$}
& {Quantity}
& {DSMC mean}
& {Benchmark}
& {$\delta A$ [\%]}
& {$U_{\mathrm{comb}}$}
& {$\lvert D_A\rvert/U_{\mathrm{comb}}$} \\
\midrule
0.023 & {$C_D$} & 0.8175 & 0.7812 & +4.64  & 0.0039 & 9.29  \\
      & {$C_L$} & 0.6582 & 0.6328 & +4.02  & 0.0029 & 8.83  \\
      & {$L/D$} & 0.8052 & 0.8136 & -1.04  & 0.0036 & 2.35  \\
\addlinespace[2pt]
0.137 & {$C_D$} & 1.0088 & 1.0248 & -1.56  & 0.0036 & 4.45  \\
      & {$C_L$} & 0.5872 & 0.5883 & -0.18  & 0.0029 & 0.38  \\
      & {$L/D$} & 0.5821 & 0.5847 & -0.45  & 0.0033 & 0.80  \\
\addlinespace[2pt]
3.146 & {$C_D$} & 1.2859 & 1.3040 & -1.39  & 0.0052 & 3.47  \\
      & {$C_L$} & 0.2541 & 0.2876 & -11.66 & 0.0015 & 22.01  \\
      & {$L/D$} & 0.1976 & 0.2177 & -9.22  & 0.0010 & 19.95  \\
\addlinespace[2pt]
8.439 & {$C_D$} & 1.3332 & 1.3475 & -1.06  & 0.0072 & 1.98  \\
      & {$C_L$} & 0.1805 & 0.1883 & -4.10  & 0.0012 & 6.54  \\
      & {$L/D$} & 0.1354 & 0.1391 & -2.65  & 0.0010 & 3.58  \\
\bottomrule
\end{tabular*}
\caption{Integrated compression-side comparison of diffuse Maxwell results with \citet{Dogra1988FlatPlate}. Benchmark values $A_{\mathrm{ref}}$ are means of five digitisations of each quantity; $L/D$ was digitised directly. Here $D_A=A_{\mathrm{DSMC}}-A_{\mathrm{ref}}$ and $\delta A=100D_A/A_{\mathrm{ref}}$. The approximate 95\% comparison half-width $U_{\mathrm{comb}}$ includes only present DSMC sampling and repeat-digitisation uncertainty \eqref{eq:benchmark_combined_uncertainty}; uncertainty in the original calculation and systematic errors are excluded. Derived columns use unrounded values.}\label{tab:baseline_integrated_metrics}
\end{table}

\begin{table}
\centering
\setlength{\tabcolsep}{7pt}
\renewcommand{\arraystretch}{1.08}
\begin{tabular}{@{}
S[table-format=1.3]
*{6}{S[table-format=1.4]}
@{}}
\toprule
& \multicolumn{2}{c}{$C_p$}
& \multicolumn{2}{c}{$C_f$}
& \multicolumn{2}{c}{$C_H$} \\[1pt]
{$Kn_\infty$}
& {NRMSE} & {NMAE}
& {NRMSE} & {NMAE}
& {NRMSE} & {NMAE} \\
\midrule
0.023 & 0.0263 & 0.0253 & 0.0415 & 0.0351 & 0.0564 & 0.0565 \\
0.137 & 0.0378 & 0.0333 & 0.0432 & 0.0375 & 0.0240 & 0.0185 \\
3.146 & 0.0191 & 0.0175 & 0.0078 & 0.0068 & 0.0187 & 0.0174 \\
8.439 & 0.0388 & 0.0365 & 0.0255 & 0.0247 & 0.0323 & 0.0311 \\
\bottomrule
\end{tabular}
\caption{Surface-profile agreement between diffuse Maxwell results and the mean of five benchmark digitisations. The dimensionless, trapezoidally weighted metrics are defined in \eqref{eq:profile_error_metrics}.}
\label{tab:baseline_surface_metrics}
\end{table}

\subsection{Surface-moment response to wall scattering}
\label{SUBSEC:surf_resp}

We first examine pressure, shear and heat transfer separately, before combining pressure and shear into drag and lift in section~\ref{SUBSEC:aerodyn_resp}. Under the common freestream normalisation, these coefficients measure wall-normal momentum, wall-parallel momentum and total gas-to-wall energy transfer. The Maxwell sweep changes the diffuse-event probability $\alpha_M$ and hence both translational and internal-energy accommodation. The CLL perturbations instead vary normal or tangential translational accommodation while retaining common internal-mode accommodation. They test distinct directions in the prescribed scattering law, not calibrated material responses.

Unless stated otherwise, responses are relative to diffuse Maxwell reflection at the same $Kn_\infty$. Profiles use 10 stations formed by blockwise area-weighted aggregation of the 30 native histories; integrated forces retain all 30 elements. The tables distinguish the signed change of the area-weighted mean, $\Delta\overline C$, from the unsigned profile discrepancy, NRMSE. The question is whether a change in one accommodation parameter remains confined to its corresponding momentum component, and whether the measured response has the magnitude prescribed by the kernel at fixed incidence.

\subsubsection{Tangential-momentum pathway}

\begin{figure}
    \centering
    \includegraphics[width=\textwidth]{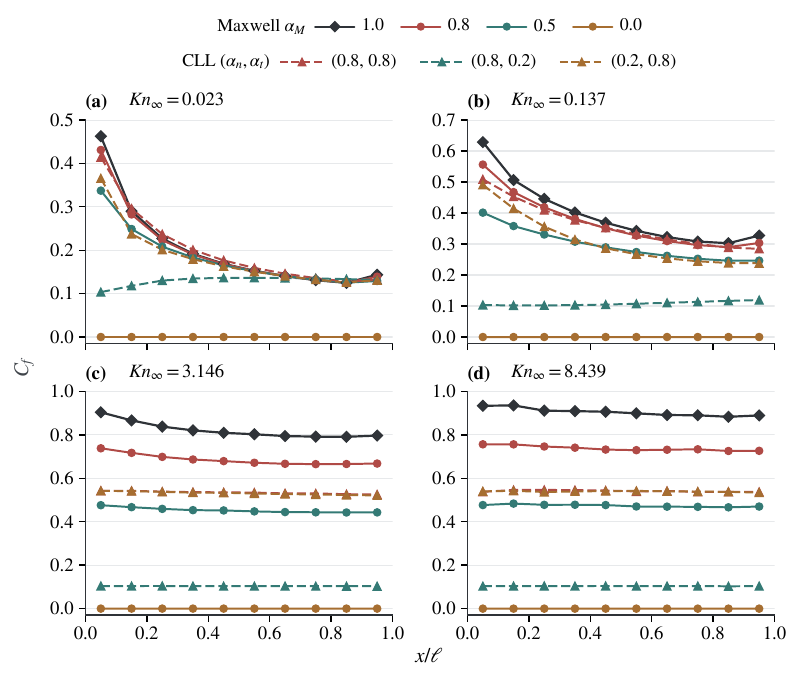}
    \caption{Compression-side profiles of the skin-friction coefficient $C_f$ for Maxwell and CLL scattering at (a) $Kn_\infty=0.023$, (b) $0.137$, (c) $3.146$ and (d) $8.439$. Markers denote time means at ten stations formed by blockwise area-weighted aggregation of the 30 native surface elements; lines are guides to the eye. Pointwise 95\% confidence intervals are tabulated in the data release and omitted here.}
    \label{fig:gsi_cf_response}
\end{figure}

The skin-friction profiles and summary metrics are shown in figure~\ref{fig:gsi_cf_response} and table~\ref{tab:gsi_cf_summary}.

\begin{table}
\centering
\setlength{\tabcolsep}{1.8pt}
\renewcommand{\arraystretch}{1.08}
\begin{tabular*}{\textwidth}{@{\extracolsep{\fill}}l
S[table-format=+3.2] S[table-format=1.3]
S[table-format=+3.2] S[table-format=1.3]
S[table-format=+3.2] S[table-format=1.3]
S[table-format=+3.2] S[table-format=1.3]@{}}
\toprule
& \multicolumn{2}{c}{$Kn_\infty=0.023$}
& \multicolumn{2}{c}{$Kn_\infty=0.137$}
& \multicolumn{2}{c}{$Kn_\infty=3.146$}
& \multicolumn{2}{c}{$Kn_\infty=8.439$} \\[1pt]
Wall model
& {$\Delta\overline{C}_f$ [\%]} & {NRMSE}
& {$\Delta\overline{C}_f$ [\%]} & {NRMSE}
& {$\Delta\overline{C}_f$ [\%]} & {NRMSE}
& {$\Delta\overline{C}_f$ [\%]} & {NRMSE} \\
\midrule
Maxwell $\alpha_M=0.8$       & -2.39 & 0.046 & -6.42 & 0.076 & -16.55 & 0.166 & -18.47 & 0.185 \\
Maxwell $\alpha_M=0.5$       & -10.40 & 0.188 & -24.95 & 0.272 & -44.90 & 0.450 & -47.68 & 0.477 \\
Maxwell $\alpha_M=0.0$       & -100.00 & 1.000 & -100.02 & 1.000 & -100.07 & 1.001 & -100.08 & 1.001 \\
\addlinespace[2pt]
CLL $(0.8,0.8)$              & -0.35 & 0.075 & -8.44 & 0.114 & -35.00 & 0.352 & -40.21 & 0.402 \\
CLL $(0.8,0.2)$              & -36.29 & 0.582 & -72.67 & 0.749 & -87.45 & 0.875 & -88.62 & 0.886 \\
CLL $(0.2,0.8)$              & -10.14 & 0.161 & -21.50 & 0.214 & -35.33 & 0.355 & -40.44 & 0.405 \\
\bottomrule
\end{tabular*}
\caption{Compression-side skin-friction response relative to diffuse Maxwell reflection at the same $Kn_\infty$. $\Delta\overline{C}_f$ is the signed percentage change in the area-weighted mean; NRMSE is the dimensionless, area-weighted profile discrepancy defined in \eqref{eq:profile_error_metrics}. Small specular shear residuals arise from mass-changing catalytic exchange.}\label{tab:gsi_cf_summary}
\end{table}

Skin friction gives the clearest initial test. Within the Maxwell family, decreasing $\alpha_M$ lowers the mean $C_f$ at every condition, but the reduction depends strongly on the surrounding flow. For $\alpha_M=0.5$, it grows from $10.40\%$ at $Kn_\infty=0.023$ to $44.90\%$ and $47.68\%$ at $3.146$ and $8.439$. Specular non-reactive impacts preserve tangential momentum exactly; the retained catalytic exchange leaves a signed residual with a maximum magnitude of approximately $0.08\%$ of the diffuse-reference shear. Thus the specular mean is effectively zero even where the intermediate Maxwell states depart strongly from a proportional response.

At the two higher conditions, the CLL pairs separate according to tangential accommodation. With $\alpha_t=0.8$, changing $\alpha_n$ from $0.8$ to $0.2$ leaves nearly coincident shear profiles: the diffuse-relative mean responses are $-35.00\%$ and $-35.33\%$ at $Kn_\infty=3.146$, and $-40.21\%$ and $-40.44\%$ at $8.439$. Reducing $\alpha_t$ instead gives $-87.45\%$ and $-88.62\%$. The separation caused by changing $\alpha_n$ is therefore only $0.33$ and $0.23$ percentage points at these conditions, compared with $9.79$ and $13.06$ points at the two lower conditions. The selected high-$Kn_\infty$ shear responses are nearly componentwise, whereas the lower-$Kn_\infty$ responses are not.

Within the imposed model, the lower-$Kn_\infty$ difference cannot be attributed to a changed tangential kernel, because both cases retain $\alpha_t=0.8$. It arises through the incident population on which that kernel acts, including its impact rate, velocities and species composition. The shear response therefore distinguishes a property of the local scattering law from a property of the coupled flow--surface system.

\subsubsection{Normal-momentum pathway}

\begin{figure}
    \centering
    \includegraphics[width=\textwidth]{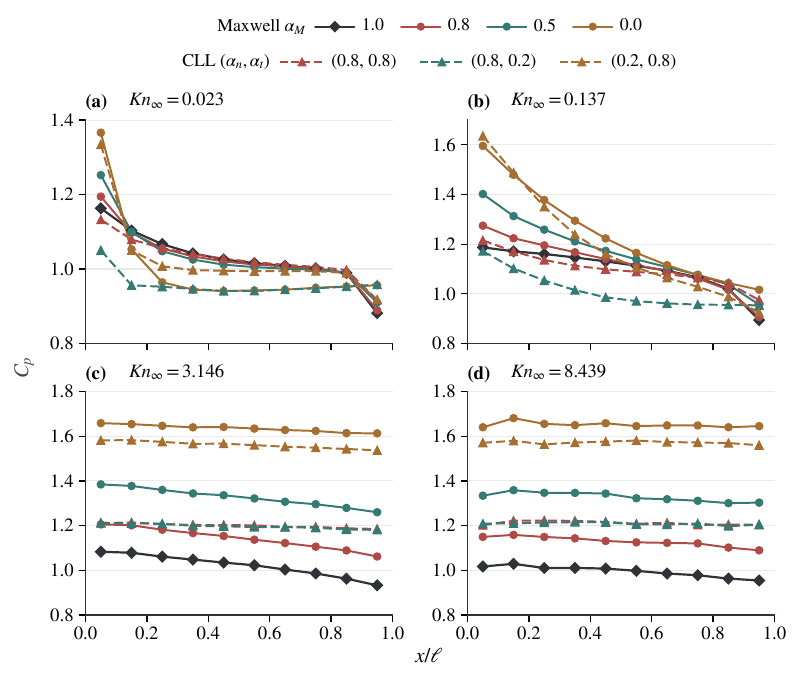}
    \caption{Compression-side profiles of the pressure coefficient $C_p$ for Maxwell and CLL scattering at (a) $Kn_\infty=0.023$, (b) $0.137$, (c) $3.146$ and (d) $8.439$. Markers denote time means at ten stations formed by blockwise area-weighted aggregation of the 30 native surface elements; lines are guides to the eye. Pointwise 95\% confidence intervals are tabulated in the data release and omitted here.}
    \label{fig:gsi_cp_response}
\end{figure}

The pressure profiles and summary metrics are shown in figure~\ref{fig:gsi_cp_response} and table~\ref{tab:gsi_cp_summary}.

\begin{table}
\centering
\setlength{\tabcolsep}{1.8pt}
\renewcommand{\arraystretch}{1.08}
\begin{tabular*}{\textwidth}{@{\extracolsep{\fill}}l
S[table-format=+3.2] S[table-format=1.3]
S[table-format=+3.2] S[table-format=1.3]
S[table-format=+3.2] S[table-format=1.3]
S[table-format=+3.2] S[table-format=1.3]@{}}
\toprule
& \multicolumn{2}{c}{$Kn_\infty=0.023$}
& \multicolumn{2}{c}{$Kn_\infty=0.137$}
& \multicolumn{2}{c}{$Kn_\infty=3.146$}
& \multicolumn{2}{c}{$Kn_\infty=8.439$} \\[1pt]
Wall model
& {$\Delta\overline{C}_p$ [\%]} & {NRMSE}
& {$\Delta\overline{C}_p$ [\%]} & {NRMSE}
& {$\Delta\overline{C}_p$ [\%]} & {NRMSE}
& {$\Delta\overline{C}_p$ [\%]} & {NRMSE} \\
\midrule
Maxwell $\alpha_M=0.8$       & +0.01 & 0.011 & +1.92 & 0.032 & +11.87 & 0.119 & +13.44 & 0.134 \\
Maxwell $\alpha_M=0.5$       & +0.42 & 0.030 & +6.30 & 0.085 & +29.88 & 0.299 & +33.47 & 0.335 \\
Maxwell $\alpha_M=0.0$       & -2.74 & 0.092 & +12.77 & 0.172 & +60.12 & 0.601 & +65.88 & 0.659 \\
\addlinespace[2pt]
CLL $(0.8,0.8)$              & -0.11 & 0.017 & +0.00 & 0.031 & +17.51 & 0.179 & +21.75 & 0.218 \\
CLL $(0.8,0.2)$              & -6.85 & 0.088 & -7.80 & 0.096 & +17.19 & 0.176 & +21.42 & 0.215 \\
CLL $(0.2,0.8)$              & -0.23 & 0.062 & +9.12 & 0.170 & +52.88 & 0.529 & +57.90 & 0.579 \\
\bottomrule
\end{tabular*}
\caption{Compression-side pressure response relative to diffuse Maxwell reflection at the same $Kn_\infty$. $\Delta\overline{C}_p$ is the signed percentage change in the area-weighted mean; NRMSE is the dimensionless, area-weighted profile discrepancy defined in \eqref{eq:profile_error_metrics}.}
\label{tab:gsi_cp_summary}
\end{table}

Pressure provides the complementary test, but its simple normal-accommodation ordering is absent at $Kn_\infty=0.023$. All mean responses remain within approximately $7\%$ of the diffuse value, while several profiles cross the reference and partly cancel on integration. The mean shifts for Maxwell $\alpha_M=0.8$, CLL $(0.8,0.8)$ and CLL $(0.2,0.8)$ are not resolved from zero by their 95\% intervals. Small mean pressure changes therefore coexist with local redistribution of the load.

A normal-accommodation ordering emerges at the higher conditions. In the Maxwell family, lower $\alpha_M$ raises pressure, with specular responses of $+60.12\%$ and $+65.88\%$ at $Kn_\infty=3.146$ and $8.439$. For incident normal energies well above the wall thermal scale, specular reversal gives a larger mean reflected-normal momentum than diffuse re-emission, increasing the impulse transferred per impact. The observed high-$Kn_\infty$ ordering follows this conditional expectation.

The CLL comparison makes this separation more precise. At fixed $\alpha_n=0.8$, changing $\alpha_t$ from $0.8$ to $0.2$ gives nearly coincident pressure responses: $+17.51\%$ and $+17.19\%$ at $Kn_\infty=3.146$, and $+21.75\%$ and $+21.42\%$ at $8.439$. Reducing $\alpha_n$ instead raises the responses to $+52.88\%$ and $+57.90\%$. The pressure separation associated with changing only $\alpha_t$ contracts from $6.74$ and $7.80$ percentage points at the two lower conditions to $0.32$ and $0.34$ points at the higher conditions. Pressure and shear thus recover complementary parameter sensitivities within the sampled CLL pairs.

The lower-$Kn_\infty$ pressure deficit of CLL $(0.8,0.2)$ is especially informative because its normal kernel is identical to that of CLL $(0.8,0.8)$. Within the imposed model, the difference must therefore enter through the incident population, rather than through a different conditional normal-scattering law. In \eqref{eq:cll_conditional_memory_moments}, reducing $\alpha_t$ doubles the coefficient multiplying incident tangential velocity in the conditional reflected mean, from $\sqrt{0.2}$ to $\sqrt{0.8}$, favouring downstream transport at fixed incidence. The separately acquired diagnostics at $Kn_\infty=0.023$ show lower near-wall density and diffuse-relative complete-side impact-rate responses of $-36.05\%$ and $-18.25\%$ for the low-$\alpha_t$ and symmetric cases, respectively (section~\ref{SUBSEC:pathway_diagnostics}). These aggregate changes are consistent with enhanced downstream transport.

\subsubsection{Energy-transfer pathway}

\begin{figure}
    \centering
    \includegraphics[width=\textwidth]{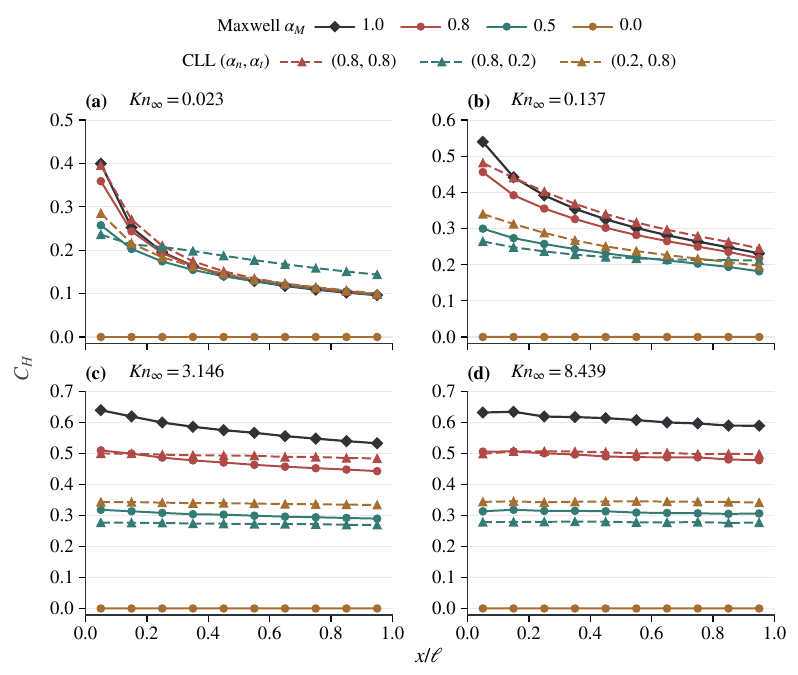}
    \caption{Compression-side profiles of the heat-transfer coefficient $C_H$ for Maxwell and CLL scattering at (a) $Kn_\infty=0.023$, (b) $0.137$, (c) $3.146$ and (d) $8.439$. $C_H$ includes translational, internal and chemical contributions; positive values denote net energy transfer from gas to wall. Markers denote time means at ten stations formed by blockwise area-weighted aggregation of the 30 native surface elements; lines are guides to the eye. Pointwise 95\% confidence intervals are tabulated in the data release and omitted here.}
    \label{fig:gsi_ch_response}
\end{figure}

The heat-transfer profiles and summary metrics are shown in figure~\ref{fig:gsi_ch_response} and table~\ref{tab:gsi_ch_summary}.

\begin{table}
\centering
\setlength{\tabcolsep}{1.8pt}
\renewcommand{\arraystretch}{1.08}
\begin{tabular*}{\textwidth}{@{\extracolsep{\fill}}l
S[table-format=+3.2] S[table-format=1.3]
S[table-format=+3.2] S[table-format=1.3]
S[table-format=+3.2] S[table-format=1.3]
S[table-format=+3.2] S[table-format=1.3]@{}}
\toprule
& \multicolumn{2}{c}{$Kn_\infty=0.023$}
& \multicolumn{2}{c}{$Kn_\infty=0.137$}
& \multicolumn{2}{c}{$Kn_\infty=3.146$}
& \multicolumn{2}{c}{$Kn_\infty=8.439$} \\[1pt]
Wall model
& {$\Delta\overline{C}_H$ [\%]} & {NRMSE}
& {$\Delta\overline{C}_H$ [\%]} & {NRMSE}
& {$\Delta\overline{C}_H$ [\%]} & {NRMSE}
& {$\Delta\overline{C}_H$ [\%]} & {NRMSE} \\
\midrule
Maxwell $\alpha_M=0.8$       & -2.68 & 0.068 & -8.80 & 0.104 & -18.29 & 0.184 & -19.32 & 0.193 \\
Maxwell $\alpha_M=0.5$       & -12.82 & 0.250 & -31.49 & 0.345 & -47.63 & 0.477 & -49.00 & 0.490 \\
Maxwell $\alpha_M=0.0$       & -100.00 & 1.000 & -100.00 & 1.000 & -99.99 & 1.000 & -99.99 & 1.000 \\
\addlinespace[2pt]
CLL $(0.8,0.8)$              & +4.05 & 0.048 & +1.58 & 0.064 & -14.58 & 0.154 & -17.73 & 0.179 \\
CLL $(0.8,0.2)$              & +7.70 & 0.342 & -32.92 & 0.383 & -52.57 & 0.528 & -54.35 & 0.544 \\
CLL $(0.2,0.8)$              & -8.05 & 0.199 & -24.75 & 0.275 & -41.21 & 0.415 & -43.60 & 0.436 \\
\bottomrule
\end{tabular*}
\caption{Compression-side total heat-transfer response relative to diffuse Maxwell reflection at the same $Kn_\infty$. $\Delta\overline{C}_H$ is the signed percentage change in the area-weighted mean; NRMSE is the dimensionless, area-weighted profile discrepancy defined in \eqref{eq:profile_error_metrics}.}
\label{tab:gsi_ch_summary}
\end{table}

Heat transfer depends on both accommodation channels and on the incident population. The local balance in \eqref{eq:surface_energy_balance} expresses energy flux as impact rate times mean exchange per incident molecule; neither factor alone determines $C_H$. The Maxwell mean reductions increase from $2.68\%$ to $19.32\%$ for $\alpha_M=0.8$ and from $12.82\%$ to $49.00\%$ for $\alpha_M=0.5$ between the lowest and highest conditions. Specular non-reactive impacts exchange no energy with the stationary wall, and the small retained catalytic residual leaves $\Delta\overline C_H\simeq-100\%$ throughout.

\PreprintBreak
At $Kn_\infty=0.023$, the CLL $(0.8,0.2)$ mean conceals a substantial spatial redistribution: $\Delta\overline C_H=+7.70\%$, yet the local contrast changes from $-40.94\%$ at $\xi=0.05$ to $+48.84\%$ at $\xi=0.95$. It crosses the diffuse profile between $\xi=0.15$ and $0.25$ and gives an NRMSE of $0.342$. The positive mean is the residual of a leading-edge reduction and a downstream enhancement, not a uniform increase. This reversal concerns the wall-model contrast, not the direction of heat flow. The complete-side factorisation in section~\ref{SUBSEC:pathway_diagnostics} assesses aggregate compensation between impact rate and energy per incident molecule; it does not by itself identify the local cause of the crossing.

At the two higher conditions, both CLL accommodation reductions lower heat transfer, but the tangential perturbation has the larger effect. Relative to CLL $(0.8,0.8)$, reducing $\alpha_t$ gives further reductions of $37.99$ and $36.62$ percentage points on the diffuse-reference scale at $Kn_\infty=3.146$ and $8.439$; reducing $\alpha_n$ gives $26.63$ and $25.87$ points. The conditional energy identities provide a reference for this asymmetry. For the equal $0.6$ accommodation decrements, the fixed-population CLL identities in \eqref{eq:cll_conditional_translational_transfer} make the ratio of the translational heat reductions equal to the incident tangential-to-normal excess-energy ratio.

The production contrast ratios are $1.4265$ and $1.4157$, with covariance-aware 95\% intervals of $1.4209$--$1.4322$ and $1.4031$--$1.4283$, respectively. Both are close in magnitude to the directed hyperthermal value $\cot^2 40^\circ=1.4203$, although that value lies just below the interval at $Kn_\infty=3.146$. At $8.439$, retaining finite-temperature mixture effects gives $\mathcal R_{\mathrm{mix}}=1.4162$; the pooled nominal-window incident events for CLL $(0.8,0.8)$ give $1.4174$ (95\% interval $1.4100$--$1.4247$), compared with the production ratio $1.4157$. The finite-temperature reference lies within both intervals. The numerical agreement provides a consistency check across constructions with different populations and energy contents.

The separately acquired compression-side component tally supports the translational interpretation. Translation contributes $99.93\%$ and $100.13\%$ of the signed total-rate contrasts for the low-$\alpha_t$ and low-$\alpha_n$ perturbations. The latter exceeds $100\%$ because the net internal and chemical contribution weakly opposes the translational reduction; the absolute-component fractions are $99.93\%$ and $99.86\%$. These are descriptive component fractions, without temporal confidence intervals. Together with the rate factorisation in section~\ref{subsubsec:diagnostic_rate_exchange}, the comparison supports an incident excess-energy explanation for the unequal heat reductions at $Kn_\infty=8.439$.

\subsubsection{Recovery of kernel-level surface responses}

\begin{figure}
    \centering
    \includegraphics[width=\textwidth]{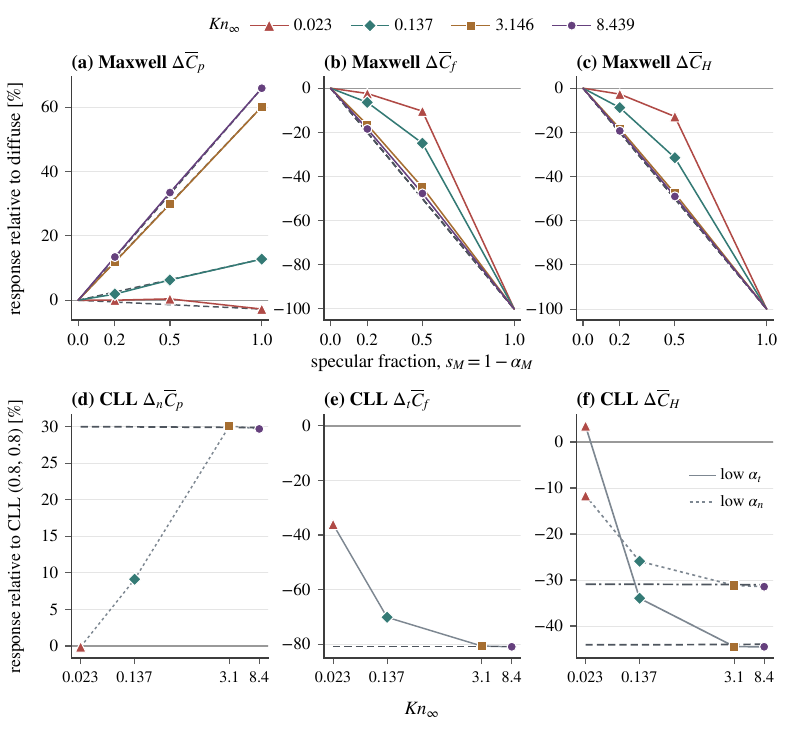}
    \caption{Compression-side area-mean responses compared with kernel references. (a--c) Maxwell pressure, skin-friction and heat-transfer responses relative to diffuse reflection versus specular fraction $s_M=1-\alpha_M$; dashed chords join separately simulated diffuse and specular endpoints, giving the affine reference for invariant incident-state flux. (d--f) CLL responses relative to $(\alpha_n,\alpha_t)=(0.8,0.8)$ versus $Kn_\infty$: (d) pressure for low $\alpha_n$, $(0.2,0.8)$; (e) skin friction for low $\alpha_t$, $(0.8,0.2)$; (f) heat transfer for both. Solid and short-dashed CLL connectors denote low $\alpha_t$ and low $\alpha_n$, respectively. Non-reactive CLL references use finite-temperature collisionless freestream-half-space moments in (d,f) and exact fixed-population tangential-momentum scaling in (e). CLL reference lines are long-dashed, except the dash-dotted low-$\alpha_n$ heat reference. Production heat transfer includes total energy exchange; CLL heat references are translational only. Markers denote production estimates; connecting lines are guides to the eye. Comparison-wise 95\% sampling intervals, including endpoint uncertainty for Maxwell affine defects, are tabulated in the data release and omitted here.}
    \label{fig:kernel_reference_recovery}
\end{figure}

The surface profiles suggest a change from coupled to nearly componentwise responses. Figure~\ref{fig:kernel_reference_recovery} tests this interpretation against the magnitudes implied by the kernels. For Maxwell reflection, fixed incident-state flux requires every surface-exchange moment to be affine in $\alpha_M$. The diagnostic in \eqref{eq:maxwell_affine_defect} compares each intermediate production response with the chord joining the separately simulated endpoints; it is not a resampling of every kernel on one common incident record. The maximum absolute shear defect over the two intermediate states, $\alpha_M=0.8$ and $0.5$, falls from $39.60$ to $25.06$, $5.13$ and $2.36$ percentage points in increasing Knudsen order. The corresponding maximum heat-transfer defects are $37.18$, $18.51$, $2.37$ and $1.00$ points. At $\alpha_M=0.5$, the contrast is immediate: the approximately $-50\%$ endpoint-affine shear and heat responses become only $-10.40\%$ and $-12.82\%$ in the self-consistent $Kn_\infty=0.023$ flow, but $-47.68\%$ and $-49.00\%$ at $8.439$.

Pressure stays much closer to its affine chord on the diffuse-reference scale, with maximum defects of $1.79$, $0.64$, $0.18$ and $0.53$ percentage points. Even these smaller defects can change the sign of a weak response: at $Kn_\infty=0.023$ and $\alpha_M=0.5$, interpolation gives $-1.37\%$ whereas production gives $+0.42\%$. Because the Maxwell kernel is affine at fixed incidence, a resolved defect diagnoses changes in incident-state flux, subject to numerical accuracy. The shear and heat defects contract strongly across the sampled sequence, but finite residuals remain at $Kn_\infty=8.439$.

CLL provides a complementary, moment-specific test. Its tangential impulse and componentwise excess-energy transfers have accommodation scalings that do not depend on the detailed form of a fixed incident distribution. The mean reflected normal speed, and hence pressure, does depend on that distribution \citep{Lord1991ExtensionsCLL,ChenEtAl2023Physisorption}. For these tests, the asymmetric cases are compared directly with the symmetric CLL case $(0.8,0.8)$ using \eqref{eq:cll_pairwise_surface_responses}; the normalisation therefore differs from that of the preceding diffuse-relative tables. For the selected parameter pairs, the exact non-reactive tangential reference and the directed cold-limit normal reference are:
\begin{equation}
\begin{aligned}
\mathcal K_{f,t}^{\mathrm{fp}}
&=
100\left[
\frac{1-\sqrt{1-0.2}}
     {1-\sqrt{1-0.8}}
-1
\right]
\simeq-80.90\%,\\
\mathcal K_{p,n}^{\mathrm{cold}}
&=
100\left[
\frac{1+\sqrt{1-0.2}}
     {1+\sqrt{1-0.8}}
-1
\right]
\simeq+30.90\%.
\end{aligned}
\label{eq:cll_directed_kernel_references}
\end{equation}
The tangential relation holds for any common non-reactive incident population with positive reference shear, independently of its species composition or detailed velocity distribution. The normal relation is a cold limit; the finite-temperature comparisons instead use the species-resolved collisionless incident-half-space reference in \eqref{eq:cll_normal_impulse_reference}.

The direct low-$\alpha_t$ shear responses are $-36.07\%$, $-70.15\%$, $-80.70\%$ and $-80.96\%$. Their departures from the fixed-population reference contract from $+44.84$ and $+10.75$ to $+0.21$ and $-0.06$ percentage points. The reference lies inside the 95\% production interval at $Kn_\infty=8.439$, whereas the $0.21$-point defect at $3.146$ remains sampling-resolved. Thus a large low-$Kn_\infty$ departure gives way to close agreement with the conditional tangential-impulse scaling. The remaining difference can include changes in impact rate and incident states as well as the small mass-changing catalytic contribution.

The normal-momentum comparison gives the same broad pattern. Finite-temperature references of $+29.99\%$, $+29.97\%$, $+29.90\%$ and $+29.87\%$ compare with production responses of $-0.12\%$, $+9.11\%$, $+30.10\%$ and $+29.69\%$, all relative to CLL $(0.8,0.8)$. The departures are $-30.11$ and $-20.86$ percentage points at the two lower conditions but only $+0.20$ and $-0.18$ points at the two higher conditions, where each reference lies within its 95\% interval. At $Kn_\infty=0.023$, the interval includes zero: the normal-accommodation perturbation produces no resolved mean-pressure change, rather than a demonstrated reversal. The contrast with the nearly $30\%$ fixed-incidence increase quantifies how strongly the coupled response can depart from the local normal-scattering expectation.

The component-energy identities test the heat-transfer response. At $Kn_\infty=0.023$, the non-reactive translational references predict low-$\alpha_t$ and low-$\alpha_n$ reductions of $-44.08\%$ and $-30.92\%$, whereas the total-production responses are $+3.51\%$ and $-11.63\%$. At the two higher conditions, all four production responses lie within $0.56$ percentage points of their corresponding references, although each defect remains resolved from zero by its 95\% interval. Production includes total energy exchange and a self-consistent incident population, while the references use non-reactive translational exchange for the collisionless freestream half-space. At $Kn_\infty=8.439$, the rate and component diagnostics assess impact-rate and non-translational contributions to that distinction; no corresponding auxiliary rate record was acquired at $3.146$.

Equal accommodation numbers are nevertheless not interchangeable between kernels. At $Kn_\infty=8.439$, Maxwell $\alpha_M=0.8$ and CLL $(0.8,0.8)$ reduce heat transfer by $19.32\%$ and $17.73\%$, but their shear reductions are $18.47\%$ and $40.21\%$, and their pressure increases are $13.44\%$ and $21.75\%$. The similar heat reductions therefore conceal distinct pressure and shear responses.

The total-energy constraint provides a separate test of common incidence across the three CLL laws and diffuse Maxwell reflection. Under the prescribed internal-mode and catalytic treatment, the affine relation in \eqref{eq:cll_total_energy_affinity} requires the diffuse heat coefficient implied by the CLL means, $D_{\mathrm{aff}}$, to equal the simulated diffuse value $D$. This constraint retains the internal and chemical contributions rather than assuming that they are negligible. Table~\ref{tab:cll_heat_plane_defect} reports the normalised residual $\mathcal E_H$ defined in \eqref{eq:cll_heat_plane_defect}, evaluated from total production heat over the complete compression side.

\begin{table}
\centering
\setlength{\tabcolsep}{9pt}
\renewcommand{\arraystretch}{1.08}
\begin{tabular}{@{}S[table-format=1.3] S[table-format=+2.2] c@{}}
\toprule
{$Kn_\infty$} & {$\mathcal E_H$ [\%]} & {95\% interval [\%]} \\
\midrule
0.023 & +6.86 & {$[6.44,\,7.29]$} \\
0.137 & +21.85 & {$[21.66,\,22.04]$} \\
3.146 & +6.96 & {$[6.70,\,7.21]$} \\
8.439 & +3.10 & {$[2.54,\,3.66]$} \\
\bottomrule
\end{tabular}
\caption{Defect in the common-incidence total-energy constraint for the complete compression side. Here $\mathcal E_H=100(D_{\mathrm{aff}}-D)/D$ compares the CLL-implied diffuse heat coefficient $D_{\mathrm{aff}}$ with its separately simulated value $D$ \eqref{eq:cll_heat_plane_defect}. All coefficients use total production heat. Comparison-wise 95\% intervals describe sampling uncertainty only, retaining numerator--denominator covariance and assuming zero cross-run covariance.}
\label{tab:cll_heat_plane_defect}
\end{table}

All four CLL-implied diffuse coefficients exceed their simulated counterparts. The normalised defect is non-monotonic across the sampled sequence, reaching its largest value, $21.85\%$, at $Kn_\infty=0.137$. For every condition, the 95\% interval for the linear residual $T_H$ remains above zero at half, primary, double and quadruple the prescribed HAC lag, including under the maximal cross-run covariance bound in section~\ref{subsubsec:cross_run_covariance_bound}. Subject to numerical accuracy and the stated wall implementation, these residuals are incompatible with common incidence across the four laws.

Figure~\ref{fig:kernel_reference_recovery} nevertheless establishes a pronounced recovery of the selected moment responses. At $Kn_\infty=0.023$, the direct CLL responses depart from their references by $19.29$--$47.59$ percentage points across the principal channels. At $3.146$ and $8.439$, pressure and shear lie within $0.21$ points of their references, and total heat transfer within $0.56$ points of its translational reference. This contraction is not produced by comparable movement of the references: the tangential reference is constant, and the normal and each heat reference vary by only approximately $0.12$ points across the matrix.

The two comparisons distinguish recovery of conditional response relations from common incidence. At $Kn_\infty=8.439$, the close agreement of $R_{\mathrm{prod}}$ with its finite-temperature reference coexists with a $3.10\%$ defect in the total-energy constraint. The ratio tests the relative sizes of two asymmetric CLL heat reductions; the constraint tests the compatibility of the three CLL values with the diffuse endpoint. Their different outcomes show why agreement of selected moments cannot identify a common incoming population. Nor does this differential recovery establish a free-molecular limit of the complete flow: the original benchmark retains transitional departures at $Kn_\infty=8.439$ \citep{Dogra1988FlatPlate}.

The following diagnostics evaluate the kernel and incident-population contributions on the same acquired non-reactive patches, then relate them to changes in encounter rate, near-wall density and outgoing angular structure. This separates the response prescribed by the local wall law from the response of the coupled flow--surface system.

\subsection{Molecular pathways and near-wall redistribution}
\label{SUBSEC:pathway_diagnostics}

The surface comparisons distinguish recovery of selected kernel responses from common incidence across wall laws. We now examine how changes in the incoming population contribute to those responses, and how they accompany the observed rate, density and angular structure. The acquired cases comprise diffuse and specular Maxwell reflection and the three CLL pairs at $Kn_\infty=0.023$, $0.137$ and $8.439$. These acquisitions are separate from the production coefficient records: complete-side rates include all encounters, event records cover non-reactive collisions in two patches, and density records contain native-cell means. No corresponding diagnostics were acquired for Maxwell $\alpha_M=0.8$ or $0.5$, or for $Kn_\infty=3.146$. Table~\ref{tab:wall_impact_diffuse_reference_scales} gives the diffuse-Maxwell rate scales.

\begin{table}
\centering
\setlength{\tabcolsep}{4pt}
\renewcommand{\arraystretch}{1.08}

\begin{tabular*}{\linewidth}{@{\extracolsep{\fill}}
S[table-format=1.3]
S[table-format=1.2e2]
S[table-format=1.3]
S[table-format=1.2e1]
@{}}
\toprule
{\(Kn_{\infty}\)}
&
{\shortstack{
\(\dot N'_{\mathrm{hit,DM}}\)\\
\([\mathrm{m^{-1}\,s^{-1}}]\)
}}
&
{\shortstack{
\(\varepsilon_{\mathrm{hit,DM}}\)\\
\([\mathrm{eV\,molecule^{-1}}]\)
}}
&
{\shortstack{
\(\dot E'_{\mathrm{wall,DM}}\)\\
\([\mathrm{W\,m^{-1}}]\)
}}
\\
\midrule
0.023 & 1.07e24 & 0.720 & 1.23e5 \\
0.137 & 1.45e23 & 1.73  & 4.01e4 \\
8.439 & 1.07e21 & 6.18  & 1.06e3 \\
\bottomrule
\end{tabular*}

\caption{Diffuse Maxwell (DM, $\alpha_M=1$) reference scales for the complete compression side. Rates are per unit span and include reactive encounters; $\varepsilon_{\mathrm{hit,DM}}$ is the mean total energy transferred per incident molecule, positive from gas to wall. Values are final-window diagnostic estimates.}
\label{tab:wall_impact_diffuse_reference_scales}
\end{table}

\subsubsection{Impact rate and energy exchange}
\label{subsubsec:diagnostic_rate_exchange}

\begin{figure}
    \centering
    \includegraphics[width=\textwidth]{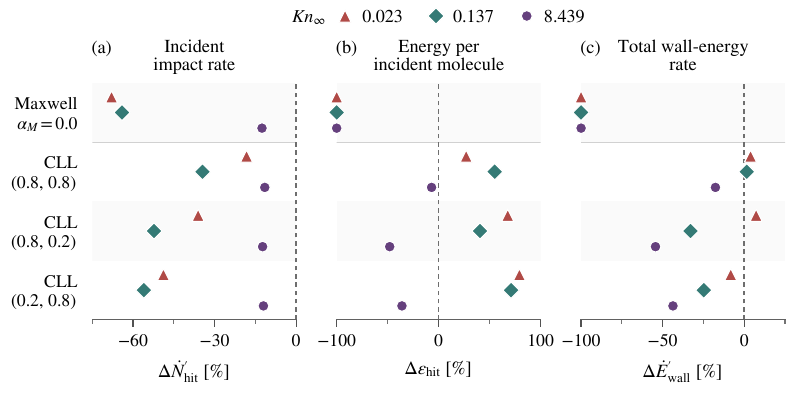}
    \caption{Signed percentage changes in wall-impact quantities relative to diffuse Maxwell reflection at the same $Kn_\infty$: (a) incident impact rate $\dot N'_{\mathrm{hit}}$; (b) mean total energy transferred per incident molecule $\varepsilon_{\mathrm{hit}}$; (c) total wall-energy rate $\dot E'_{\mathrm{wall}}$. Rates cover the complete compression side per unit span and include reactive encounters, with $\dot E'_{\mathrm{wall}}=\dot N'_{\mathrm{hit}}\varepsilon_{\mathrm{hit}}$. Energy transfer is positive from gas to wall. Markers denote final-window estimates from separate diagnostic runs, without temporal confidence intervals.}
    \label{fig:wall_impact_mechanisms}
\end{figure}

Figure~\ref{fig:wall_impact_mechanisms} separates the total wall-energy response into impact rate and mean energy transferred per incident molecule through \eqref{eq:wall_energy_factorisation}. The product allows compensation: fewer impacts need not imply less heating.

At $Kn_\infty=0.023$, the impact rate is strongly wall-model dependent: specular Maxwell changes $\dot N'_{\mathrm{hit}}$ by $-67.89\%$, while CLL $(0.8,0.8)$, $(0.8,0.2)$ and $(0.2,0.8)$ give $-18.25\%$, $-36.05\%$ and $-48.83\%$, respectively. Both asymmetric CLL cases therefore reduce incidence more than the symmetric case.

Among the three sampled conditions, each CLL law has its largest diffuse-relative impact-rate deficit at $Kn_\infty=0.137$. The three deficits span $-56.02\%$ to $-34.45\%$, compared with $-48.83\%$ to $-18.25\%$ at $Kn_\infty=0.023$. Yet the separation among the CLL responses narrows from $30.59$ to $21.57$ percentage points, and then to $0.79$ points at $Kn_\infty=8.439$. Thus the intermediate state has a larger relative depletion but a smaller spread among the three CLL laws. Each contrast is normalised by its own condition's diffuse reference; these are not comparisons of absolute deficit rates.

The CLL energy responses show why impact rate alone does not predict heating. At $Kn_\infty=0.023$, $18$--$49\%$ fewer impacts are offset by $27$--$79\%$ more energy transferred per incident molecule, leaving the diffuse-relative total wall-energy response between $-8.3\%$ and $+7.3\%$. At $Kn_\infty=0.137$, $34$--$56\%$ fewer impacts combine with $41$--$71\%$ greater per-molecule transfer, leaving $-32.9\%$ to $+1.6\%$. At $Kn_\infty=8.439$, the impact-rate responses differ by only $0.79$ percentage points, whereas per-molecule transfer spans $41.1$ points. For the asymmetric contrasts relative to CLL $(0.8,0.8)$, the order-symmetric algebraic allocation in \eqref{eq:wall_energy_symmetric_allocation} assigns $98.4$--$98.5\%$ of the total-rate reductions to the per-molecule term. Each CLL law's diffuse-relative total-energy response becomes more negative across the sampled states, while compensation between encounter number and exchange gives way to exchange differences at nearly common rates. This factorisation distinguishes encounter number from average exchange, but the latter still depends on both the scattering kernel and the incident population.

\subsubsection{Kernel and incident-population contributions}
\label{subsubsec:kernel_population_contributions}

\begin{figure}
    \centering
    \includegraphics[width=0.85\textwidth]{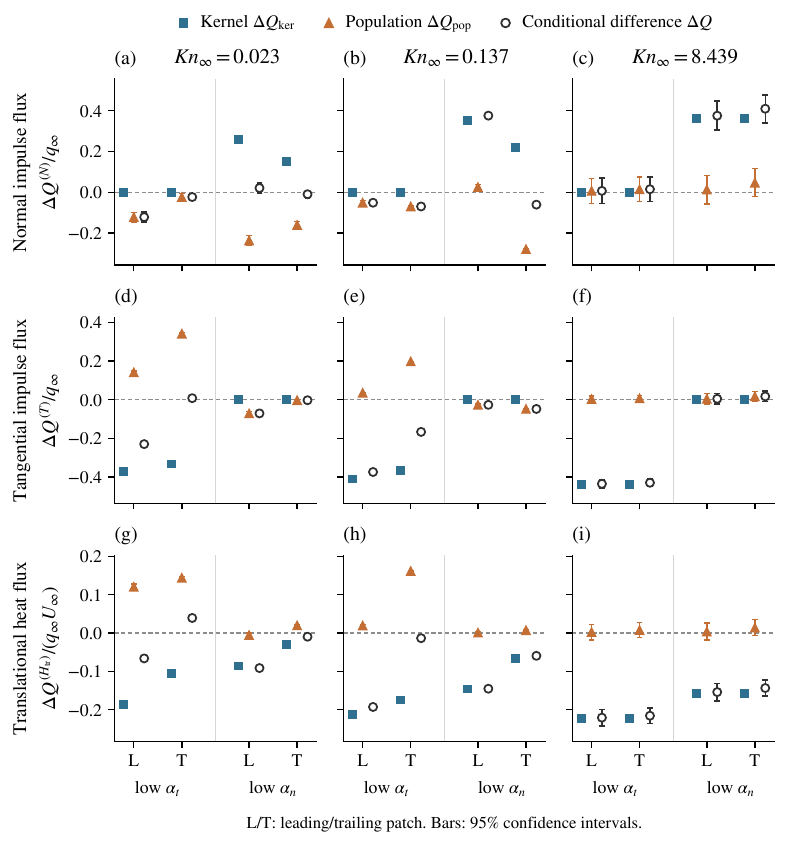}
    \caption{Kernel and incident-population contributions to non-reactive patch-mean responses relative to CLL $(\alpha_n,\alpha_t)=(0.8,0.8)$. Columns correspond to $Kn_\infty=0.023$, $0.137$ and $8.439$; rows show normal momentum, tangential momentum and translational energy. Low-$\alpha_t$ and low-$\alpha_n$ groups denote CLL $(0.8,0.2)$ and $(0.2,0.8)$, each at the leading (L) and trailing (T) patches. Squares and triangles denote the order-symmetric allocations $\Delta Q_{\mathrm{ker}}$ and $\Delta Q_{\mathrm{pop}}$ in \eqref{eq:kernel_population_allocation}; open circles denote their sum, the conditional response. Momentum and energy flux differences are normalised by $q_\infty$ and $q_\infty U_\infty$, respectively; vertical scales are common within each row. Bars show comparison-wise 95\% sampling intervals accounting for temporal correlation and covariance between allocations; cross-run covariance is assumed zero.}
    \label{fig:kernel_population_split}
\end{figure}

Cross-kernel evaluation separates these two influences by applying each conditional wall law to each acquired incident population. For either asymmetric CLL case relative to CLL $(0.8,0.8)$, \eqref{eq:kernel_population_allocation} allocates the conditional change into $\Delta Q_{\mathrm{ker}}$ and $\Delta Q_{\mathrm{pop}}$ by averaging the two orders of changing kernel and population. Figure~\ref{fig:kernel_population_split} shows these contributions for normal momentum, tangential momentum and translational energy at both patches. All numerical contributions below are differences of dimensionless non-reactive patch coefficients, not percentages or full-side production changes. Their sum gives the conditional response; the separately recorded outgoing exchanges retain the wall-scattering sampling remainder in \eqref{eq:conditional_observed_closure}.

At $Kn_\infty=0.023$, a small pressure response conceals substantial opposing contributions. For reduced normal accommodation, pooling the two patches gives a kernel contribution of $+0.20405$ and a population contribution of $-0.19840$. Their sum is $+0.00565$, with a 95\% interval of $[-0.00899,0.02029]$. The unresolved net change therefore does not indicate a weak conditional normal-scattering response: the population change nearly cancels it. Conversely, reducing tangential accommodation leaves the conditional normal impulse unchanged, so its kernel pressure contribution is identically zero. The pooled pressure reduction, $-0.07249$ with interval $[-0.08627,-0.05871]$, enters entirely through the acquired incidence within this conditional comparison.

\PreprintBreak
The trailing-patch response to reduced tangential accommodation goes beyond cancellation. The kernel and population contributions are $-0.33333$ and $+0.34117$ for shear, leaving $+0.00784$ with a 95\% interval of $[0.00482,0.01086]$. For translational heat transfer they are $-0.10484$ and $+0.14395$, leaving $+0.03912$ with interval $[0.03717,0.04106]$. In both moments, the population contribution overturns the conditional reduction. The recorded outgoing velocities give the same positive signs, with observed contrasts of $+0.00797$ and $+0.03882$; their differences from the conditional contrasts are not resolved from zero. Both conditional net increases remain positive under the prescribed lag variations and the maximal cross-run covariance sensitivity. These local increases do not imply increased full-side shear or total production heating, which have different spatial and energy support.

The balance also varies along the plate. At $Kn_\infty=0.137$, the population contribution to the low-$\alpha_n$ pressure response reinforces the kernel contribution at the leading patch but opposes and exceeds it at the trailing patch (figure~\ref{fig:kernel_population_split}b). For reduced tangential accommodation, population changes partially offset the conditional shear and heat reductions at both patches without reversing their net signs. Population feedback is therefore neither a spatially uniform attenuation nor a correction of one sign for all moments.

Departure from a freestream-based reference involves a further distinction. For the low-$\alpha_n$ pressure comparison in the trailing patch at $Kn_\infty=0.023$, the matched non-reactive freestream predicts a $+29.99\%$ response. Applying the same kernel change to the acquired symmetric-CLL population held fixed gives only $+14.12\%$; allowing incidence to change with the wall gives an unresolved $-0.97\%$ response (95\% interval $[-2.73\%,0.79\%]$). The ordered split in \eqref{eq:reference_population_split} assigns $-15.87$ percentage points to the baseline-incidence difference and $-15.09$ to the additional population change. Both comparisons matter: incidence is already modified in the baseline flow before the wall perturbation modifies it further.

At $Kn_\infty=8.439$, the kernel contributions lie much closer to the net conditional changes, but the population terms are not precisely enough determined to establish their absence. For example, the pooled low-$\alpha_n$ pressure comparison gives $\Delta Q_{\mathrm{ker}}/q_\infty=0.36251$ and $\Delta Q_{\mathrm{pop}}/q_\infty=0.02999$, with a 95\% interval of $[-0.01913,0.07912]$ for the latter. This is consistent with a much more nearly componentwise response, while permitting a finite population correction.

The two-patch incident records provide a complementary moment-level comparison. Across the CLL cases, the pooled non-reactive excess-energy ratio $\mathcal R_{t/n}$ \eqref{eq:incident_component_energy_ratio} narrows from $1.27$--$3.31$ at $Kn_\infty=0.023$ to $1.34$--$2.17$ at $Kn_\infty=0.137$ and $1.42$--$1.43$ at $Kn_\infty=8.439$. The last range is close in magnitude to the directed hyperthermal value $\cot^2 40^\circ=1.4203$. Encounter rate and the partition of incident excess energy therefore become less wall-model dependent together. At $Kn_\infty=8.439$, the pooled low-$\alpha_t$ value is $1.4293$, whose Fieller interval $[1.4219,1.4368]$ excludes the matched reference $1.4162$; the symmetric and low-$\alpha_n$ pooled intervals include it. The corresponding delta-method intervals give the same inclusion decisions. Complete moment matrices, ordered allocations and observed--conditional closure checks accompany the analysis records \citep{Naevdal2026DataRelease}.

\subsubsection{Spatial redistribution}
\label{subsubsec:diagnostic_spatial_redistribution}

\begin{figure}
    \centering
    \includegraphics[width=\textwidth]{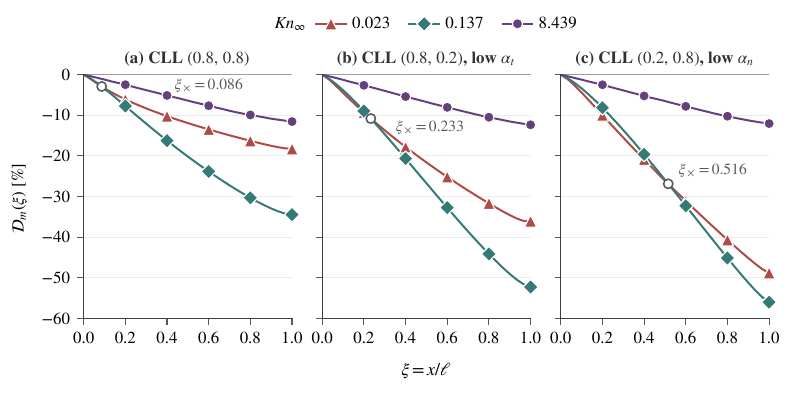}
    \caption{Cumulative compression-side impact-rate contrasts relative to diffuse Maxwell reflection: (a) CLL $(0.8,0.8)$; (b) $(0.8,0.2)$; (c) $(0.2,0.8)$. The contrast $\mathcal D_m$, defined in \eqref{eq:cumulative_impact_rate_contrast}, uses the complete-side diffuse impact rate at the same $Kn_\infty$ as its denominator and is displayed in per cent; $100\mathcal D_m(1)$ equals the percentage response in figure~\ref{fig:wall_impact_mechanisms}(a). Lines join the 30 native downstream element-edge values without smoothing. Open circles mark the first crossings of the $Kn_\infty=0.023$ and $0.137$ curves for $\xi>0$, interpolated within native elements. These final-window estimates include reactive encounters and have no temporal confidence intervals.}
    \label{fig:cumulative_impact_rate_contrast}
\end{figure}

The moment allocations quantify changes in exchange but do not locate the surrounding density adjustment. The cumulative contrast $\mathcal D_m$ in \eqref{eq:cumulative_impact_rate_contrast} first locates where each complete-side incident-rate deficit accumulates (figure~\ref{fig:cumulative_impact_rate_contrast}); the native density contrasts then provide its two-dimensional context.

The additional intermediate-condition deficit accumulates downstream rather than appearing as a uniform leading-edge offset. The $Kn_\infty=0.137$ cumulative contrast first becomes more negative than its $Kn_\infty=0.023$ counterpart at $x/\ell=0.086$, $0.233$ and $0.516$ for CLL $(0.8,0.8)$, low $\alpha_t$ and low $\alpha_n$, respectively. These crossings locate where one accumulated relative deficit overtakes another; they do not mark the onset of a clearance mechanism. All native CLL-minus-diffuse element contrasts are negative, so the cumulative deficits grow monotonically in magnitude rather than resulting from cancellation between plate regions.

At $Kn_\infty=8.439$, specular Maxwell and the three CLL laws have nearly equal complete-side impact rates, $11.55$--$12.57\%$ below diffuse Maxwell. Their absolute point estimates lie within $1\%$ of the finite-temperature collisionless reference $\ell n_\infty\sum_sX_sJ_{1,s}\simeq9.401\times10^{20}\,\mathrm{m^{-1}\,s^{-1}}$, whereas the diffuse rate is $14.1\%$ higher. For the CLL cases, half of each diffuse-relative deficit accumulates only by $x/\ell=0.452$--$0.462$, and $75\%$ by $x/\ell=0.682$--$0.696$. The deficit therefore extends over $O(\ell)$ rather than being confined to the leading edge. Agreement in integrated encounter number alone does not establish a collisionless incident-state distribution.

The sampled ordering is consistent with a collisional contribution to wall-law-dependent incidence. In collisionless flat-plate flow, reflected molecules cannot alter subsequent incidence through intermolecular collisions; in the slip regime, accommodation continues to enter through Knudsen-layer corrections, velocity slip and temperature jump \citep{LuanEtAl2024CLLSlip}. The three diagnostic states also differ in freestream temperature and composition, so they neither locate an interior maximum nor establish continuum-side screening.

\begin{figure}
    \centering
    \includegraphics[width=\textwidth]{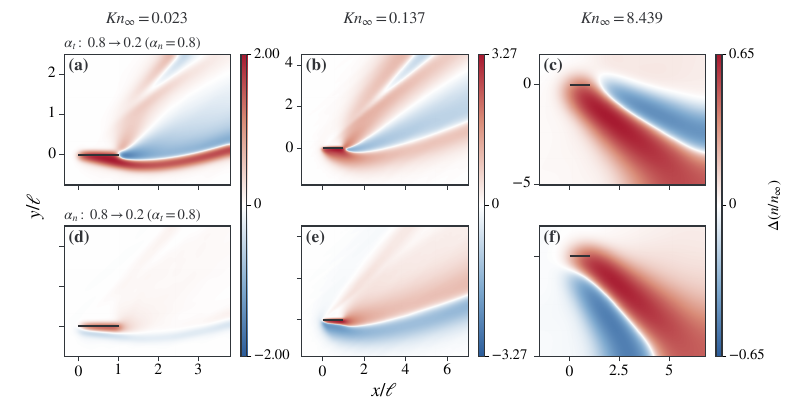}
    \caption{Contrasts in mean total-mixture number density over the final diagnostic interval. Columns correspond to $Kn_\infty=0.023$, $0.137$ and $8.439$; panels (a--c) show $\Delta_t=[n_{0.8,0.8}-n_{0.8,0.2}]/n_\infty$ and (d--f) show $\Delta_n=[n_{0.8,0.8}-n_{0.2,0.8}]/n_\infty$. Positive (red) and negative (blue) values indicate greater density under the symmetric and perturbed CLL laws, respectively. Black lines mark the plate. Both rows share a zero-centred inverse-hyperbolic-sine colour scale within each column; colour and axis limits differ between columns. Rasterisation and smoothing are for display only; quantitative morphology uses unsmoothed native-cell means.}
    \label{fig:density_difference_n_over_ninf}
\end{figure}

The density contrasts in figure~\ref{fig:density_difference_n_over_ninf} show where the CLL perturbations redistribute the surrounding gas. Quantitative morphology uses the unsmoothed native final-interval fields and definitions in section~\ref{subsec:selected_diagnostics}.

The red compression-side band in panel~(a) denotes lower density under CLL $(0.8,0.2)$. Over the common registered support at each condition, $\mathcal L_1(\Delta_n)/\mathcal L_1(\Delta_t)$ is $0.192$, $0.968$ and $0.979$ at $Kn_\infty=0.023$, $0.137$ and $8.439$. Thus the integrated magnitude of the normal-accommodation density contrast is only $19.2\%$ of the tangential contrast at the lowest condition, but is already comparable to it at the intermediate state. Their shapes nevertheless remain distinct. Vectors from the negative-lobe to the positive-lobe centroids are separated by $99.5^\circ$ at $Kn_\infty=0.137$ and $156.7^\circ$ at $8.439$, the latter $23.3^\circ$ from opposition. These vectors describe the signed density pattern, not molecular motion. Their angular separation varies by at most $1.4^\circ$ under the prescribed integration-support contraction.

The conditional CLL moments distinguish two kinematic tendencies. At fixed incident state, reducing $\alpha_t$ from $0.8$ to $0.2$ doubles the coefficient multiplying incident tangential velocity in the reflected mean, from $\sqrt{0.2}$ to $\sqrt{0.8}$ \eqref{eq:cll_conditional_memory_moments}. Reducing $\alpha_n$ from $0.8$ to $0.2$ instead reduces the conditional mean normal translational-energy transfer to one quarter \eqref{eq:cll_conditional_translational_transfer}; for a common population with positive mean normal excess energy, more normal kinetic energy is retained after reflection. These are fixed-incidence relations, not ratios between the self-consistent production states. They favour downstream drift for low $\alpha_t$ and wall-normal departure for low $\alpha_n$.

Intermolecular collisions couple reflected molecules to subsequent incidence by modifying both approaching and previously reflected populations \citep{Lord1992IncompleteAccommodation}. The density and rate changes are consistent with altered streamwise or wall-normal transport, while the cross-kernel comparison quantifies the accompanying change in wall exchange. Neither calculation resolves how the population contribution divides between deflection of approaching molecules and re-impact of reflected molecules. 

\begin{figure}[t]
  \centering
  \includegraphics[width=\linewidth]{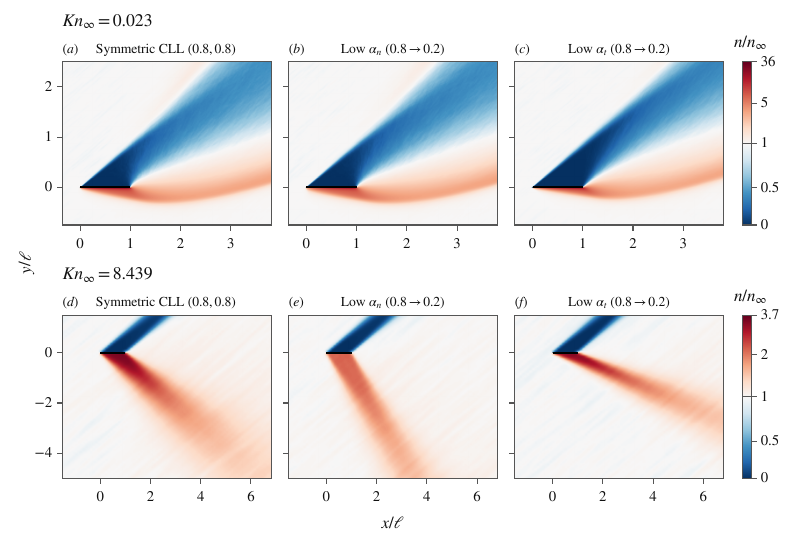}
  \caption{Normalised total-mixture number density, $n/n_\infty$, from movie~2 at (a--c) $Kn_\infty=0.023$ and (d--f) $8.439$. Columns show symmetric CLL, $(\alpha_n,\alpha_t)=(0.8,0.8)$; reduced normal accommodation, $(0.2,0.8)$; and reduced tangential accommodation, $(0.8,0.2)$. Each snapshot averages 32 realisations over four source bins: $66~\si{\micro\second}$ centred at $3.861~\mathrm{ms}$ in (a--c), and $516~\si{\micro\second}$ centred at $30.188~\mathrm{ms}$ in (d--f). Times are measured from the simulation start in (a,d) and the most recent wall-law switch in the remaining panels; columns are not chronological. Spatial limits and nonlinear colour scales are shared within each row but differ between rows, with white at $n/n_\infty=1$. Black lines mark the plate. Plate-aware Gaussian filtering is applied only for display (section~\ref{subsec:movie_density_acquisition}). These finite-time snapshots do not establish post-switch stationarity.}  \label{fig:cll_movie_snapshots}
\end{figure}

Figure~\ref{fig:cll_movie_snapshots} shows density snapshots from the switch-and-return sequence in movie~2, in which the wall law is changed without reinitialising the flow (section~\ref{subsec:movie_density_acquisition}). At $Kn_\infty=0.023$, the normal-accommodation switch gives a weaker visible rearrangement (a,b) than the tangential-accommodation switch, which produces a narrower, more nearly plate-parallel dense region on the compression side (a,c). At $Kn_\infty=8.439$, the near-wall density enhancement becomes steeper under reduced $\alpha_n$ (d,e) and shallower under reduced $\alpha_t$ (d,f).

In the early displayed frames after each switch at $Kn_\infty=8.439$, the new near-wall enhancement coexists with a more distant remnant of the preceding density pattern, rather than the whole field reorienting at once. The return switches produce opposite qualitative rearrangements, without establishing exact reversibility. These are observations of ensemble-mean total density, not tracked reflected populations or a reconstruction of the molecular feedback path. The two movie panels use different physical clocks and averaging intervals and therefore do not establish a common propagation or relaxation time.

\subsubsection{Angular structure and conditional variance}
\label{subsubsec:diagnostic_angular_variance}

\begin{figure}
    \centering
    \includegraphics[width=\textwidth]{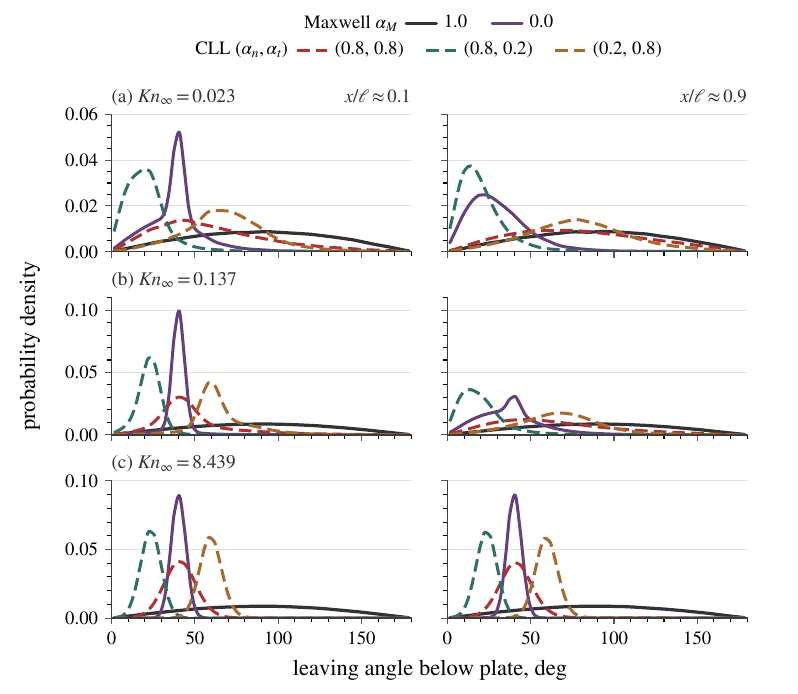}
    \caption{Probability densities (per degree) of the projected leaving angle for non-reactive compression-side encounters. Rows (a--c) correspond to $Kn_\infty=0.023$, $0.137$ and $8.439$; left and right columns show the leading ($0.075\leq x/\ell\leq0.125$) and trailing ($0.875\leq x/\ell\leq0.925$) patches. The angle in \eqref{eq:selected_event_angle} runs from the downstream tangent ($0^\circ$) through the outward normal ($90^\circ$) to the upstream tangent ($180^\circ$). Each curve averages over its model's own incident population. Display curves use $3^\circ$ histograms with one-bin Gaussian smoothing, unit normalisation and shape-preserving cubic interpolation; angular statistics use the underlying events.}
    \label{fig:outgoing_angle_distribution}
\end{figure}

Angular distributions are established observables in molecular-beam and molecular-dynamics studies \citep{Murray2017GasSurfaceScattering,ChenEtAl2023Physisorption,ApparPoovathingal2026VLEOGSIModel}. Here, however, each distribution averages over its own self-consistent incident population rather than a common prescribed beam. At $Kn_\infty=8.439$, the non-diffuse laws produce narrow lobes with model-dependent directions and widths at both patches (figure~\ref{fig:outgoing_angle_distribution}). At $Kn_\infty=0.023$, the distributions are broader and more position dependent; the specular population broadens and shifts towards shallower departure downstream. Because specular reflection adds no conditional angular spread, that variation is inherited from incidence. Diffuse Maxwell remains broad and centred near wall-normal departure because it removes incident directional memory.

The CLL ordering is consistent with the conditional component kernels \citep{Lord1991ExtensionsCLL,SPARTA24Sep2025}: low $\alpha_t$ favours downstream drift and shallow departure, whereas low $\alpha_n$ favours normal-energy retention and steeper departure. At $Kn_\infty=8.439$, the pooled leading--trailing mean angles are $23.4^\circ$, $41.4^\circ$ and $59.8^\circ$ for CLL $(0.8,0.2)$, $(0.8,0.8)$ and $(0.2,0.8)$, respectively. The $|\Delta(n/n_\infty)|$-weighted distance of the density contrast from the compression-face line is likewise greater for low $\alpha_n$, $2.371\ell$, than for low $\alpha_t$, $2.177\ell$. The shallow-to-steep CLL ordering persists at all three diagnostic conditions: directional distinctions remain visible even where the macroscopic normal and tangential responses are strongly coupled.

Conditional resampling of the exact SPARTA kernel at each recorded incident state separates angular scatter at fixed incidence from dispersion of the conditional mean angle \eqref{eq:outgoing_angle_variance}. Each law is applied to its own incident record, so this is a within-model decomposition, not a common-input causal separation. Across all 30 populations, $V_{\mathrm{ker}}+V_{\mathrm{inc}}$ matches the nested-resampling total variance within $0.01\%$, and that total differs from the observed outgoing variance by at most $1.6\%$. Resampling standard errors describe finite-$M$ Monte Carlo uncertainty only; observed-moment intervals use the event-slice Bartlett--HAC acquisition covariance specified in section~\ref{subsec:selected_diagnostics}.

Specular reflection provides the direct incidence diagnostic: $V_{\mathrm{ker}}$ vanishes to floating-point precision, so its outgoing angular variance is inherited entirely from the incident directions. Leading-to-trailing variances increase from $262$ to $337~\mathrm{deg}^2$ at $Kn_\infty=0.023$, from $58.5$ to $379~\mathrm{deg}^2$ at $0.137$, and from $11.4$ to $14.2~\mathrm{deg}^2$ at $8.439$. The corresponding ratios are $1.28$ (95\% interval $1.22$--$1.35$), $6.49$ ($6.41$--$6.57$) and $1.25$ ($1.14$--$1.36$). All three intervals exclude unity. The intermediate state combines a comparatively directed leading population with a broad trailing population, whereas the lowest and highest states have, respectively, large and small variances at both patches. The two sampled patches therefore reveal a strong change in incidence at $Kn_\infty=0.137$.

For diffuse Maxwell reflection, the conditional reference is exactly $V_D=(180/\upi)^2(\upi^2/4-2)\simeq1534.3873~\mathrm{deg}^2$. The resampled $V_{\mathrm{ker}}$ approaches this value, while the finite-$M$-corrected incident estimates cluster near zero, from $-0.041$ to $0.016~\mathrm{deg}^2$, and are retained without clipping. CLL contains both components. Its estimated kernel fraction $V_{\mathrm{ker}}/(V_{\mathrm{ker}}+V_{\mathrm{inc}})$ increases in every model--patch pair between the sampled endpoints, from $20$--$66\%$ at $Kn_\infty=0.023$ to $58$--$83\%$ at $8.439$. This increase concerns the share of angular variance supplied by conditional scatter; it neither requires an increase in absolute $V_{\mathrm{ker}}$ nor counts molecules whose wall-imposed direction survives collisions. Both terms are evaluated on the model's own incident population. The corresponding component values and conditional Monte Carlo standard errors are provided in the release. Distribution width is therefore not, by itself, a measure of accommodation: specular width is incident-population variance, diffuse width is kernel variance, and CLL width combines the two.

The results concern constant-coefficient CLL probes, not calibrated material or roughness models. Material-specific incidence dependence, conditional cross-component correlations and roughness-induced shadowing or multiple reflections remain outside the sweep \citep{WangEtAl2021DataBasedKernel,ApparPoovathingal2026VLEOGSIModel,AntonEtAl2025WaveScatteringRoughness,ParkJun2025VelocityCorrugation}. At the macroscopic scale, isolated roughness can also alter hypersonic instability growth and transition onset \citep{VanDenEynde2016RoughnessTransition}; that process is distinct from the molecular accommodation changes considered here. Within this scope, the diagnostics distinguish changes in reflected direction from changes in the population and exchange at the wall. Their consequences for drag and lift follow from force projection.

\subsection{Geometric projection into drag and lift}
\label{SUBSEC:aerodyn_resp}

The diagnostics distinguish the conditional exchange prescribed by the wall from the population on which it acts. The aerodynamic consequence depends on how normal and tangential momentum transfer combine. Define the integrated plate-normal and plate-tangential coefficients
\begin{equation*}
C_N=\int_0^1 C_p(\xi)\,\mathrm{d}\xi,
\qquad
C_T=\int_0^1 C_f(\xi)\,\mathrm{d}\xi.
\end{equation*}
Equation~\eqref{eq:integrated_coefficients} can then be written as
\begin{equation}
\begin{bmatrix}
C_D\\[1pt]
C_L
\end{bmatrix}
=
\begin{bmatrix}
\sin\alpha & \cos\alpha\\
\cos\alpha & -\sin\alpha
\end{bmatrix}
\begin{bmatrix}
C_N\\[1pt]
C_T
\end{bmatrix}.
\label{eq:plate_force_projection}
\end{equation}
The transformation is orthogonal and therefore preserves response magnitudes and angles. A pure normal-force change and a pure tangential-force change consequently map to orthogonal directions in the $(C_D,C_L)$ plane: pressure raises drag and lift, whereas positive shear raises drag and subtracts from lift.

\begin{figure}
    \centering
    \includegraphics[width=0.85\textwidth]{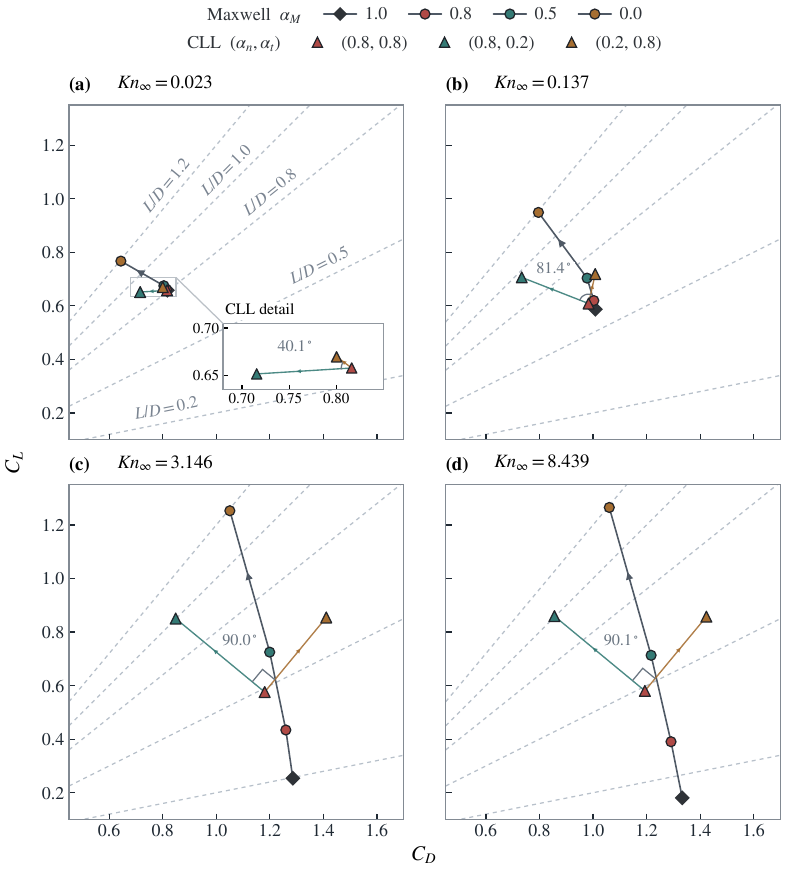}
    \caption{Integrated compression-side aerodynamic coefficients at $40^\circ$ incidence for (a) $Kn_\infty=0.023$, (b) $0.137$, (c) $3.146$ and (d) $8.439$. Symbols denote production means; sampling uncertainty is not shown. Maxwell arrows follow decreasing $\alpha_M$ along the sampled states. CLL arrows show finite changes from $(0.8,0.8)$ to $(0.8,0.2)$ and $(0.2,0.8)$ for reduced tangential and normal accommodation, respectively. Angle labels give their separation; corner markers in (c,d) indicate near-orthogonality. The inset enlarges the CLL states in (a). Equal data-unit scaling, including in the inset, preserves response angles. Dashed rays mark constant $L/D$.}
    \label{fig:gsi_integrated_polar}
\end{figure}

The coefficients and signed responses relative to diffuse Maxwell reflection are reported in table~\ref{tab:gsi_integrated_summary} and shown in figure~\ref{fig:gsi_integrated_polar}. Horizontal and vertical displacements in each panel correspond to changes in drag and lift, while rays from the origin mark constant $L/D$. All non-reference integrated responses are resolved from zero except the $C_D$ and $C_L$ changes of CLL $(0.8,0.8)$ at $Kn_\infty=0.023$ and the $C_D$ change of CLL $(0.2,0.8)$ at $Kn_\infty=0.137$.

\begin{table}
\centering
\setlength{\tabcolsep}{1.45pt}
\renewcommand{\arraystretch}{1.04}
\begin{tabular*}{\textwidth}{@{\extracolsep{\fill}}
S[table-format=1.3]
l
S[table-format=1.4]
S[table-format=+3.2]
S[table-format=1.4]
S[table-format=+3.2]
S[table-format=1.4]
S[table-format=+3.2]
@{}}
\toprule
{$Kn_\infty$}
& {Wall model}
& {$C_D$}
& {$\Delta C_D$ [\%]}
& {$C_L$}
& {$\Delta C_L$ [\%]}
& {$L/D$}
& {$\Delta(L/D)$ [\%]} \\
\midrule
0.023 & Maxwell $\alpha_M=1.0$ & 0.8175 & 0.00   & 0.6582 & 0.00    & 0.8052 & 0.00 \\
      & Maxwell $\alpha_M=0.8$       & 0.8139 & -0.44  & 0.6614 & +0.49   & 0.8127 & +0.94 \\
      & Maxwell $\alpha_M=0.5$       & 0.8041 & -1.64  & 0.6751 & +2.57   & 0.8396 & +4.28 \\
      & Maxwell $\alpha_M=0.0$       & 0.6437 & -21.26 & 0.7671 & +16.55  & 1.1918 & +48.02 \\
      & CLL $(0.8,0.8)$              & 0.8162 & -0.16  & 0.6578 & -0.06   & 0.8059 & +0.09 \\
      & CLL $(0.8,0.2)$              & 0.7156 & -12.46 & 0.6515 & -1.01   & 0.9104 & +13.07 \\
      & CLL $(0.2,0.8)$              & 0.8002 & -2.12  & 0.6696 & +1.74   & 0.8369 & +3.94 \\
\addlinespace[2pt]
0.137 & Maxwell $\alpha_M=1.0$ & 1.0088 & 0.00   & 0.5872 & 0.00    & 0.5821 & 0.00 \\
      & Maxwell $\alpha_M=0.8$       & 1.0029 & -0.59  & 0.6197 & +5.52   & 0.6179 & +6.15 \\
      & Maxwell $\alpha_M=0.5$       & 0.9777 & -3.08  & 0.7036 & +19.82  & 0.7196 & +23.63 \\
      & Maxwell $\alpha_M=0.0$       & 0.7961 & -21.09 & 0.9489 & +61.59  & 1.1919 & +104.77 \\
      & CLL $(0.8,0.8)$              & 0.9833 & -2.53  & 0.6087 & +3.66   & 0.6190 & +6.35 \\
      & CLL $(0.8,0.2)$              & 0.7337 & -27.27 & 0.7063 & +20.28  & 0.9626 & +65.38 \\
      & CLL $(0.2,0.8)$              & 1.0081 & -0.07  & 0.7185 & +22.36  & 0.7128 & +22.45 \\
\addlinespace[2pt]
3.146 & Maxwell $\alpha_M=1.0$ & 1.2859 & 0.00   & 0.2541 & 0.00    & 0.1976 & 0.00 \\
      & Maxwell $\alpha_M=0.8$       & 1.2596 & -2.04  & 0.4343 & +70.93  & 0.3448 & +74.49 \\
      & Maxwell $\alpha_M=0.5$       & 1.1994 & -6.73  & 0.7250 & +185.34 & 0.6045 & +205.92 \\
      & Maxwell $\alpha_M=0.0$       & 1.0506 & -18.30 & 1.2529 & +393.11 & 1.1925 & +503.53 \\
      & CLL $(0.8,0.8)$              & 1.1806 & -8.19  & 0.5759 & +126.67 & 0.4878 & +146.89 \\
      & CLL $(0.8,0.2)$              & 0.8483 & -34.03 & 0.8505 & +234.73 & 1.0026 & +407.42 \\
      & CLL $(0.2,0.8)$              & 1.4106 & +9.70  & 0.8544 & +236.26 & 0.6057 & +206.53 \\
\addlinespace[2pt]
8.439 & Maxwell $\alpha_M=1.0$ & 1.3332 & 0.00   & 0.1805 & 0.00    & 0.1354 & 0.00 \\
      & Maxwell $\alpha_M=0.8$       & 1.2911 & -3.16  & 0.3905 & +116.31 & 0.3025 & +123.37 \\
      & Maxwell $\alpha_M=0.5$       & 1.2167 & -8.74  & 0.7131 & +295.02 & 0.5861 & +332.85 \\
      & Maxwell $\alpha_M=0.0$       & 1.0607 & -20.45 & 1.2652 & +600.82 & 1.1928 & +780.92 \\
      & CLL $(0.8,0.8)$              & 1.1935 & -10.48 & 0.5804 & +221.49 & 0.4863 & +259.12 \\
      & CLL $(0.8,0.2)$              & 0.8557 & -35.82 & 0.8595 & +376.09 & 1.0044 & +641.76 \\
      & CLL $(0.2,0.8)$              & 1.4232 & +6.75  & 0.8573 & +374.90 & 0.6024 & +344.87 \\
\bottomrule
\end{tabular*}
\caption{Integrated compression-side coefficients and signed percentage changes relative to diffuse Maxwell reflection at the same $Kn_\infty$. $L/D=C_L/C_D$ and percentage changes use unrounded retained means.}
\label{tab:gsi_integrated_summary}
\end{table}

Along the diffuse branch, integrated normal loading remains nearly constant while shear grows strongly with rarefaction. Reconstructing the plate-fixed components from \eqref{eq:plate_force_projection}, $C_N$ remains within approximately $0.995$--$1.098$ over the four conditions, whereas $C_T$ rises from $0.203$ to $0.905$. The growing shear contribution raises drag and cancels an increasing fraction of the normal-force contribution to lift, moving the diffuse branch towards larger $C_D$ and smaller $C_L$.

Within the Maxwell family, decreasing $\alpha_M$ moves every condition upward and to the left. In the specular limit, $C_D$ falls by approximately $18$--$21\%$ across the matrix, while the lift increase grows from $16.55\%$ at $Kn_\infty=0.023$ to $600.82\%$ at $Kn_\infty=8.439$. The percentage increase at high rarefaction is amplified by the small diffuse-reference lift, but the absolute change is also large, from $C_L=0.1805$ to $1.2652$.

Along the fully specular Maxwell branch ($\alpha_M=0$), similar drag reductions conceal a reversal in the pressure response. At $Kn_\infty=0.023$, integrated pressure changes by only $-2.74\%$, so the drag decrease is generated predominantly by removal of shear. The pressure response is positive from $Kn_\infty=0.137$ onward and reaches $60.12\%$ and $65.88\%$ at the two highest conditions; there, additional pressure drag is more than offset by the collapse of skin friction. A drag-only comparison would miss this change in force composition.

The same fully specular branch supplies a geometric consistency check. For a shear-free resultant,
\begin{equation}
C_T=0
\quad\Longrightarrow\quad
\frac{L}{D}
=
\frac{C_N\cos\alpha}{C_N\sin\alpha}
=
\cot\alpha.
\label{eq:specular_geometric_limit}
\end{equation}
For $\alpha=40^\circ$, $\cot\alpha=1.1918$. The relation is fixed by force direction rather than rarefaction. The four fully specular values span $1.1918$--$1.1928$, remaining within $0.09\%$ of the geometric value and providing a close internal check on the force transformation and the effectively shear-free wall response.

The two finite CLL perturbations test whether the integrated response retains separate momentum channels. At $Kn_\infty=8.439$, the retained mean displacements in plate-fixed coordinates are
\begin{align*}
(\Delta C_N,\Delta C_T)_{(0.8,0.2)-(0.8,0.8)}
&=(-0.0034,-0.4382),\\
(\Delta C_N,\Delta C_T)_{(0.2,0.8)-(0.8,0.8)}
&=(+0.3598,-0.0021).
\end{align*}
The low-$\alpha_t$ perturbation is almost purely tangential, whereas the low-$\alpha_n$ perturbation is almost purely normal. At $Kn_\infty=3.146$ and $8.439$, the complementary-component fractions are below $0.8\%$ as point estimates, with upper endpoints of the covariance-aware two-sided 95\% intervals below $2.1\%$. Because the force transformation is orthogonal, it preserves rather than creates this separation: the corresponding polar displacements lie within $0.12^\circ$ of $90^\circ$, while the upper endpoints of the two-sided 95\% intervals for angular departure remain below $0.8^\circ$.

The high-condition geometry also remains close to componentwise under the covariance sensitivity in section~\ref{subsubsec:cross_run_covariance_bound}. Across the half, primary, double and quadruple HAC lags, holding the estimated within-run covariances fixed gives maximal-covariance upper endpoints below $2.62\%$ for the complementary fractions and $1.18^\circ$ for departure from orthogonality.

At $Kn_\infty=0.023$, this componentwise structure is lost. The nominal tangential perturbation changes the normal and tangential coefficients by comparable amounts, $(-0.0694,-0.0730)$, whereas the nominal normal perturbation appears mainly through a tangential change, $(-0.0012,-0.0199)$. Their directions are separated by $40.1^\circ$, rather than approximately $90^\circ$ at $Kn_\infty=3.146$ and $8.439$. Orthogonal projection cannot create this coupling: the non-orthogonality is already present in the plate-fixed response. It is the integrated counterpart of the cross-component effects established by the surface comparisons and examined through the incident-population diagnostics.

By $Kn_\infty=0.137$, the aerodynamic split is already conspicuous: CLL $(0.8,0.2)$ and $(0.2,0.8)$ increase lift by $20.28\%$ and $22.36\%$, respectively, while changing drag by $-27.27\%$ and $-0.07\%$. The latter drag response remains unresolved, with a 95\% interval of $[-0.185\%,+0.037\%]$.

A limiting symmetry explains why the asymmetric CLL cases can produce similar lift without similar drag. Consider a common non-reactive collimated beam as the freestream and wall thermal speeds tend to zero, within the free-molecular flat-plate setting \citep{CercignaniLampis1972FlatPlate}. Write $s_\alpha=\sin\alpha$, $c_\alpha=\cos\alpha$, $\beta_n=\sqrt{1-\alpha_n}$ and $\beta_t=\sqrt{1-\alpha_t}$. To retain the prescribed coefficient normalisation, let $\zeta=\sum_s X_sm_s/m_\infty$ be the ratio of the species-summed beam mass to the reference molecular mass. The cold-beam impulses give $C_N^{\mathrm{cb}}=2\zeta s_\alpha^2(1+\beta_n)$ and $C_T^{\mathrm{cb}}=2\zeta s_\alpha c_\alpha(1-\beta_t)$. Projection through \eqref{eq:plate_force_projection} yields
\begin{equation}
\begin{aligned}
C_L^{\mathrm{cb}}&=2\zeta s_\alpha^2c_\alpha(\beta_n+\beta_t),\\
C_D^{\mathrm{cb}}&=2\zeta s_\alpha\left(1+s_\alpha^2\beta_n-c_\alpha^2\beta_t\right).
\end{aligned}
\label{eq:cll_cold_force_symmetry}
\end{equation}
The incident beam contributes no lift relative to its own direction; the transverse reflected momentum depends on the sum of the two retention factors. Exchanging the accommodation coefficients therefore leaves $C_L^{\mathrm{cb}}$ invariant, whereas $C_D^{\mathrm{cb}}(a,b)-C_D^{\mathrm{cb}}(b,a)=2\zeta\sin\alpha[\sqrt{1-a}-\sqrt{1-b}]$. This symmetry holds at general non-degenerate incidence under the stated beam assumptions, not only at $40^\circ$. For the selected pair it gives $C_L^{\mathrm{cb}}\simeq0.8493$ for both laws.

\PreprintBreak
The production forces approach this equal-lift pattern through different normal and tangential contributions. At $Kn_\infty=8.439$, the direct lift difference between the asymmetric CLL cases, evaluated from the unrounded retained means, is $0.0021$, with a 95\% interval of $[-0.0019,0.0061]$, and is not resolved from zero. CLL $(0.8,0.2)$ combines a normal-force lift contribution of $0.926$ with an opposing shear contribution of $0.066$, whereas CLL $(0.2,0.8)$ combines $1.204$ with $0.347$, giving $C_L=0.859$ and $0.857$, respectively. Their drag coefficients are $0.856$ and $1.423$: the latter is approximately $66\%$ greater, despite the similar lift. The corresponding $L/D$ values are $1.004$ and $0.602$. Similar lift therefore does not imply similar molecular exchange or aerodynamic performance.

Finite temperature breaks the ideal equality even at common collisionless incidence. Using \eqref{eq:finite_temperature_force_coefficients}, the CLL $(0.8,0.2)$ minus $(0.2,0.8)$ reference lift difference is $+0.00512$ at $Kn_\infty=8.439$; both this value and the cold-beam value zero lie inside the production interval. The comparison therefore does not distinguish these two limiting descriptions statistically. At $Kn_\infty=3.146$, the production lifts differ by only $0.46\%$, but their signed difference, $-0.00390$ with interval $[-0.00565,-0.00214]$, is resolved and opposite in sign to the finite-temperature reference $+0.00548$. The exchange symmetry explains the near-equal-lift tendency, not exact equality of the simulated forces. Departures can reflect modified incidence, finite thermal effects and the difference between non-reactive references and the reactive production loads.

The CLL response geometry changes little between the two highest conditions: across the three laws, the largest change in $L/D$ is $0.0033$, and the largest relative change in $C_D$ or $C_L$ is approximately $1.1\%$, measured relative to its value at the lower condition. Most of these changes are sampling-resolved, but none exceeds approximately $2.3\%$ of the corresponding range across all seven wall models at $Kn_\infty=3.146$. Statistical resolution therefore does not imply substantial further physical evolution. The near-collapse is consistent with approach to a high-rarefaction response, but two conditions do not establish an asymptote.

The force projection completes the argument. Population changes can cancel or reverse a conditional wall-law tendency, while a limiting exchange symmetry can preserve lift despite different normal and tangential exchanges. At known incidence, the pair $(C_D,C_L)$ uniquely determines the integrated plate-fixed components, but neither a single coefficient nor, in general, the pair identifies the underlying scattering law and incident distribution. Similar encounter rates can coexist with different loads, and similar lift with different drag. Recovery of selected kernel-level responses therefore constrains the molecular-to-aerodynamic relation without uniquely determining the complete flow.

\section{Conclusions}
\label{SEC:conclu}

The macroscopic response to accommodation is a property of the coupled flow--surface system, not of the scattering kernel alone. The four-condition inclined-flat-plate study combines controlled DSMC comparisons with fixed-incidence references and cross-kernel evaluation of acquired populations. Together, these distinguish the conditional exchange prescribed by the wall from changes in the population on which it acts, and show that the latter can alter the sign as well as the magnitude of the response.

At $Kn_\infty=0.023$, Maxwell shear and heat transfer depart from their endpoint-affine references by up to $40\%$ of the diffuse-reference values. The non-reactive CLL patch analysis shows how a direct wall-law tendency can be obscured or overturned. Under the symmetric allocation, the population change nearly cancels the pressure increase associated with reduced normal accommodation when the two patches are pooled. For reduced tangential accommodation, the population contribution outweighs the negative kernel contributions in the trailing patch, producing net increases in both shear and translational heat transfer relative to CLL $(0.8,0.8)$; the recorded exchanges show the same response signs. A weak net response therefore need not indicate a weak conditional perturbation. Differences from a freestream reference also contain two effects: modified baseline incidence and the additional population change associated with the wall perturbation.

The complementary diagnostics show why neither encounter rate nor distribution width alone identifies the response. At the highest diagnostic condition, nearly common CLL encounter rates coexist with distinct outgoing directions and exchange per molecule. The density contrasts likewise differ in shape even when their integrated magnitudes become comparable. Specular angular width is inherited from incidence, diffuse width is supplied by the kernel and CLL combines both. A recognisable directional signature of scattering can therefore persist even where the macroscopic momentum responses are strongly coupled.

At $Kn_\infty=3.146$ and $8.439$, the principal CLL pressure and shear responses approach their componentwise kernel references and produce nearly orthogonal displacements in the drag--lift plane. This recovery is selective. The total-energy constraint required by common incidence across the three CLL cases and diffuse Maxwell retains a sampling-resolved residual at every condition, with the largest normalised defect at $Kn_\infty=0.137$. At $8.439$, the CLL-implied diffuse heat coefficient still exceeds the separately simulated value by $3.10\%$, despite close agreement of the asymmetric CLL heat-response ratio with its finite-temperature reference. Recovery of a selected response relation therefore does not establish common incidence across wall laws or a free-molecular limit of the complete flow.

A limiting symmetry completes the aerodynamic interpretation. For a common non-reactive beam with negligible freestream and wall thermal speeds, lift depends on the sum of the normal and tangential CLL velocity-retention factors, whereas drag weights them differently. Exchanging the accommodation coefficients therefore preserves lift but not drag. At $Kn_\infty=8.439$, CLL $(0.2,0.8)$ produces approximately $66\%$ more drag than CLL $(0.8,0.2)$, while their lift difference is unresolved. Finite temperature and modified incidence can break the ideal equality, and the small production lift difference at $Kn_\infty=3.146$ remains resolved. The symmetry explains the near-equal-lift tendency without imposing equality on the simulations: similar lift is assembled from different normal and tangential momentum exchanges.

These conclusions concern a reconstructed benchmark and constant-coefficient scattering probes under the prescribed catalytic treatment, not material-calibrated predictions. Temperature and composition vary between benchmark conditions, and the detailed population decomposition covers only the acquired non-reactive patches. A density-only continuation and comparisons with material-constrained kernels would test the broader applicability of the response structure. The present analysis identifies a more discriminating basis for that comparison: assess the conditional molecular exchange, the incidence sustained by the coupled flow and several momentum and energy observables together. Agreement in a load is not, by itself, identification of the wall physics that produced it.

\section*{Declarations and acknowledgements}
\phantomsection
\addcontentsline{toc}{section}{Declarations and acknowledgements}
\markboth{Declarations and acknowledgements}{}
\begin{bmhead}[Supplementary material]
The supplementary material comprises movies~1 and~2.
\end{bmhead}

\begin{bmhead}[Acknowledgements]
E. N{\ae}vdal thanks Astrid N{\ae}vdal, Ingrid N{\ae}vdal and Thorvald oude Egbrink for access to additional computing resources used in part of the production campaign. He also thanks Thorvald oude Egbrink for insightful discussions, Tuana Asu Okt\"{u}rk for careful proofreading and sustained support, and Tor Harald N{\ae}vdal and Silvia Mart\'{\i}nez for their encouragement and practical support throughout the study. S. A. Altmeyer is a Serra H\'{u}nter Fellow.
\end{bmhead}

\begin{bmhead}[Use of generative AI]
OpenAI's ChatGPT 5.6 Sol (accessed July--August 2026) was used for manuscript editing and code assistance. The authors reviewed all AI-assisted material and take full responsibility for the manuscript.
\end{bmhead}

\begin{bmhead}[Funding statement]
This work was supported by grants PID2023-150014OB-C21 and PID2023-150029NB-I00, funded by MICIU/AEI/10.13039/501100011033 and by ERDF/EU.
\end{bmhead}

\begin{bmhead}[Declaration of interests]
The authors report no conflict of interest.
\end{bmhead}

\begin{bmhead}[Data availability statement]
Simulation inputs, digitised reference data, retained production and selected-diagnostic records, processed results, and analysis and figure-generation code are available in the primary numerical archive on Zenodo \citep{Naevdal2026DataRelease}. Movies~1 and~2 accompany the article and are also deposited as separate files in the same record, alongside a higher-quality master version of movie~1. The archived movie~1 workflow records provenance for new visualisation runs; the submitted movie's original simulation frames and contemporaneous generation records are unavailable. For movie~2, the numerical archive includes the producer, inputs, configuration and validation code, but excludes the underlying density histories and run-specific acquisition, processing and rendering records, which are retained separately.
\end{bmhead}

\begin{bmhead}[Author ORCIDs]
E. N{\ae}vdal, \url{https://orcid.org/0000-0001-6549-4353};\newline
S. A. Altmeyer, \url{https://orcid.org/0000-0001-5964-0203}.
\end{bmhead}

\begin{bmhead}[Author contributions]
E. N{\ae}vdal conceived the study, developed the methodology and software, conducted the investigation and validation, performed the formal analysis, curated the data, prepared the visualisations, wrote the original draft and led the revision. S. A. Altmeyer supervised the research, secured funding and contributed to critical review and revision. Both authors approved the final manuscript.
\end{bmhead}
\bibliographystyle{jfm_dsmc}
\bibliography{references}
\end{document}